\documentclass[preprint,aps,prd,nofootinbib]{revtex4-1}
\usepackage{array}
\usepackage{adjustbox}
\usepackage{graphicx} 
\usepackage{amsmath}
\usepackage{amssymb}
\usepackage{bbold}
\usepackage{color}
\usepackage{hyperref}
\usepackage{amsthm}
\usepackage{multirow}
\usepackage{makecell}
\usepackage{tikz}
\usetikzlibrary{quantikz2}
\usepackage{bm}
\usepackage[normalem]{ulem}

\def\bea{\begin{eqnarray}}
\def\eea{\end{eqnarray}}
\def\<{\langle}
\def\>{\rangle}
\def\tr{\text{tr}}

\def\ordr{\mathcal{O}}

\def\be{\begin{equation}}
\def\ee{\end{equation}}

\newcommand{\eq}[1]{Eq.~(\ref{eq:#1})}
\newcommand{\tab}[1]{Tab.~\ref{tab:#1}}
\newcommand{\fig}[1]{Fig.~\ref{fig:#1}}

\begin{document}

\title{Quantum  Magic in High Energy Collisions}

\author{\vspace{0.5cm}Ying-Ying Li$^{\,a}$, Ian Low$^{\,b,c}$, Yi-Lin Wang$^{\,a, d, e}$ and Zhewei Yin$^{\,a}$}
\affiliation{\vspace{0.5cm}
\mbox{$^a$Institute of High Energy Physics, Chinese Academy of Sciences, Beijing 100049, China}\\
\mbox{$^b$Department of Physics and Astronomy, Northwestern University, Evanston, IL 60208, USA}\\ 
\mbox{$^c$High Energy Physics Division, Argonne National Laboratory, Lemont, IL 60439, USA}\\
\mbox{$^d$Interdisciplinary Center for Theoretical Study,}\\
\mbox{University of Science and Technology of China, Hefei, Anhui 230026, China}\\
\mbox{$^e$Peng Huanwu Center for Fundamental Theory, Hefei, Anhui 230026, China}
}

\begin{abstract}

Quantum magic, or nonstabilizerness, is a  quantum resource associated
with computational advantage in quantum systems. In high energy collisions,
Quantum Electrodynamics (QED) is  inefficient at generating magic
 while the weak mixing angle, a fundamental
constant of nature, sits near a value
that minimizes magic production in charged-lepton scattering. These
observations were made in the laboratory (lab) basis, in which spin is projected
along the incoming beam axis. An alternative choice is the helicity basis, in
which spin is projected along the direction of motion of each particle. The
transformation between these two bases is, in general, not a Clifford operation
and therefore can change the amount of magic. We present a detailed study of
magic production in both bases for QED and electroweak processes, and compare
these results with the basis-invariant non-local magic. In the
ultra-relativistic limit, magic production is generally smaller in the
helicity basis due to helicity selection rules, while the lab basis
generally yields less magic in the non-relativistic regime. We provide circuit realizations of the ultra-relativistic  Bhabha amplitudes
using linear combinations of unitaries and show that  the lab basis construction contains  a larger $T$-gate count at
generic scattering angles. Interestingly, in both bases the physical weak
mixing angle lies close to the value that minimizes magic production.

\end{abstract}

\maketitle

\tableofcontents

\section{Introduction}

 Entanglement has long been regarded as the most prominent feature distinguishing quantum from classical physics, and it is often cited as the primary resource enabling quantum computers to outperform classical ones. From the computational point of view, however, entanglement alone is not sufficient to guarantee a quantum advantage. The Gottesman-Knill theorem~\cite{Gottesman:1998se,Aaronson:2004xuh} states that quantum circuits built solely from Clifford gates---the Hadamard, phase, and controlled-NOT gates---can be efficiently simulated on a classical computer, even when the states involved are highly entangled. An additional ingredient is therefore required to access the full computational power of quantum mechanics. This ingredient is non-stabilizerness, or ``magic,'' introduced by Bravyi and Kitaev~\cite{Bravyi:2004isx}, which quantifies the resource that a non-Clifford operation, such as the $T$-gate, must supply for universal quantum computation.

Characterizing and quantifying magic has consequently become an important task. Among the available measures, the stabilizer R\'enyi entropy (SRE)~\cite{Leone:2021rzd} has gained wide use due to its ease of computation and its operational meaning as a resource monotone. In this work we adopt the second-order SRE, $M_2$, as defined in \eq{M2def} below, which vanishes for all stabilizer states and is invariant under Clifford operations. For two-qubit systems, the maximal value attainable is $M_2 = \log(16/7) \approx 0.827$~\cite{Liu:2025frx}, providing a natural yardstick against which to measure the magic generated by a physical process.

Interest in magic, moreover, now extends well beyond quantum computing, where it has become a useful diagnostic of ``quantumness'' across a broad range of physical systems. It has been investigated in quantum phases of matter~\cite{Ellison:2020dkj}, conformal field theories~\cite{White:2020zoz}, the physics of black holes and information scrambling~\cite{Hayden:2007cs,Leone:2022afi}, and models of quantum gravity~\cite{Cepollaro:2024qln}, as well as in the ground states and quench dynamics of quantum many-body systems~\cite{Oliviero:2022euv,Rattacaso:2023kzm,Gu:2024ure}, in the vacuum of quantum field theories~\cite{Nystrom:2024oeq,Cepollaro:2024sod}, and in curved spacetime~\cite{Yang:2025zrl}.

In high-energy physics, quantum-information concepts have likewise opened a new line of inquiry into the structure of fundamental interactions \cite{Fang:2024ple,Afik:2025ejh}. A prominent theme is the connection between the suppression of entanglement generated in scattering and the emergence of enhanced symmetries, first observed in low-energy nucleon-nucleon scattering~\cite{Beane:2018oxh} and subsequently explored in a variety of settings~\cite{Low:2021ufv,Liu:2022grf,Carena:2023vjc,Liu:2023bnr,Hu:2024hex,Kowalska:2024kbs,McGinnis:2025brt,Busoni:2025dns,Carena:2025wyh,Hu:2025lua,McGinnis:2025iab,McGinnis:2025xgt,Li:2026kha,Low:2026oyf,Low:2026evp,Low:2026kxb}, alongside the discovery of an area law for entanglement in particle scattering~\cite{Aoude:2024xpx,Low:2024mrk,Low:2024hvn}. Entanglement has also been used to constrain Standard Model parameters, addressing the structure of electroweak interactions \cite{Cervera-Lierta:2017tdt,Liu:2025pny,Liu:2025iwh,Cao:2026aye} or the pattern of flavor mixing~\cite{Thaler:2024anb}. More recently the emphasis has broadened from entanglement to magic, with studies of non-stabilizerness in nuclear forces~\cite{Robin:2024oqc,Li:2026udy}, dense neutrino systems~\cite{Chernyshev:2024pqy}, top-quark production and the top sector~\cite{White:2024nuc,Aoude:2025jzc}, models of dark matter \cite{Alvarado:2026owd}, gluon and graviton scattering~\cite{Gargalionis:2025iqs,Nunez:2025xds,Gargalionis:2026onv}, spin-0 decays~\cite{Viaux:2026skf}, and deep-inelastic scattering at the Electron-Ion Collider~\cite{Cheng:2025zaw}, together with efforts on obtaining such measures at colliders~\cite{Durupt:2025wuk,Cheng:2026ktp} and broader surveys of quantum complexity in nuclear and high-energy phenomenology~\cite{Robin:2026lqp}. 

Closest to the present work, this line of inquiry has been turned directly on the question of whether the fundamental interactions are special from a computational standpoint. In Ref.~\cite{Liu:2025qfl}, two of the authors of this work initiated a study of magic production in tree-level QED, computing the $M_2$ generated in $2\to2$ scattering of electrons and muons starting from the 60 two-qubit stabilizer states, which themselves carry zero magic. The final-state magic was found to be governed by only a handful of angular patterns, in most cases falling well below the maximal value, with several processes generating no magic at all; the lone exception reaching $\log(16/7)$ was low-energy $\mu^+\mu^- \to e^+e^-$ in the limit $m_e/m_\mu \to 0$. This suggested that QED, despite readily producing maximally entangled states, is an inefficient mechanism for generating quantum advantage. This program is subsequently extended to the electroweak sector \cite{Liu:2025bgw}, showing that minimizing the magic produced in M{\o}ller scattering with respect to the weak mixing angle $s_W^2 = \sin^2\theta_W$ yields a value in striking agreement with the empirical $\hat{s}_W^2(m_Z)$ at the sub-percent level, reinforcing the notion that the Standard Model tends to generate minimal quantum resources.

 A central subtlety underlies these results. Unlike the entanglement of a bipartite pure state, which is invariant under local unitaries, the magic measured by the SRE is built from the expectation values of Pauli strings and is therefore \emph{not} invariant under a general, non-Clifford rotation of the single-qubit basis. The amount of magic ascribed to a scattering final state thus depends on the reference frame and the axis along which the fermion spins are projected to define the qubits. The analyses of Refs.~\cite{Liu:2025qfl,Liu:2025bgw} were carried out in the ``lab basis,''  in which the spins are projected onto the beam ($z$) axis in the center-of-mass frame. A natural and physically distinct alternative is the ``helicity basis,'' in which spins are projected onto the direction of each particle's three-momentum, which is the natural choice for massless particles. This frame dependence was explicitly flagged as an open issue in Ref.~\cite{Liu:2025qfl}, and it raises an immediate question: are the physical conclusions drawn from magic production---the apparent inefficiency of fundamental forces in generating magic, and the magic-minimizing determination of $s_W^2$---robust against the choice of basis, or are they artifacts of a particular frame?

In this work we address this question by systematically studying the basis dependence of magic in two-qubit systems, focusing on the transformation between the lab and helicity bases for $2\to2$ charged-lepton scattering. We first analyze, in Sec.~\ref{sec:basis}, how much magic a change of basis can generate on its own: a single rotation $R_y(\theta)$ acting on the stabilizer states produces only four distinct angular patterns and can generate at most $\log(16/9)$, never saturating the two-qubit maximum. Since the lab-to-helicity map is precisely such a rotation combined with a Clifford operation, part of the difference in magic between the two bases is purely kinematic in origin. We then compute the magic generated for all 60 stabilizer initial states in both bases, for tree-level QED processes in the high-energy and non-relativistic limits (Sec.~\ref{sec:qed}) and for electroweak processes near and above the weak scale (Sec.~\ref{sec:ew}). We find that the magic is generally smaller in the helicity basis at high energies---a feature traceable to the helicity selection rules~\cite{Cheung:2015aba, Azatov:2016sqh, Craig:2019wmo} that suppress many amplitudes in the massless limit---while the lab basis tends to yield smaller magic in the non-relativistic regime. These observations have a direct bearing on the basis-independent, or non-local, magic~\cite{Robin:2025ymq}, which is expected to track the smaller of the two. Importantly, we find that the magic-minimizing value of $s_W^2$ remains close to the empirical weak mixing angle in both bases, increasingly so at higher energies, demonstrating that the electroweak result of Ref.~\cite{Liu:2025bgw} is not an artifact of the lab basis. Finally, we discuss in Sec.~\ref{sec:lcu-circuit} the quantum-circuit realization of the scattering matrices within the Linear Combination of Unitaries framework, where the comparatively sparse helicity-basis amplitude translates into a reduced $T$-gate cost, illustrating a concrete link between the magic content of a process and the resources required to simulate it. Two appendices list the two-qubit stabilizer states and the relevant scattering amplitudes for processes considered in this work.

\section{A Lightning Review on Quantum magic in two-qubit systems}
\label{sec:magic}


We begin with a brief review of magic in quantum information, following Refs.~\cite{Leone:2021rzd,Liu:2025qfl}. We denote the single-qubit Pauli matrices by $\{I, X, Y, Z\} \equiv \{\sigma_0, \sigma_1, \sigma_2, \sigma_3\}$ and define the computational basis $\{|0\rangle, |1\rangle\}$ for a qubit using the eigenstate of $Z$: 
\bea
Z|0\> =+ |0\>, \quad Z|1\> = -|1\>.
\eea
Then Pauli matrices are given by
\bea
I = \left( \begin{array}{cc}
    1 & 0 \\
    0 & 1
\end{array} \right),\quad X = \left( \begin{array}{cc}
    0 & 1 \\
    1 & 0
\end{array} \right), \quad Y = \left(\begin{array}{cr}
    0 & -i \\
    i & 0
\end{array} \right),\quad Z = \left(\begin{array}{cr}
    1 & 0 \\
    0 & -1
\end{array} \right).\label{eq:pmdef}
\eea
In the computational basis the eigenstates of $X$ and $Y$ gates are
\bea
\label{eq:Xeig}
|\pm\>&=&\frac1{\sqrt{2}}(|0\>\pm |1\>) \ , \qquad X|\pm\> = \pm |\pm\>\ , \\
\label{eq:Yeig}
|\!\!\pm\! i\>&=&\frac1{\sqrt{2}}(|0\>\pm i |1\>) \ ,\! \qquad Y|\!\!\pm\! i\> = \pm |\!\!\pm\! i\>\ ,
\eea
The $n$-qubit Pauli group $\mathcal{G}_n$ consists of tensor products of $n$ Pauli matrices dressed with a phase,
\bea
\mathcal{G}_n = \left\{ \phi\, P_1 \otimes P_2 \otimes \cdots \otimes P_n \ \big| \ P_i \in \{I, X, Y, Z\}, \ \phi \in \{\pm 1, \pm i\} \right\},
\eea
where the phase $\phi$ is needed for the Pauli strings to close under multiplication. Any two elements of $\mathcal{G}_n$ either commute or anticommute, and every element squares to $\pm I$.

A stabilizer state $|\psi\>$ is a common $+1$ eigenstate of a maximal abelian subgroup $\mathcal{S} \subset \mathcal{G}_n$, called its stabilizer group. For an $n$-qubit system $\mathcal{S}$ has $2^n$ elements but only $n$ independent generators, and $-I \notin \mathcal{S}$. The unitary operations that preserve the Pauli group under conjugation, $\mathcal{C}_n = \{ U \ | \ U \mathcal{G}_n U^\dagger = \mathcal{G}_n \}$, form the Clifford group and is generated by the Hadamard ($H$), phase ($S$) , and controlled-NOT (CNOT) gates. For two-qubit systems these ``Clifford gates'' are
\be
 H=\frac{1}{\sqrt{2}}
 \begin{pmatrix}1&1\\1&-1\end{pmatrix}
 \ ,\quad S=\begin{pmatrix}1&0\\0&i\end{pmatrix}\ ,\quad \operatorname{CNOT}
=
\begin{pmatrix}
1&0&0&0\\
0&1&0&0\\
0&0&0&1\\
0&0&1&0
\end{pmatrix} \ .
\ee
Acting repeatedly with Clifford gates on a computational-basis product state generates all stabilizer states. The Gottesman-Knill theorem~\cite{Gottesman:1998se,Aaronson:2004xuh} guarantees that any circuit built solely from stabilizer-state preparation, Clifford gates, and Pauli measurements can be simulated efficiently on a classical computer, even when the intermediate states are highly entangled. Accessing the full power of quantum computation therefore requires a non-Clifford operation such as the $T$-gate, 
\be
T=
\begin{pmatrix}
1 & 0\\
0 & e^{i\pi/4}
\end{pmatrix} \ ,
\ee
whose effect can equivalently be supplied by injecting non-stabilizer ``magic'' states~\cite{Bravyi:2004isx}. Magic thus quantifies the genuinely non-classical resource that entanglement alone does not capture. Universal quantum computation requires the presence of either $T$-gate or magic states. 

To quantify magic we adopt the second-order stabilizer R\'enyi entropy (2-SRE)~\cite{Leone:2021rzd}, which vanishes for all stabilizer states. The SRE is invariant under Clifford operations and  additive over tensor products. It is defined for systems of $n$ qubits as
\bea
    M_2 \left(\left|\psi\right\rangle\right) 
    &=& - \log\, \sum_{P\in \mathcal{P}_n} \frac1{2^n}  \left\langle\psi\vert P \vert \psi \right\rangle^{4} \ ,\label{eq:M2def}
\eea
where $\mathcal{P}_n = \left\{P_1 \otimes P_2\otimes\dots\otimes P_n \right\}$ is the $n$-qubit Pauli string, with $  P_i \in 
     \{I, X, Y, Z\}$. By definition, $M_2 = 0$ for any of the 60 stabilizer states $|\psi_{\text{s}} \>_i$, $ i =1, 2, \cdots, 60$ in 2-qubit systems, which we show in Appendix \ref{sec:state-list}. They are the common eigenvectors of maximal abelian subgroups $\mathcal{S}_2$ of $\mathcal{P}_2$. Throughout this work, a two-qubit stabilizer state is denoted by its stabilizer group elements. Signs inside the brackets are assigned independently; e.g. $\langle\pm XX,\pm YY\rangle$ denotes all four sign combinations. The subscript $\< \cdots \>_\pm$ instead denotes a simultaneous sign flip of both generators, e.g. $\langle XX,YY\rangle_{\pm}
\equiv
\left\{
\langle XX,YY\rangle,\,
\langle -XX,-YY\rangle
\right\}$.
The following sequence, whose element appears in Eq. (\ref{eq:M2def}),
\begin{align}
    \rm{spec}(\psi)&\equiv\{\langle \psi|P|\psi\rangle, P\in \mathcal{P}_n \},\label{eq:specdef}
\end{align}
is referred to as the Pauli spectrum of the pure state $|\psi\rangle$. 
For two-qubit systems the maximal attainable value of the 2-SRE is $M^{\rm max}_2 = \log(16/7) \approx 0.827$~\cite{Liu:2025frx}; further discussions on maximal magic and its relation to entanglement can be found in \cite{Roman:2026mcy,Knipfer:2026ixm}.
%

\section{Lab v.s. Helicity Basis}
\label{sec:basis}

\begin{figure}
    \centering
    \includegraphics[width=0.65\linewidth]{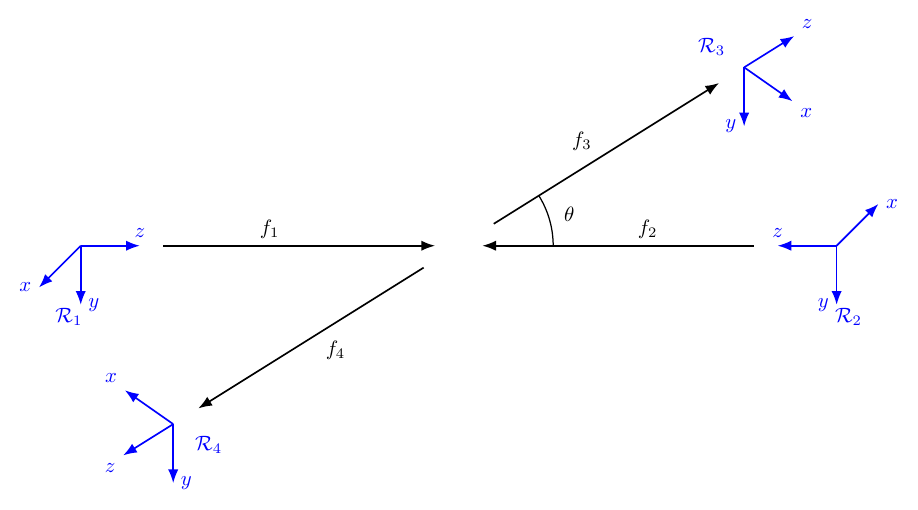}
    \caption{The ``helicity basis'' $\mathcal{R}_i$ for the $i$-th particle projects the spin in the direction of its own motion, whereas in the ``lab basis'' all spins are defined in $\mathcal{R}_1$ and projected in the direction of motion of particle 1.}
    \label{fig:codsys}
\end{figure}

The most common physical realization of a qubit is a spin-1/2 fermion, with the computational basis $\{|0\>, |1\> \}$ defined as the eigenbasis of the spin projection along the $z$-axis of a coordinate system $\mathcal{R}$. The  choice of $\mathcal{R}$ is usually motivated by the physical setting under consideration. In 2-to-2 scatterings of spin-1/2 fermions, there are two natural choices for the coordinate system in which to define the spin projections: the ``lab basis $\mathcal{R}_1$,'' where all spins are projected along the direction of motion of particle 1, and the ``helicity basis $\mathcal{R}_i$,'' where each particle's spin is projected along its own direction of motion. See Fig.~\ref{fig:codsys}.

The lab basis is commonly adopted for massive particles, since the helicity is not a well-defined quantum number. For massless particles, on the other hand, the helicity is a conserved quantum number and the helicity basis is often a convenient choice of basis.  The coordinate transformation between the lab and the helicity bases is not a Clifford operation and, as a consequence, the amount of quantum magic computed in the two reference frames are not the same.  In the following we investigate in detail the change in the SRE of stabilizer states due to coordinate transformations between the lab and the helicity bases. 

Let us first consider a single fermion in a general coordinate system $\mathcal{R}$ and define the computational basis using the eigenstates $|\!\!\pm\!z \>$ of the spin operator $\hat{s}_z $ in the $z$-direction of $\mathcal{R}$:
\bea
| 0 \>_{\mathcal{R}} \equiv |\!\!+\!z\>, \qquad |1 \>_{\mathcal{R}} \equiv |\!\!-\!z\>.\label{eq:1qbob}
\eea
The eigenstates of $X$ and $Y$ are defined in the computational basis in Eqs.~(\ref{eq:Xeig}) and (\ref{eq:Yeig}).\footnote{We have chosen a phase convention in Eq.~(\ref{eq:1qbob}). See a recent discussion in Ref.~\cite{Chu:2026yxm} on the effect of the phase convention.} 
Then a general (pure) spin state $|\psi\>$ for the fermion can be represented  in the reference frame ${\cal R}$ as
\bea
\label{eq:genphi}
|\psi\> = \cos \frac{\theta}2\ | 0 \>_{\mathcal{R}} + e^{i \phi} \sin \frac{\theta}2\ | 1 \>_{\mathcal{R}} \ .
\eea

Next we perform a general rotation of ${\cal R}$ around a unit vector $\hat{\bm{n}}$ by an angle $\alpha$  to arrive at  a different coordinate system $\mathcal{R}'$. The action of the rotation is given by the unitary operator
\bea
R_{n}  (\alpha) = \exp \left(i\frac{\bm{\sigma}\cdot \hat{\mathbf{n}}}{2} \alpha \right) = \cos\frac{\theta}2\ \bm{1} + i \sin\frac{\theta}2 \bm{\sigma}\cdot\hat{\bm{n}}\ .\label{eq:rotdef}
\eea
The two computational bases in ${\cal R}$ and  ${\cal R}^\prime$ are related by
\bea
|a \>_{{\cal R}^\prime} =  R_{n}  (\alpha)\ |a\>_{\cal R} \ , \quad a = 0 , 1 \ .
\eea
where $|0\>_{\mathcal{R}'} = |\!\!+\!z'\>$ and  $|1\>_{\mathcal{R}'} = |\!\!-\!z'\>$ are the eigenstates of the spin operator $\hat{s}_{z'}$ in $\mathcal{R}'$. The same state $|\psi\>$ will thus admit a different representation in $\mathcal{R}'$.

More explicitly, if we rotate ${\cal R}$ around $\hat{\bm{n}}_\phi
= \hat{\bm{x}} \sin \phi  - \hat{\bm{y}} \cos \phi $ by $\theta$,  the state $|0\>_{\cal R}$ becomes
\bea
|0\>_{\cal R}\ \to\  R_{n_\phi}(\theta)\ |0\>_{\cal R'}=  \cos \frac{\theta}2\ |0\>_{{\cal R}^\prime} + e^{i\phi} \sin\frac{\theta}2\ |1\>_{{\cal R}^\prime} .
\eea
Comparing with Eq.~(\ref{eq:genphi}) we see that, starting from $|\!\!+\!z\>$ in the ${\cal R}$ frame, one can arrive at any spinor in the ${\cal R}^\prime$ via rotation. This is the reflection of the statement that a general coordinate transformation is not a Clifford operation and, starting from a zero magic state, one can arrive at any possible values of magic (for a singlet qubit) by rotation -- the quantum magic depends on how one defines the computational basis. 

The fact that the value of magic depends on the choice of reference frame to realize a physical qubit should not come as a surprise. In quantum mechanics the expectation value of a unitary operator depends on the choice of a basis to express the wave function and the 2-SRE defined in Eq.~(\ref{eq:M2def}) computes the magic using the Pauli spectrum -- the expectation values -- of the quantum state. Indeed, most physical measurements require a choice of reference frame and are not invariant under coordinate transformations.



Next we consider realizing two distinguishable qubits using two spin-1/2 fermions $A$ and $B$. In the most general case $A$ and $B$ could have their own reference frames ${\cal R}_A$ and ${\cal R}_B$, respectively, and project their spins to different axes. The two-qubit computational basis is $|a b\>_{\mathcal{R}_A\otimes \mathcal{R}_B} \equiv |a \>_{\mathcal{R}_A} \otimes |b \>_{\mathcal{R}_B}$, $a,b = 0,1.$

The first physical scenario we analyze is when $A$ and $B$ share a common reference frame and project their spins to the same direction, ${\cal R}_A={\cal R}_B={\cal R}$, which is realized in the ``lab basis'' adopted in Refs.~\cite{Liu:2025qfl,Liu:2025bgw}. A rotation of $\mathcal{R}$ around some axis $\hat{\bm{n}}$ by angle $\alpha$ is represented by the following unitary operator:
\begin{align}
 R_n^{\otimes 2} (\alpha)  &={e}^{i\alpha\,{\bm{\sigma} \cdot \hat{\bm{n}}}/{2}}\otimes {e}^{i\alpha\,{\bm{\sigma} \cdot \hat{\bm{n}}}/{2}}\ .
\end{align}
Starting with a certain two-qubit state $\psi_{AB}$, one cannot reach all possible two-qubit states by  $R_n^{\otimes 2} (\alpha)$ because it rotates the two qubits uniformly.

Throughout this work, a two-qubit stabilizer state is denoted by its two stabilizer group elements, see . Signs inside the brackets are chosen independently; e.g., $\langle\pm XX,\pm YY\rangle$ denotes all four sign combinations. A subscript $\pm$ instead denotes a simultaneous sign flip of both generators, $\langle P_1,P_2\rangle_\pm\equiv\{\langle P_1,P_2\rangle,\langle-P_1,-P_2\rangle\}$.

\begin{figure}[htbp]
    \centering

    \begin{minipage}{0.3\textwidth}
        \centering
        \includegraphics[width=\linewidth]{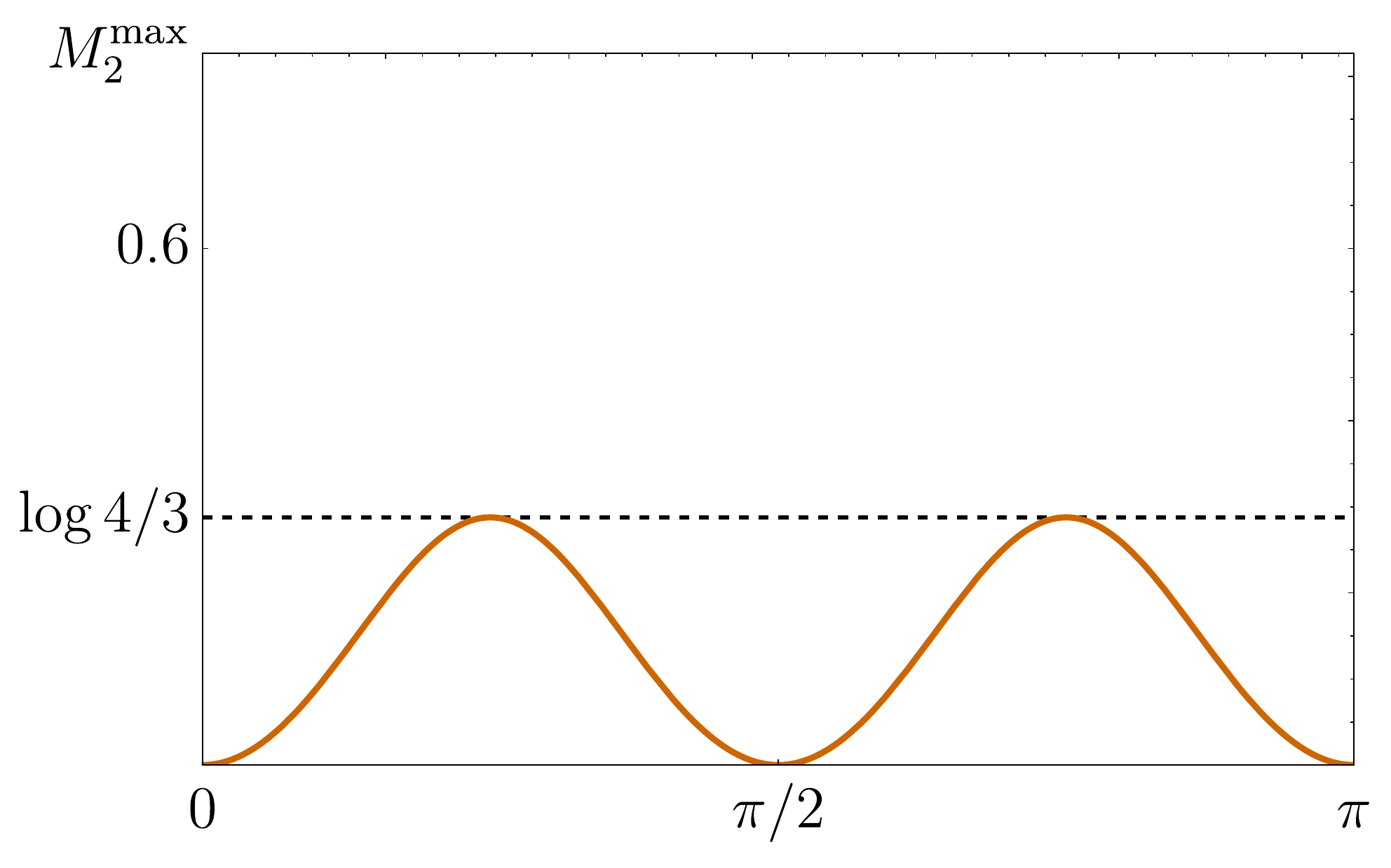}
    \end{minipage}
    \hfill
    \begin{minipage}{0.3\textwidth}
        \centering
        \includegraphics[width=\linewidth]{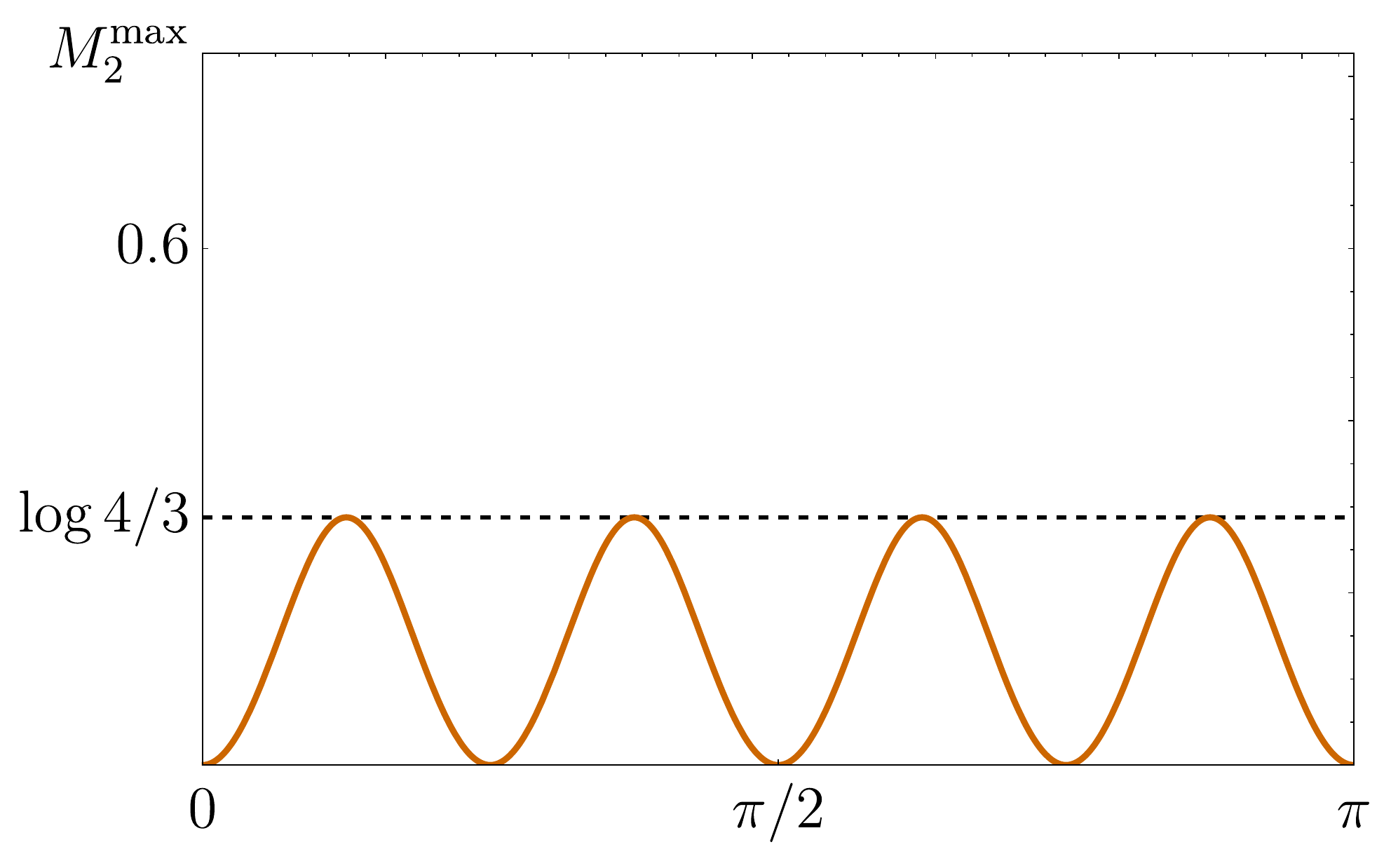}
    \end{minipage}
    \hfill
    \begin{minipage}{0.3\textwidth}
        \centering
        \includegraphics[width=\linewidth]{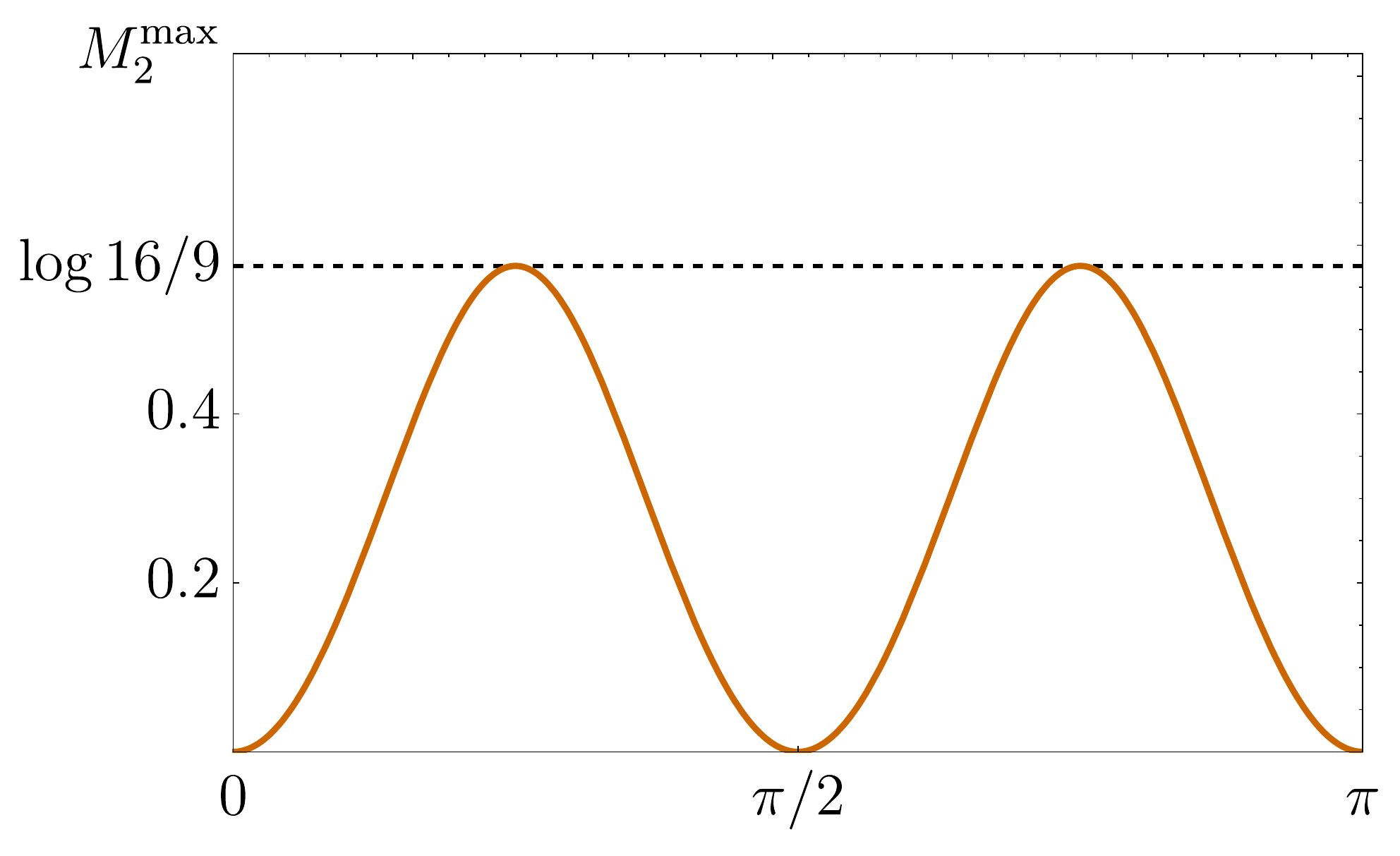}
    \end{minipage}

    \caption{The non-zero angular distributions generated by the rotation $R_y^{\otimes 2} (\theta)$, where we plot $M_2$ versus $\theta$.}
    \label{fig:rot}
\end{figure}

As a warm-up exercise, which will be useful for our discussions later, we can now compute how much magic can be generated by going from $\mathcal{R}$ to $\mathcal{R}'$ through the rotation $R_y^{\otimes 2}(\theta)$. The 60 stabilizer states $\{ |\psi_{\text{s}} \>_i \}$ in $\mathcal{R}$ will have 4 different kinds of magic dependence on $\theta$ in $\mathcal{R}'$, as presented in Table \ref{tab:rotyss}. We see that 8 of the states remains to be stabilizer states in $\mathcal{R}'$. For the three non-zero distributions of $M_2$, which we plot in Fig. \ref{fig:rot}, two of them reaches a largest value of $\log (4/3)$, while the other reaches $\log (16/9)$; none of them can saturate the maximal magic $\log (16/7)$ for two qubit systems. 
\begin{table}
    \centering
    \begin{tabular}{|c|c|}
    \hline
    $\exp [-M_2 ( R_y^{(2)} (\theta) |\psi_{\text{s}}\>_i) ]$    & $|\psi_{\text{s}}\>_i$ \\
    \hline
    1     & $\langle \pm IY,\pm YI\rangle$, $\langle -XX,\pm YY\rangle$, $\langle XZ,-YY\rangle$, $\langle -XZ,YY\rangle$\\
    \hline
    $\frac{1}{8}(7+\cos{4\theta})$ & $\langle \pm IX,\pm YI\rangle$,$\langle \pm IY,\pm XI\rangle$, $\langle \pm IY,\pm ZI\rangle$, $\langle \pm IZ,\pm YI\rangle$\\
    \hline
    $\frac{1}{8}(7+\cos{8\theta})$ & $\langle XX, \pm YY\rangle$, $\langle XZ,YY\rangle$, $\langle -XZ,-YY\rangle$\\
    \hline
    $\frac{1}{64}(7+\cos{4\theta})^2$ & $\langle \pm IX,\pm XI\rangle$, $\langle \pm IX,\pm ZI\rangle$, $\langle \pm IZ,\pm XI\rangle$, $\langle \pm IZ,\pm ZI\rangle$, \\
    & $\langle \pm XY, \pm YX\rangle$, $\langle \pm XZ,\pm YX\rangle$, $\langle\pm XY, \pm YZ\rangle$, $\langle \pm XX,\pm YZ\rangle$
    \\
    \hline
    \end{tabular}
    \caption{The magic generated by the rotation $R_y^{\otimes 2} (\theta)$ for the 60 stabilizer states in the original basis.}
    \label{tab:rotyss}
\end{table}

Now, as we would like to study states generated by  $2\to 2$ scatterings $f_1 f_2 \to f_3 f_4$ of spin-$1/2$ fermions, we will be particularly interested in the transformation between the lab and the helicity bases.
To define them, it is useful to define coordinate systems $\mathcal{R}_i$ associated with $f_i$, in the center-of-mass (CM) frame, as shown in Fig. \ref{fig:codsys}: The $z$-axis in $\mathcal{R}_i$ is always in the direction of the momentum of $f_i$, while the $y$-axis is always normal to the scattering plane. In the lab basis $\textbf{L}$, the computational bases for all initial and final state particles are defined according to the same coordinate frame $\mathcal{R}_1$, while in the helicity basis $\textbf{H}$, each of the particles in the scattering process uses its own distinct frame $\mathcal{R}_i$ to define the computational basis. 


For the  initial particles $f_1$ and $f_2$, the relation between the two bases is simple: Going from $\textbf{L}$ to $\textbf{H}$ amounts to changing the coordinate frame for $f_2$ from $\mathcal{R}_1$ to $\mathcal{R}_2$, which is a rotation of $\pi$ among their commonly shared $y$-axis. The relation between a pure state $\psi$ in the different bases  is thus given by
\bea
|{\psi}\rangle_{\textbf{H}}&=\hat{R}_\text{i} \, |{\psi}\rangle_{\textbf{L}}, \quad \hat{R}_\text{i} \equiv I\otimes {R}_{y}(\pi), \quad
    R_y(\pi) =  {e}^{i\pi {\sigma_y}/{2}} = i\, Y .
\eea
When computing the Pauli spectrum in Eq. (\ref{eq:specdef}), and consequently the SRE in Eq. (\ref{eq:M2def}), $\hat{R}_\text{i}$ in the above will sandwich the Pauli string elements: $P\rightarrow(\text{I}\otimes Y)P(\text{I}\otimes Y)=P_1 \otimes(Y P_2 Y)$. This is a 2-qubit Clifford operation, and its effect is a reshuffling of the Pauli spectrum. Consequently, $\hat{R}_\text{i}$ will not change magic, and an initial stabilizer state in $\textbf{L}$ is also an initial stabilizer state in $\textbf{H}$. 
Using $YXY=-X$ and $YZY=-Z$, the correspondence between the 60 initial states in the two bases is shown in Table~\ref{tab:correspondence}, where we denote each stabilizer state by its unique stabilizer group and list only those states which differ in the two bases.

\begin{table*}[t]
\centering

\begin{minipage}{0.48\textwidth}
\centering
\begin{tabular}{|c|c|}
\hline
$\mathcal{S}_{\rm H}$ & $\mathcal{S}_{\rm L}$\\
\hline
$\langle \pm IX,\pm XI\rangle$ & $\langle \mp IX,\pm XI\rangle$\\
\hline
$\langle \pm IX,\pm YI\rangle$ & $\langle \mp IX,\pm YI\rangle$\\
\hline
$\langle \pm IX,\pm ZI\rangle$ & $\langle \mp IX,\pm ZI\rangle$\\
\hline
$\langle \pm IZ, \pm XI\rangle$ & $\langle \mp IZ, \pm XI\rangle$\\
\hline
$\langle \pm IZ, \pm YI\rangle$ & $\langle \mp IZ, \pm YI\rangle$\\
\hline
$\langle \pm IZ, \pm ZI\rangle$ & $\langle \mp IZ, \pm ZI\rangle$\\
\hline
\end{tabular}
\end{minipage}
\hfill
\begin{minipage}{0.48\textwidth}
\centering
\begin{tabular}{|c|c|}
\hline
$\mathcal{S}_{\rm H}$ & $\mathcal{S}_{\rm L}$\\
\hline
$\langle \pm XX,\pm YY\rangle$ & $\langle \mp XX,\pm YY\rangle$\\
   \hline
   $\langle \pm XY,\pm YX\rangle$ & $\langle \pm XY,\mp YX\rangle$\\
   \hline
   $\langle \pm XZ,\pm YY\rangle$ & $\langle \mp XZ,\pm YY\rangle$\\
   \hline
   $\langle \pm XZ, \pm YX\rangle$ & $\langle \mp XZ, \mp YX\rangle$\\
   \hline
   $\langle \pm XY,\pm YZ\rangle$ & $\langle \pm XY,\mp YZ\rangle$\\
   \hline
   $\langle \pm XX,\pm YZ\rangle$ & $\langle \mp XX,\mp YZ\rangle$\\
   \hline
\end{tabular}
\end{minipage}
\caption{The correspondence between stabilizer states related by $\hat{R}_\text{i}$, which takes the helicity basis for initial particles to the lab basis. We denote each stabilizer state by its stabilizer group and list only the states which differ in the two bases.}
\label{tab:correspondence}

\end{table*}

For the 2-qubit final state $\text{f}$, the coordinate transformation for going from $\textbf{L}$ to $\textbf{H}$ is the rotation along the $y$-axis, by an angle $\theta$ for $f_3$ and $\theta + \pi$ for $f_4$:
\begin{align}
    |\text{f}\rangle_{\textbf{H}}& = \hat{R}_\text{f}\ |\text{f}\rangle_{\textbf{L}},
    \quad \hat{R}_\text{f} = R_y(\theta)\otimes R_y(\theta + \pi) =  \text{e}^{i\frac{\sigma_y}{2}\theta} \otimes \text{e}^{i\frac{\sigma_y}{2}(\theta+\pi)}=\hat{R}_\text{i} R_y^{\otimes 2} (\theta).
\end{align}
As we have discussed earlier, the $\hat{R}_\text{i}$ appearing in the equation above will sandwich the Pauli strings first in the computation of magic, thus it will not change magic, and the effect of $\hat{R}_\text{f}$ on the SRE is thus equivalent to a pure rotation $R_y^{\otimes 2} (\theta)$. Under such a coordinate transformation, the final state magic will in general get modified.

The basis dependence discussed above motivates a basis-\emph{independent} characterization of the magic carried by a two-qubit state, obtained by minimizing $M_2$ over all local (single-qubit) reparameterizations of the computational basis. This is the \emph{non-local magic}~\cite{Robin:2025ymq,Gargalionis:2026onv}, which we define as
\bea
M_2^{\text{NL}} (|\psi\>) \equiv \min_{R_\text{A} \otimes R_\text{B}} M_2 \left[ (R_\text{A} \otimes R_\text{B}) |\psi\> \right], \label{eq:M2NL}
\eea
where $R_\text{A}$ and $R_\text{B}$ are independent single-qubit rotations of the form $R_n (\alpha)$ in \eq{rotdef}. Since redefining the computational basis of each qubit acts precisely as such a rotation, $M_2^{\text{NL}}$ is by construction invariant under local basis changes---including the lab-to-helicity transformation discussed above---and returns the smallest magic attainable across all choices of frame. It therefore provides a common, frame-independent baseline against which the lab- and helicity-basis results of the following sections can be compared: the magic measured in either basis can never fall below $M_2^{\text{NL}}$.

For a two-qubit \emph{pure} state, the minimization in \eq{M2NL} does not need to be carried out explicitly, as it is fixed entirely by the entanglement of $|\psi\>$~\cite{Gargalionis:2026onv,Busoni:2026lvp}. The result can be expressed directly through the reduced density matrix $\rho_\text{A} = \tr_\text{B} |\psi\>\<\psi|$ of either qubit. In terms of the concurrence \cite{Hill:1997pfa,Wootters:1997id}
\bea
\chi = \sqrt{2 \left( 1 - \tr \rho_\text{A}^2 \right)},
\eea
the non-local magic reads \cite{Gargalionis:2026onv,Busoni:2026lvp}
\bea
M_2^{\text{NL}} (|\psi\>) = - \log \left( 1 - \chi^2 + \chi^4 \right). \label{eq:M2NLchi}
\eea
Equivalently, it is set by the \emph{anti-flatness} of the entanglement spectrum~\cite{Robin:2025ymq},
\bea
F_\text{A} (|\psi\>) = \tr \rho_\text{A}^3 - \left( \tr \rho_\text{A}^2 \right)^2, \label{eq:antiflat}
\eea
with $M_2^{\text{NL}} = -\log (1 - 4F_\text{A})$. Both statements express the same fact: the non-local magic of a two-qubit pure state is a function of its entanglement alone, vanishing for product states ($\chi = 0$) and for maximally entangled states ($\chi = 1$)---consistent with all 60 stabilizer states carrying $M_2^{\text{NL}} = 0$---and peaking at $M_2^{\text{NL}} = \log (4/3)$ for $\chi^2 = 1/2$ in between.

\section{Magic production in QED processes for different bases}

\label{sec:qed}

\begin{figure}[t]
\centering
\includegraphics[width=0.8\textwidth]{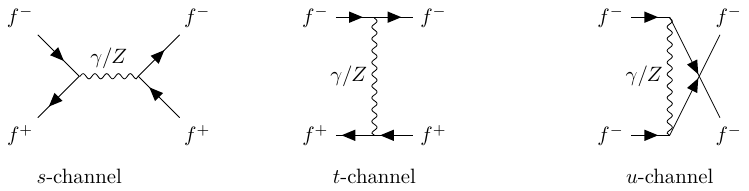}
\caption{
Representative $s/t/u$-channel Feynman diagrams in the fermionic $2\to 2$ scattering in electroweak theories.}
\label{fig:feynman_diagram}
\end{figure}

We study the basis dependence of magic generation in fermionic $2 \to 2$ scattering processes in tree-level QED, considering both the ultra-relativistic and non-relativistic limits. The ultra-relativistic limit corresponds to 
$|\boldsymbol{p}|/m \gg 1$, while the non-relativistic limit corresponds to
$|\boldsymbol{p}|/m \ll 1$, 
where $\boldsymbol{p}$ and $m$ denote the three-momentum and mass of the final-state particles, respectively. In particular, we focus on four representative processes: \textbf{Process (A)} Bhabha scattering, $e^+ e^- \to e^+ e^-$, which receives contributions from both the $s$- and $t$-channel amplitudes; \textbf{Process (B)} M{\o}ller scattering, $e^- e^- \to e^- e^-$, which proceeds via the $t$-channel and $u$-channel amplitudes; \textbf{Process (C)} $e^- \mu^- \to e^- \mu^-$, which proceeds only via the $t$-channel process; and \textbf{Process (D)} $e^+ e^- \to \mu^+ \mu^-$ and $\mu^+ \mu^- \to e^+ e^-$, which occur exclusively through the $s$-channel. The representative $s/t/u$-channel Feynman diagrams are shown in \fig{feynman_diagram}.

In the center-of-mass frame with total energy $\sqrt{s}$, the two incoming particles carry equal and opposite momenta along the $z$ direction. We evaluate the magic $M_{2, i} \left(\theta\right)$ of the final state generated from each stabilizer state $\left|\psi_s\right\rangle_i$, as a function of the scattering angle $\theta$, both in the lab basis and in the helicity basis. We also compare the final state magic averaged over the scattering angle for each stabilizer state $|\psi_{\text{s}}\>_{i}$
\bea
\<M_{2,i}\> &\equiv& \frac1{4\pi} \int  M_{2, i} \left(\theta\right)\sin\theta\, d\theta\,d\phi\,,
\label{eq:averageOtheta}
\eea
as well as the average over all the 60 stabilizer states taken as the initial state,
\bea
\<{\cal M}_2\> &\equiv& \frac{1}{60} \sum_{i=1}^{60}   \<M_{2,i}\>,
\label{eq:averageM}
\eea
in both bases.
\subsection{Ultra-relativistic Limit}
\label{sec:high-energy-limit}
\begin{table}
\centering
\begin{tabular}{|c|c|c|c|}
\hline
\multicolumn{2}{|c|}{$\mathcal{S}_{\rm H}$} & $e^+e^-\to e^+e^-$ & $e^-e^-\to e^-e^-$ \\
\hline
G1&\adjustbox{valign=c}{%
  \begin{tabular}{@{}c@{}}
    $\langle IX,XI\rangle_\pm,\langle XX,YZ\rangle_\pm$
  \end{tabular}
}
&
\adjustbox{valign=c}{%
  \includegraphics[width=0.28\linewidth]{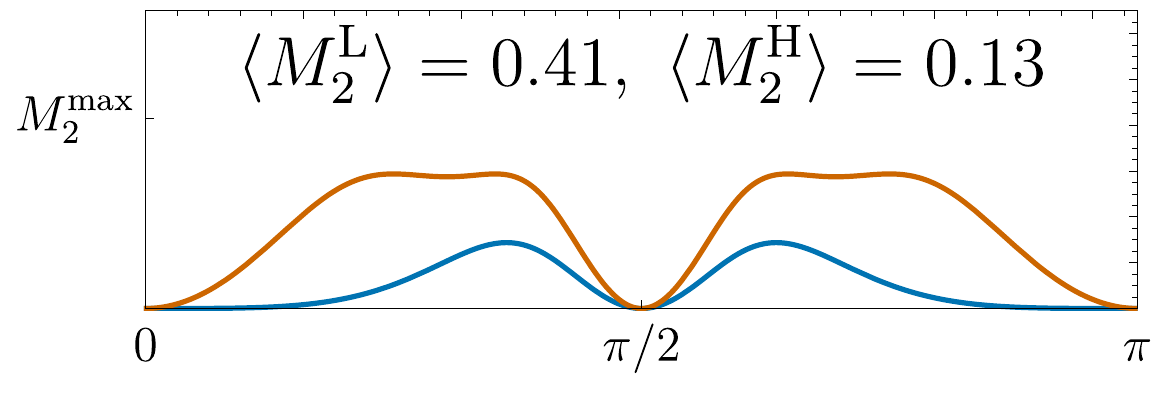}}
&
\adjustbox{valign=c}{%
  \includegraphics[width=0.28\linewidth]{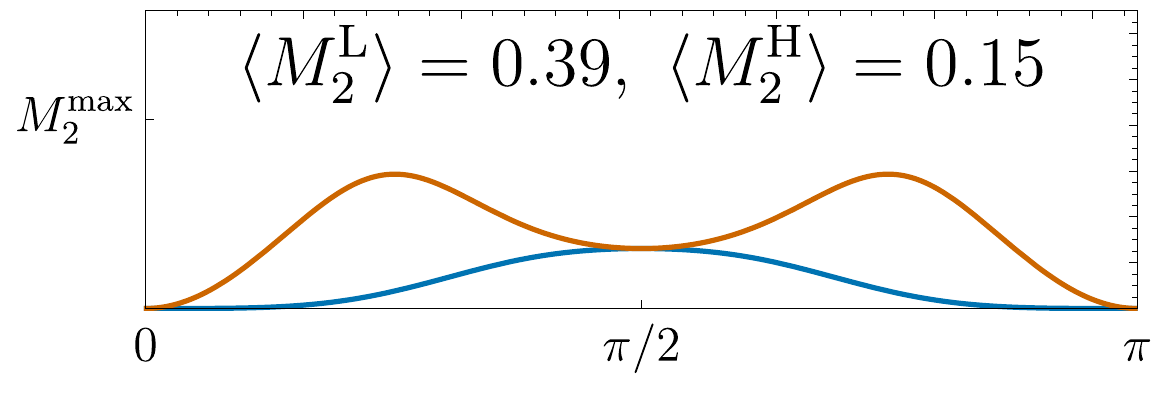}
}
\\
\hline
G2&\adjustbox{valign=c}{%
  \begin{tabular}{@{}c@{}}
    $\langle IX,-XI\rangle_\pm,\langle XX,-YZ\rangle_\pm$
  \end{tabular}
}
&
\adjustbox{valign=c}{%
  \includegraphics[width=0.28\linewidth]{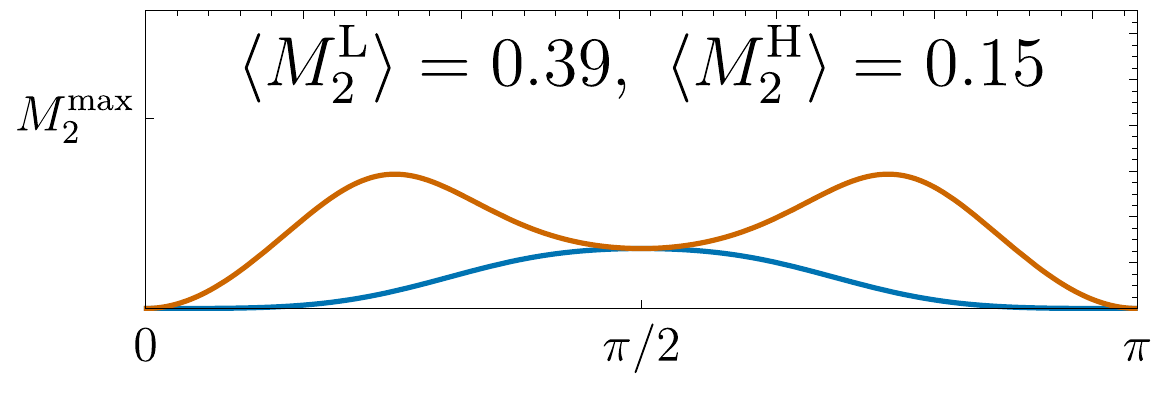}}
&
\adjustbox{valign=c}{%
  \includegraphics[width=0.28\linewidth]{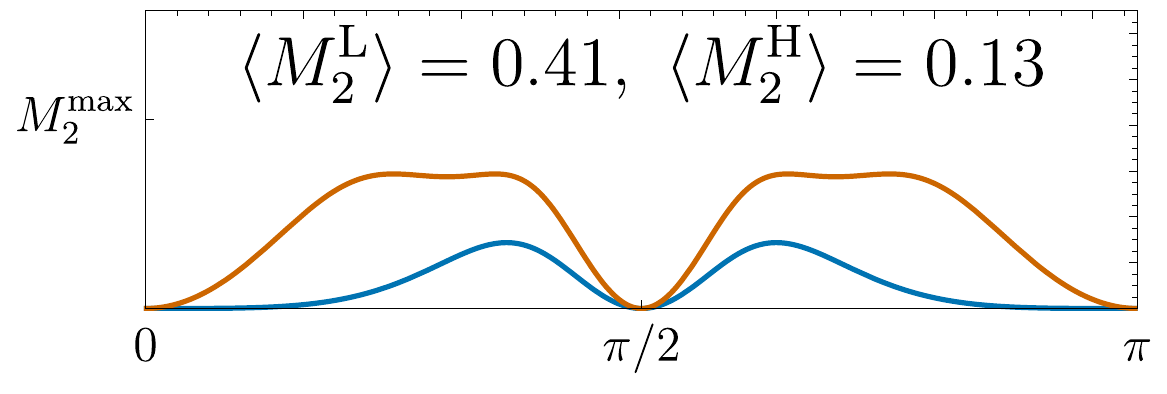}
}
\\
\hline

G3&\adjustbox{valign=c}{%
  \begin{tabular}{@{}c@{}}
    $\langle \pm IX,\pm YI\rangle,\langle \pm IY,\pm XI\rangle$\\
    $\langle \pm IY,\pm ZI\rangle,\langle \pm IZ,\pm YI\rangle$
  \end{tabular}
}
&
\adjustbox{valign=c}{%
  \includegraphics[width=0.28\linewidth]{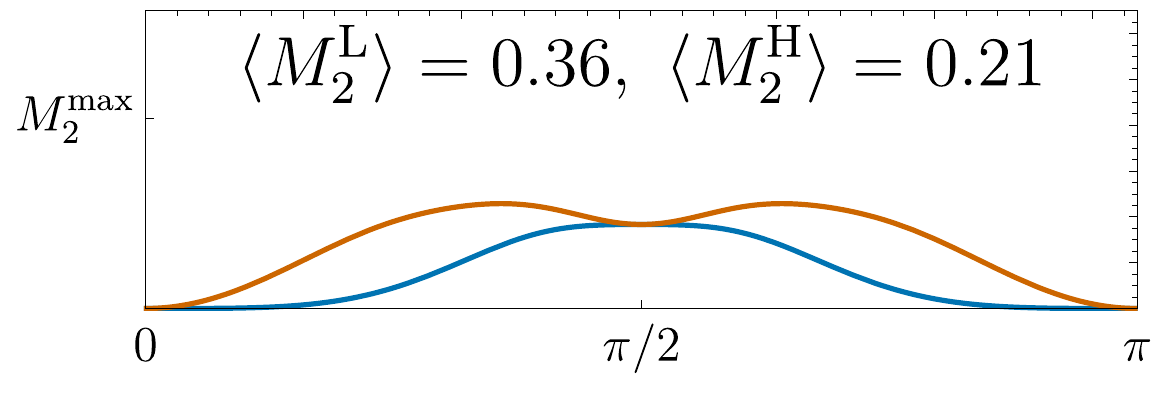}}
&
\adjustbox{valign=c}{%
  \includegraphics[width=0.28\linewidth]{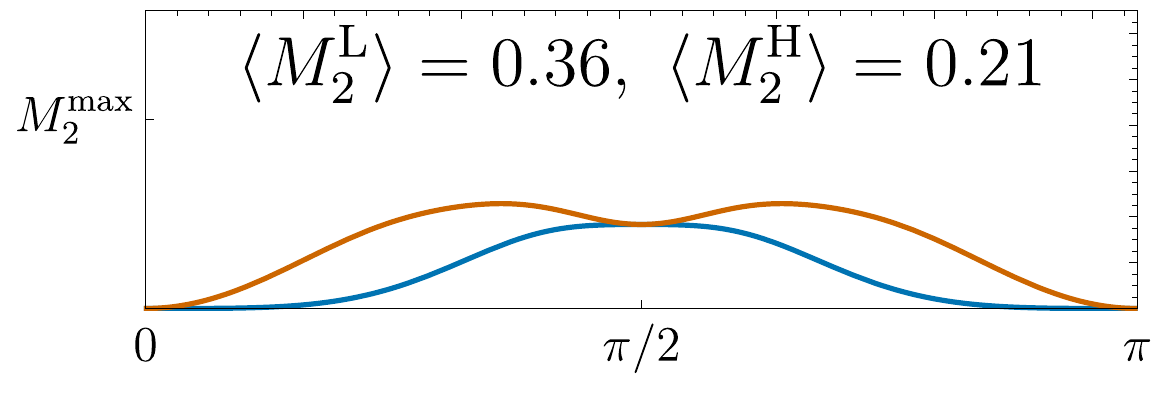}
}
\\
\hline

G4&\adjustbox{valign=c}{%
  \begin{tabular}{@{}c@{}}
    $\langle IX,ZI\rangle_\pm,\langle IZ,XI\rangle_\pm$\\
    $\langle \pm XZ,-YX\rangle,\langle XY,\pm YZ\rangle$
  \end{tabular}
}
&
\adjustbox{valign=c}{%
  \includegraphics[width=0.28\linewidth]{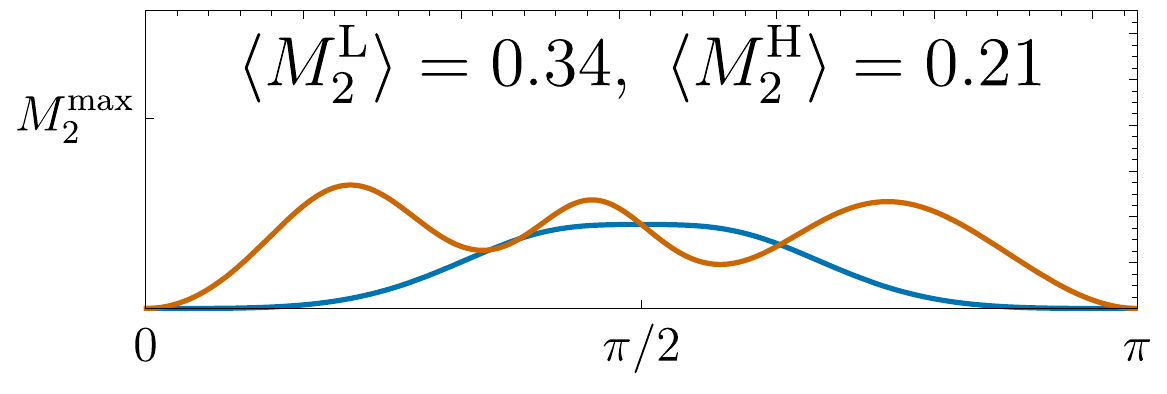}}
&
\adjustbox{valign=c}{%
  \includegraphics[width=0.28\linewidth]{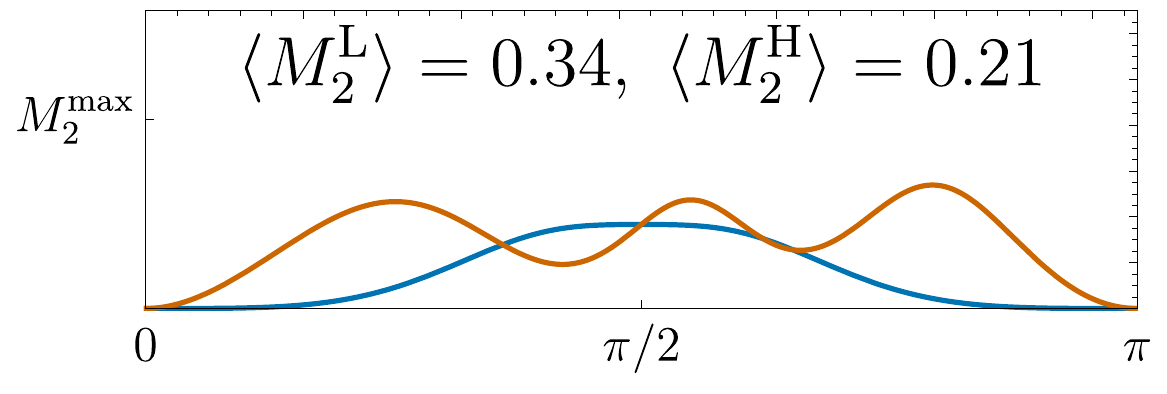}
}
\\
\hline

G5&\adjustbox{valign=c}{%
  \begin{tabular}{@{}c@{}}
    $\langle IX,-ZI\rangle_\pm,\langle IZ,-XI\rangle_\pm$\\
    $\langle \pm XZ,YX\rangle,\langle -XY,\pm YZ\rangle$
  \end{tabular}
}
&
\adjustbox{valign=c}{%
  \includegraphics[width=0.28\linewidth]{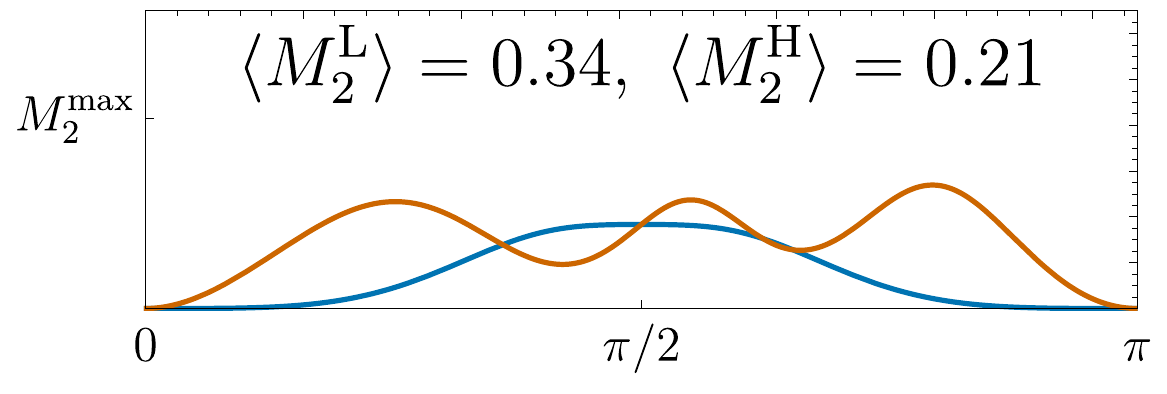}}
&
\adjustbox{valign=c}{%
  \includegraphics[width=0.28\linewidth]{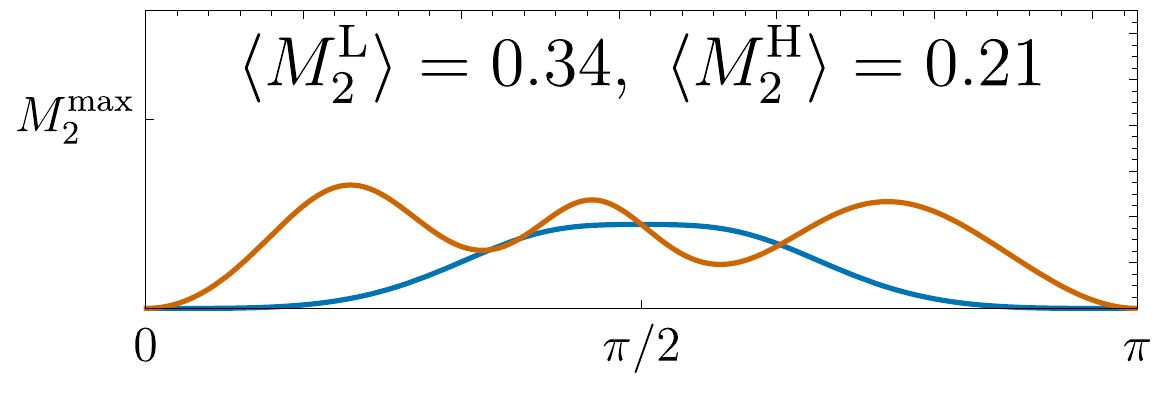}
}
\\
\hline

G6&\adjustbox{valign=c}{%
  \begin{tabular}{@{}c@{}}
    $\langle IY,YI\rangle_\pm$
  \end{tabular}
}
&
\adjustbox{valign=c}{%
  \includegraphics[width=0.28\linewidth]{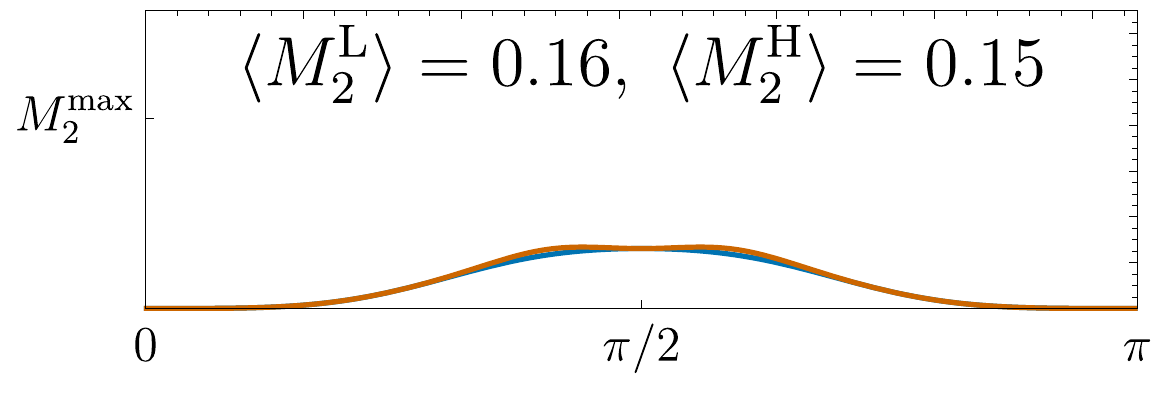}}
&
\adjustbox{valign=c}{%
  \includegraphics[width=0.28\linewidth]{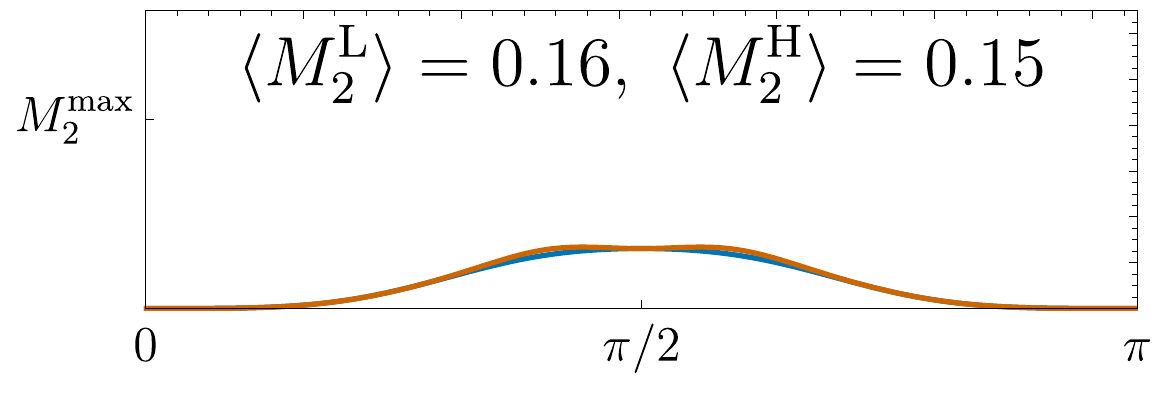}
}
\\
\hline
G7&\adjustbox{valign=c}{%
  \begin{tabular}{@{}c@{}}
    $\langle IY,-YI\rangle_\pm,\langle XZ,-YY\rangle_\pm$
  \end{tabular}
}
&
\adjustbox{valign=c}{%
  \includegraphics[width=0.28\linewidth]{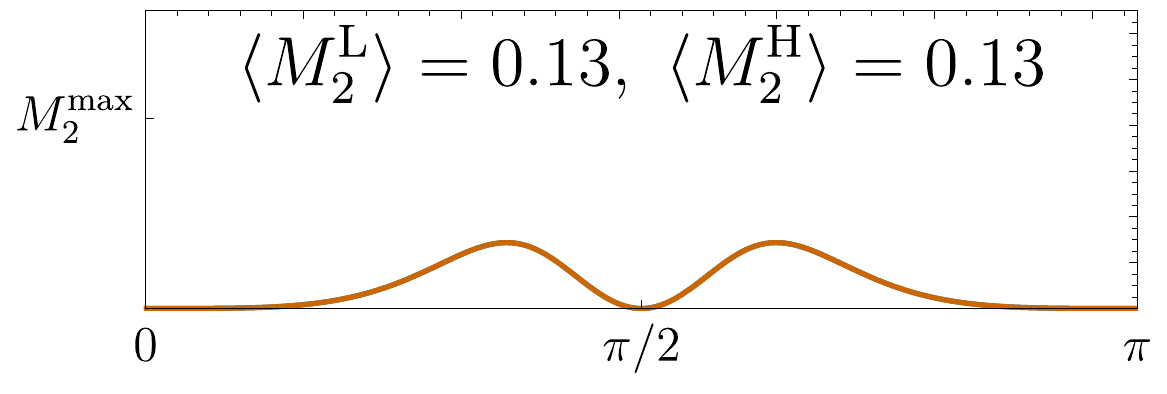}}
&
\adjustbox{valign=c}{%
  \includegraphics[width=0.28\linewidth]{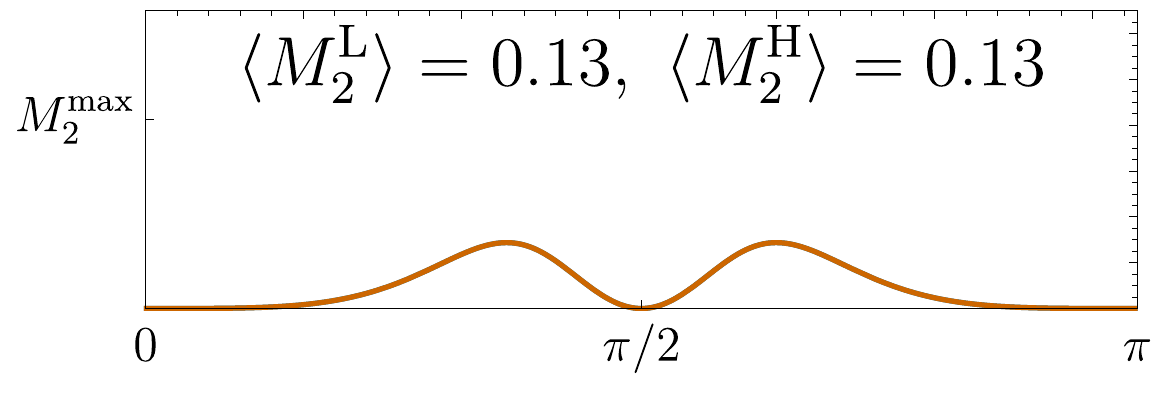}
}
\\
\hline
G8&\adjustbox{valign=c}{%
  \begin{tabular}{@{}c@{}}
   $\langle IZ,ZI\rangle_\pm,\langle XY,YX\rangle_\pm$
  \end{tabular}
}
&
\adjustbox{valign=c}{%
  \includegraphics[width=0.28\linewidth]{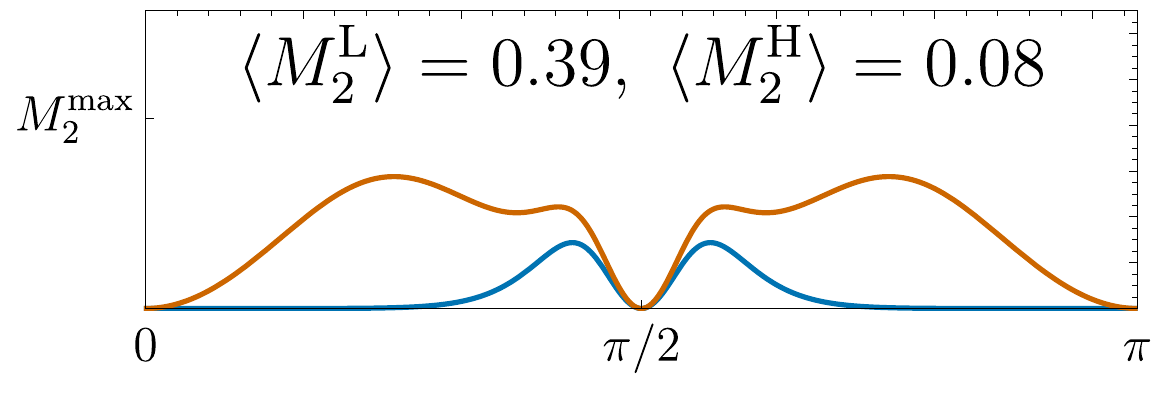}}
&
\adjustbox{valign=c}{%
  \includegraphics[width=0.28\linewidth]{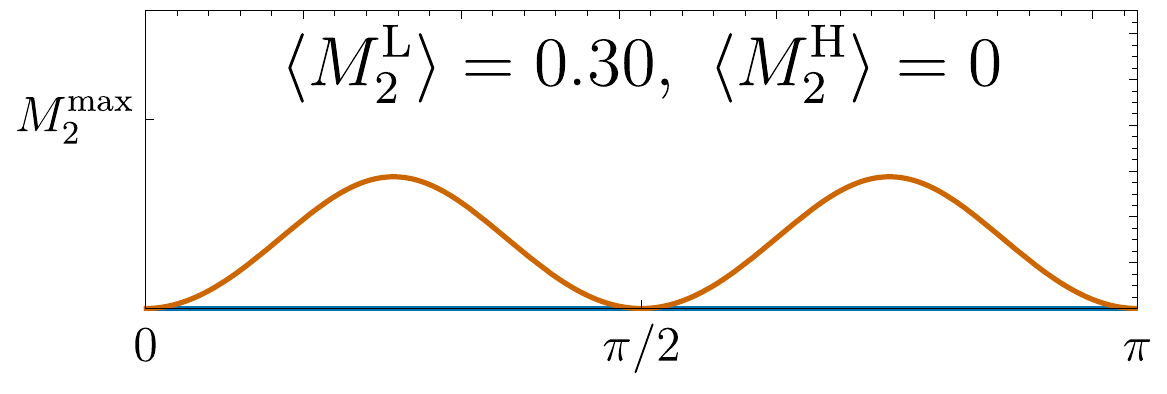}
}
\\
\hline
G9&\adjustbox{valign=c}{%
  \begin{tabular}{@{}c@{}}
   $\langle IZ,-ZI\rangle_\pm,\langle XY,-YX\rangle_\pm$
  \end{tabular}
}
&
\adjustbox{valign=c}{%
  \includegraphics[width=0.28\linewidth]{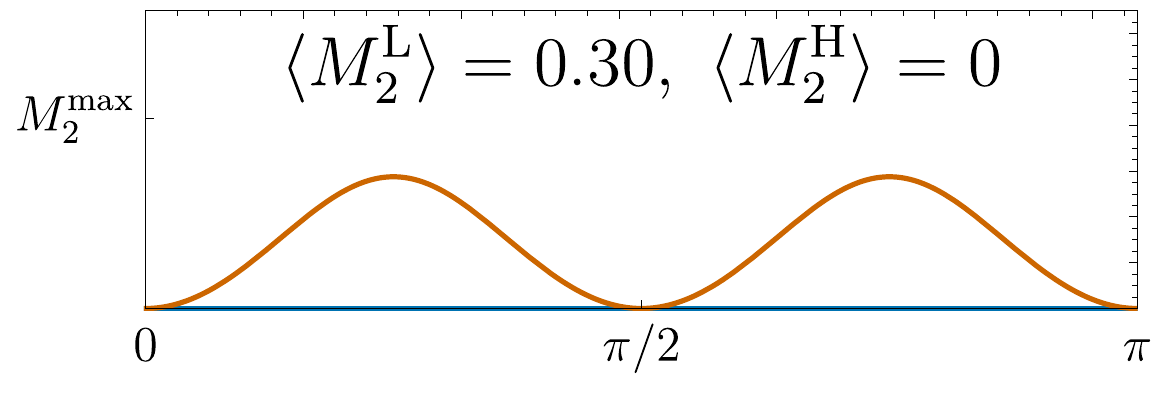}}
&
\adjustbox{valign=c}{%
  \includegraphics[width=0.28\linewidth]{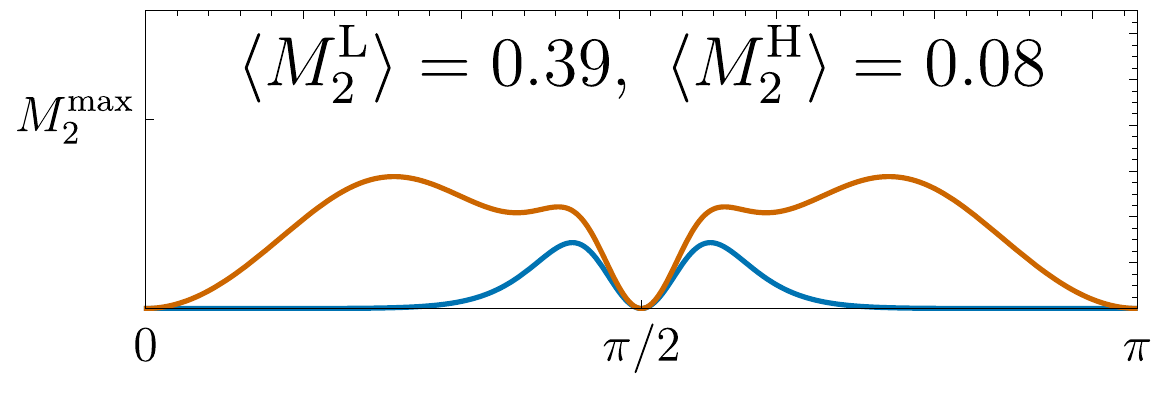}
}
\\
\hline
G10&\adjustbox{valign=c}{%
  \begin{tabular}{@{}c@{}}
   $\langle XX,\pm YY\rangle$
  \end{tabular}
}
&
\adjustbox{valign=c}{%
  \includegraphics[width=0.28\linewidth]{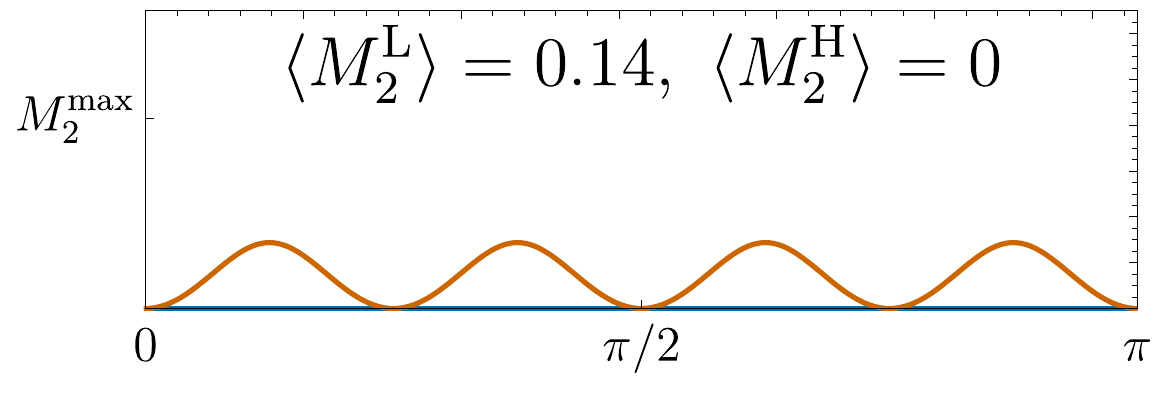}}
&
\adjustbox{valign=c}{%
  \includegraphics[width=0.28\linewidth]{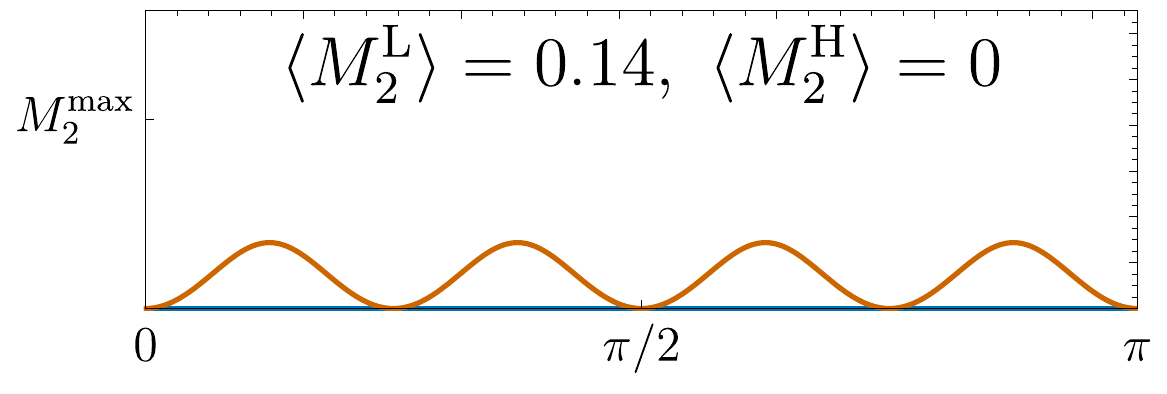}
}
\\
\hline

G11& $\langle -XX,\pm YY\rangle$
& \multicolumn{2}{c|}{$M_2^{\rm L}=M_2^{\rm H}=0$}
\\
\hline

G12&\adjustbox{valign=c}{%
  \begin{tabular}{@{}c@{}}
   $\langle XZ,YY\rangle$
  \end{tabular}
}
&
\adjustbox{valign=c}{%
  \includegraphics[width=0.28\linewidth]{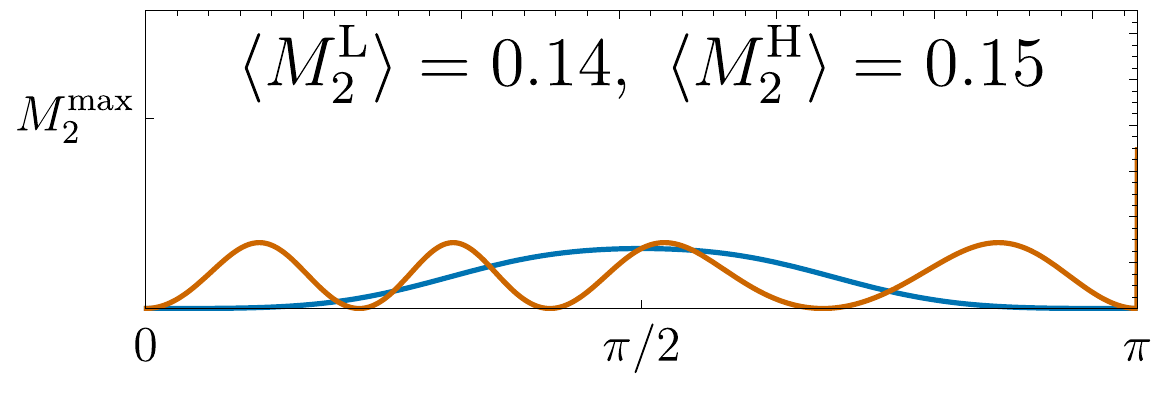}}
&
\adjustbox{valign=c}{%
  \includegraphics[width=0.28\linewidth]{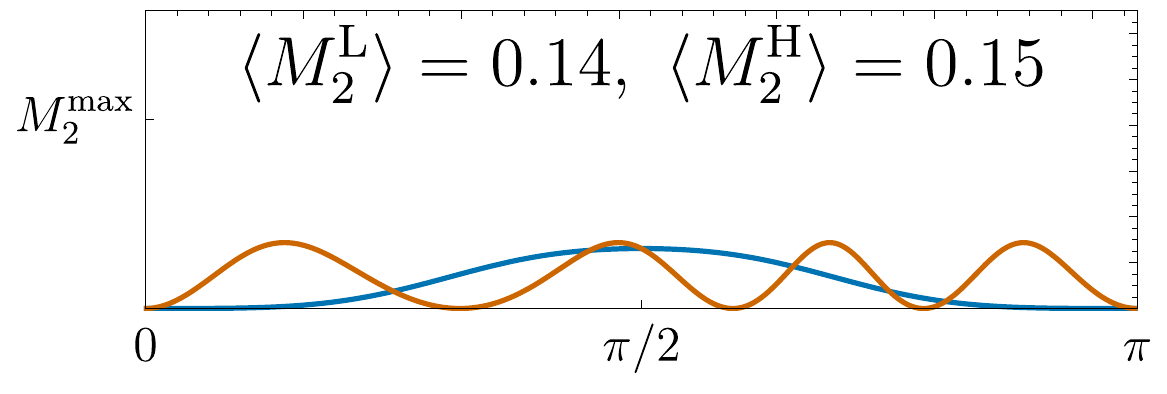}
}
\\
\hline
G13&\adjustbox{valign=c}{%
  \begin{tabular}{@{}c@{}}
  $\langle -XZ,-YY\rangle$
  \end{tabular}
}
&
\adjustbox{valign=c}{%
  \includegraphics[width=0.28\linewidth]{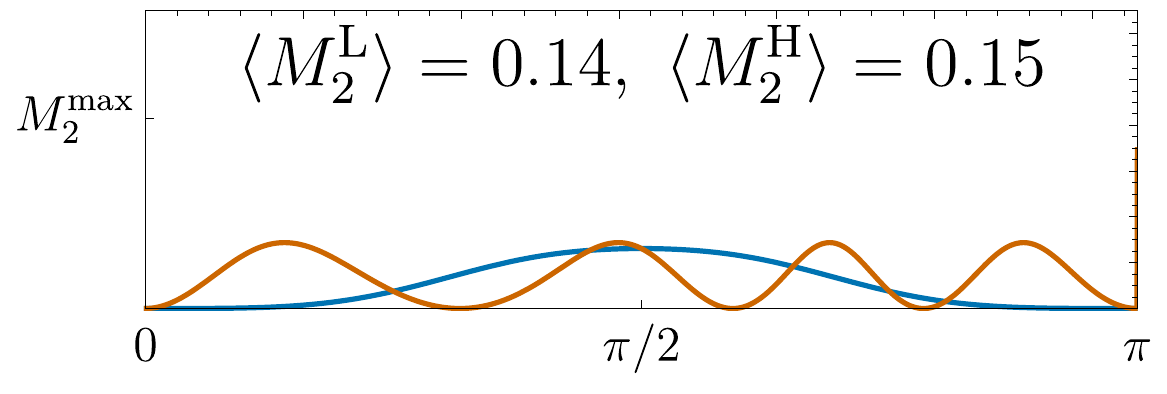}}
&
\adjustbox{valign=c}{%
  \includegraphics[width=0.28\linewidth]{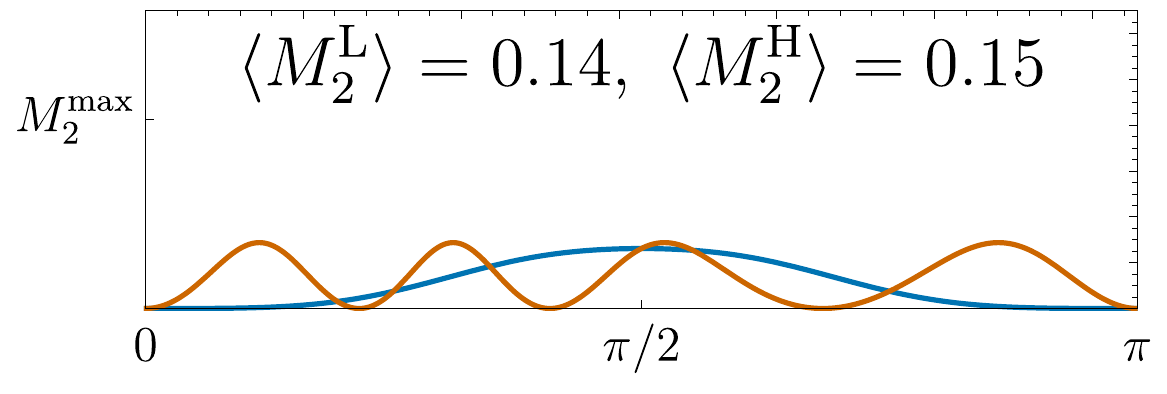}
}
\\
\hline
\end{tabular}
\caption{Magic $M_2$ versus the scattering angle $\theta$ in the helicity basis (blue) and lab basis (orange) for Bhabha and M{\o}ller scattering in the ultra-relativistic limit for different stabilizer states. The largest value of the magic can be reached is $\log (9/5)$ for the stabilizer states in G1 (G2) for Bhabha scattering (M{\o}ller scattering). }
\label{tab:magic-distribution-eeee}
\end{table}

\begin{table}
\centering
\begin{tabular}{|c|c|c|}
\hline
\multicolumn{2}{|c|}{$\mathcal{S}_{\rm H}$} & $e^-\mu^-\to e^-\mu^-$\\
\hline
G1&\adjustbox{valign=c}{%
  \begin{tabular}{@{}c@{}}
    $\langle IX,XI\rangle_\pm$,
    $\langle XX,YZ\rangle_\pm$
  \end{tabular}
}
&
\adjustbox{valign=c}{%
  \includegraphics[width=0.28\linewidth]{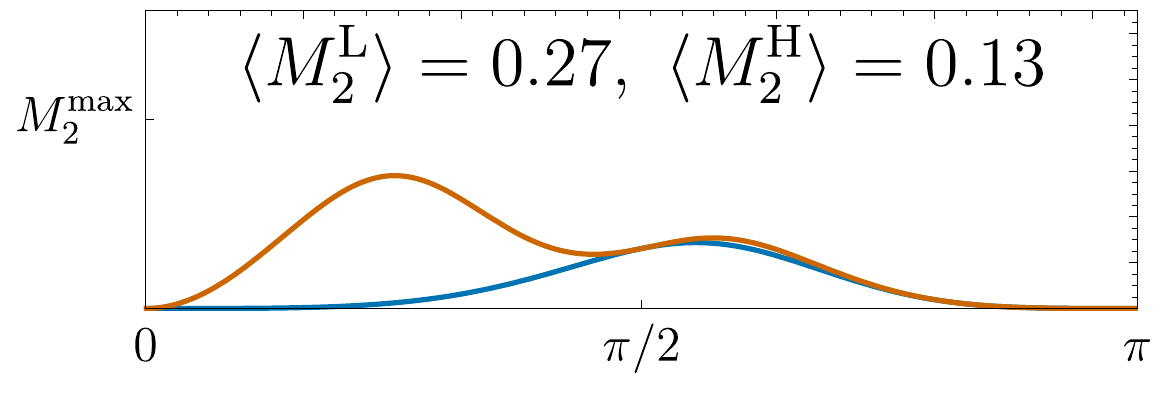}
}
\\
\hline
G2&\adjustbox{valign=c}{%
  \begin{tabular}{@{}c@{}}
    $\langle IX,-XI\rangle_\pm,\langle XX,-YZ\rangle_\pm$
  \end{tabular}
}
&
\adjustbox{valign=c}{%
  \includegraphics[width=0.28\linewidth]{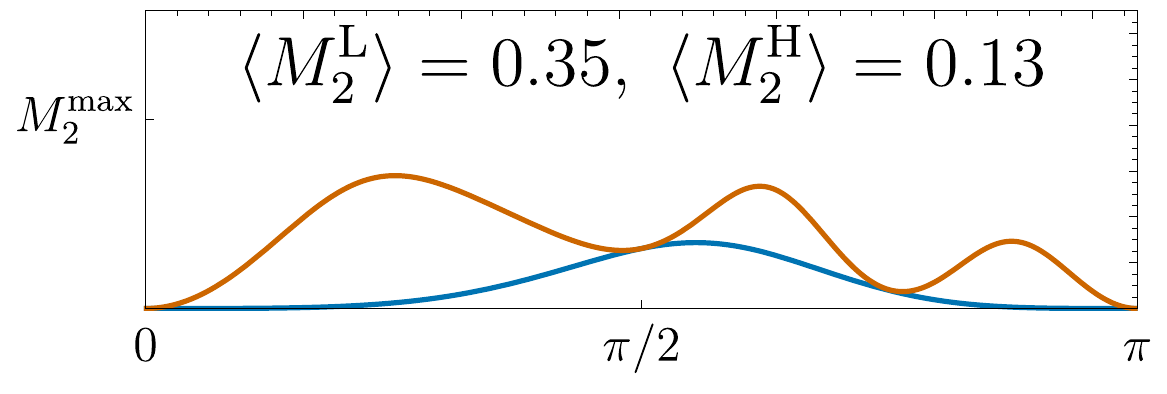}
}
\\
\hline

G3&\adjustbox{valign=c}{%
  \begin{tabular}{@{}c@{}}
    $\langle \pm IX,\pm YI\rangle,\langle \pm IY,\pm XI\rangle$\\
    $\langle \pm IY,\pm ZI\rangle,\langle \pm IZ,\pm YI\rangle$
  \end{tabular}
}
&
\adjustbox{valign=c}{%
  \includegraphics[width=0.28\linewidth]{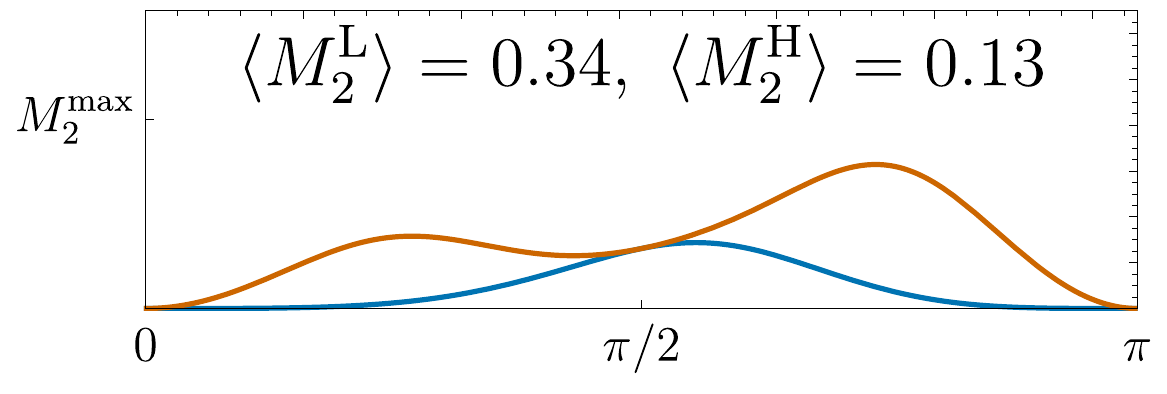}
}
\\
\hline

G4&\adjustbox{valign=c}{%
  \begin{tabular}{@{}c@{}}
    $\langle IX,ZI\rangle_\pm,\langle IZ,XI\rangle_\pm$\\
    $\langle \pm XZ,-YX\rangle,\langle XY,\pm YZ\rangle$
  \end{tabular}
}
&
\adjustbox{valign=c}{%
  \includegraphics[width=0.28\linewidth]{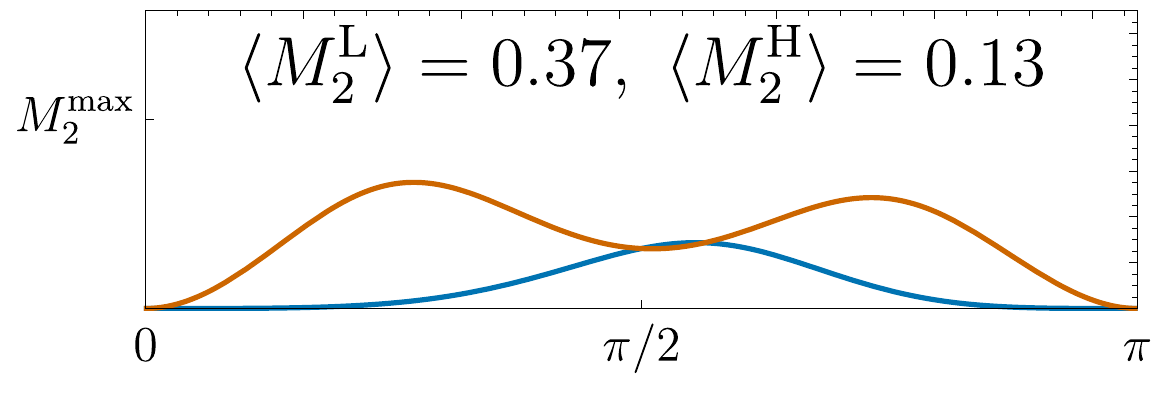}
}
\\
\hline

G5&\adjustbox{valign=c}{%
  \begin{tabular}{@{}c@{}}
    $\langle IX,-ZI\rangle_\pm,\langle IZ,-XI\rangle_\pm$\\
    $\langle \pm XZ,YX\rangle,\langle -XY,\pm YZ\rangle$
  \end{tabular}
}
&
\adjustbox{valign=c}{%
  \includegraphics[width=0.28\linewidth]{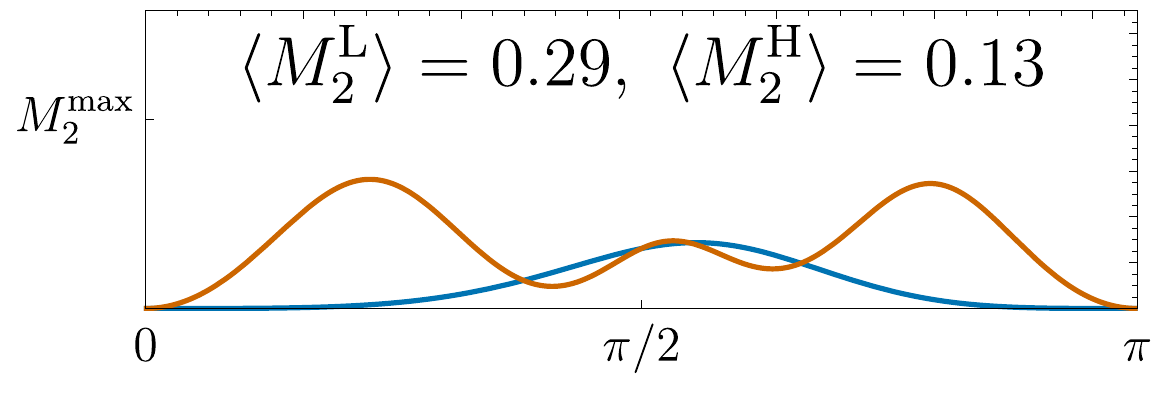}
}
\\
\hline

G6&\adjustbox{valign=c}{%
  \begin{tabular}{@{}c@{}}
    $\langle IY,YI\rangle_\pm$
  \end{tabular}
}
&
\adjustbox{valign=c}{%
  \includegraphics[width=0.28\linewidth]{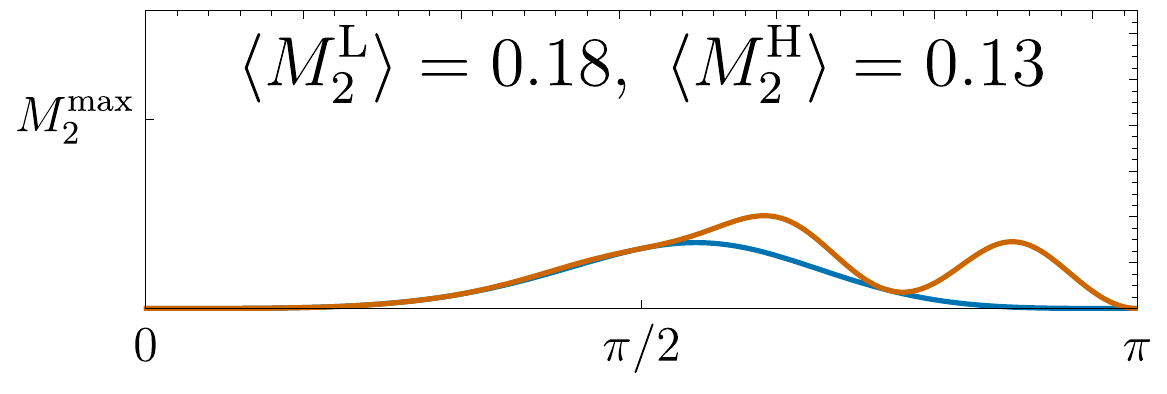}
}
\\
\hline
G7&\adjustbox{valign=c}{%
  \begin{tabular}{@{}c@{}}
    $\langle IY,-YI\rangle_\pm,\langle XZ,-YY\rangle_\pm$
  \end{tabular}
}
&
\adjustbox{valign=c}{%
  \includegraphics[width=0.28\linewidth]{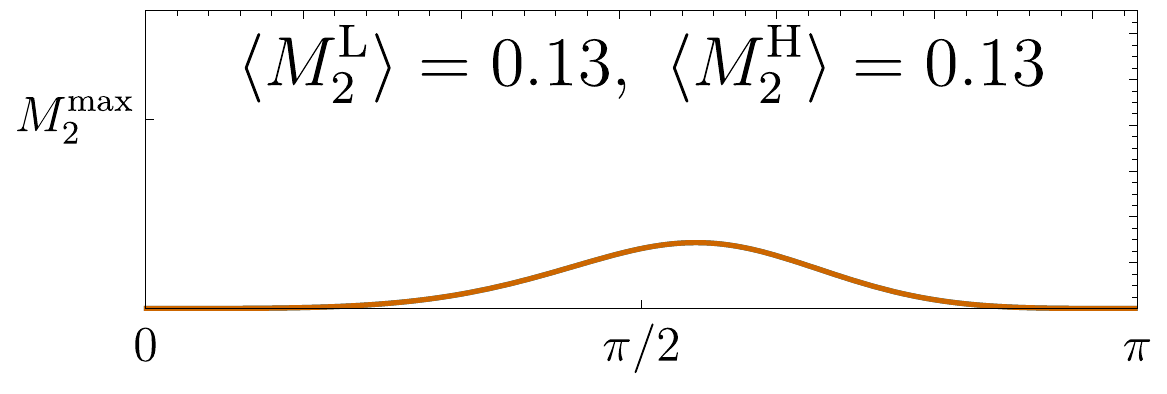}
}
\\
\hline
G8&\adjustbox{valign=c}{%
  \begin{tabular}{@{}c@{}}
    $\langle\pm IZ,\pm ZI\rangle,\langle\pm XY,\pm YX\rangle$
  \end{tabular}
}
&
\adjustbox{valign=c}{%
  \includegraphics[width=0.28\linewidth]{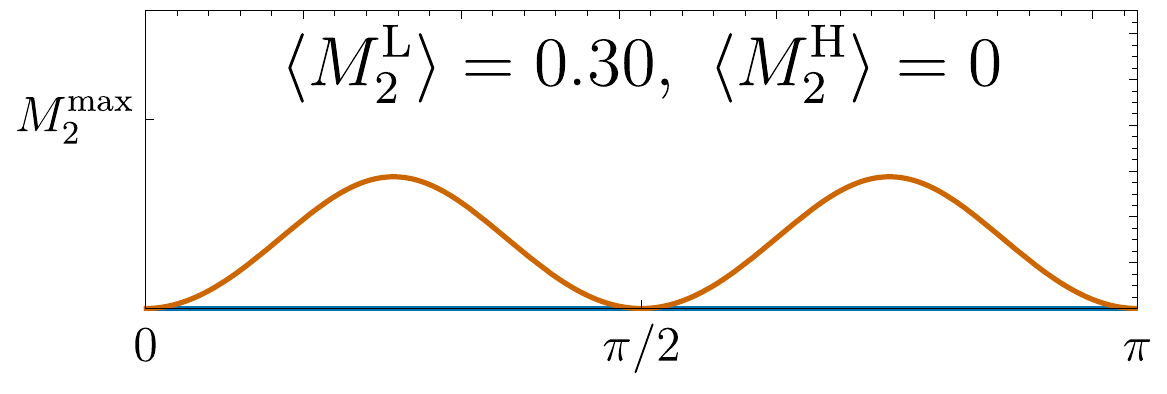}
}
\\
\hline
G9&\adjustbox{valign=c}{%
  \begin{tabular}{@{}c@{}}
    $\langle XX,\pm YY\rangle$
  \end{tabular}
}
&
\adjustbox{valign=c}{%
  \includegraphics[width=0.28\linewidth]{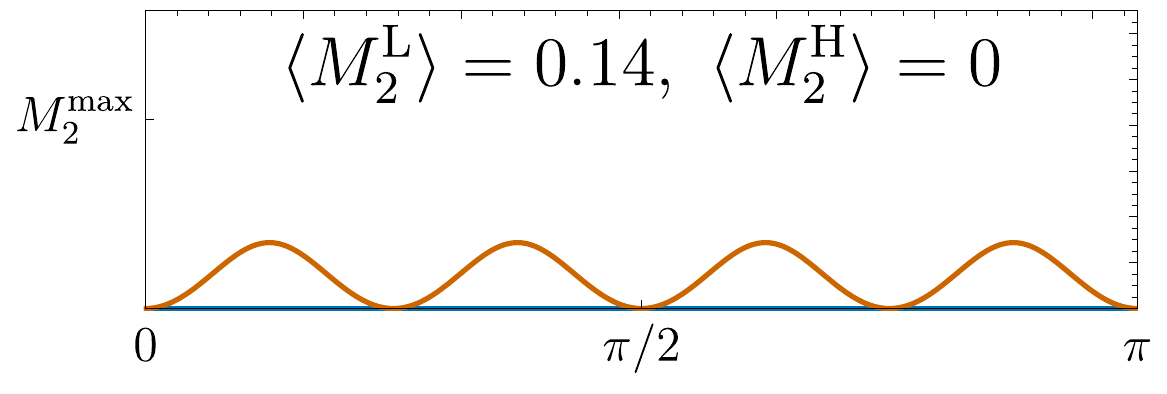}
}
\\
\hline
G10&
    $\langle -XX,\pm YY\rangle$
&
$M^{\rm L}_2= M_2^{\rm H}=0$
\\
\hline
G11&\adjustbox{valign=c}{%
  \begin{tabular}{@{}c@{}}
    $\langle XZ,YY\rangle$
  \end{tabular}
}
&
\adjustbox{valign=c}{%
  \includegraphics[width=0.28\linewidth]{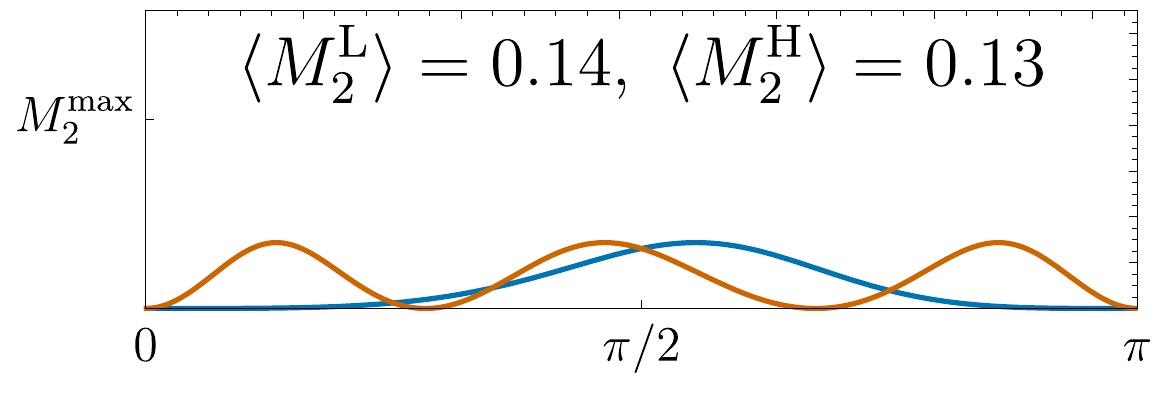}
}
\\
\hline
G12&\adjustbox{valign=c}{%
  \begin{tabular}{@{}c@{}}
    $\langle -XZ,-YY\rangle$
  \end{tabular}
}
&
\adjustbox{valign=c}{%
  \includegraphics[width=0.28\linewidth]{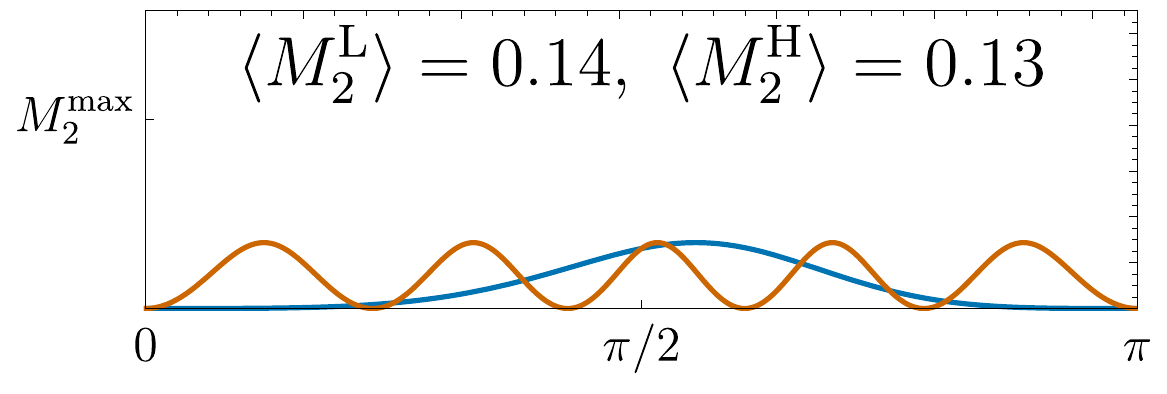}
}
\\
\hline
\end{tabular}
\caption{Magic $M_2$ versus the scattering angle $\theta$ in the helicity basis (blue) and lab basis (orange) for $e^-\mu^- \to e^-\mu^-$ scattering in the ultra-relativisitc limit for different the stabilizer states. The largest value of the magic can be reached is $\sim 0.628$ for the stabilizer states in G3. }
\label{tab:magic-distribution-emem}
\end{table}

\begin{table}
\centering
\begin{tabular}{|c|c|c|}
\hline
\multicolumn{2}{|c|}{$\mathcal{S}_{\rm H}$} & $e^+e^-\to \mu^+\mu^-$
\\
\hline

G1&\adjustbox{valign=c}{%
  \begin{tabular}{@{}c@{}}
    $\langle IX,XI\rangle_\pm,
    \langle IY,-YI\rangle_\pm,
    \langle -XX,YY\rangle_\pm$\\
    $\langle XZ,-YY\rangle_\pm,
    \langle XX,YZ\rangle_\pm$
  \end{tabular}
}
&
\adjustbox{valign=c}{%
  \begin{tabular}{@{}c@{}}
    $M_2^{\rm L}=M_2^{\rm H}=0$
  \end{tabular}
}
\\
\hline

G2&\adjustbox{valign=c}{%
  \begin{tabular}{@{}c@{}}
    $\langle IX,-XI\rangle_\pm,
    \langle IY,YI\rangle_\pm,
    \langle IZ,-ZI\rangle_\pm,
    \langle XX,YY\rangle_\pm$\\
    $\langle XY,-YX\rangle_\pm,
    \langle XZ,YY\rangle_\pm,
    \langle XX,-YZ\rangle_\pm$
  \end{tabular}
}
&
\adjustbox{valign=c}{%
  \includegraphics[width=0.28\linewidth]{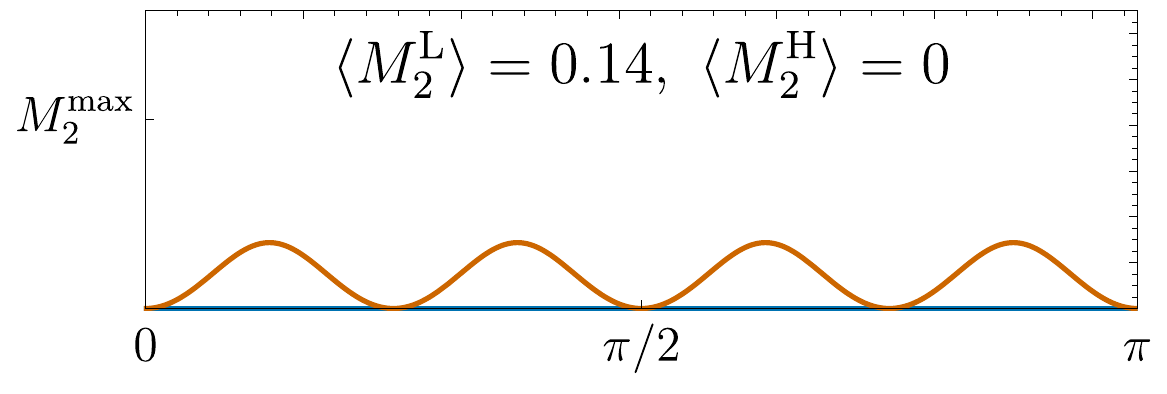}
}
\\
\hline

G3&\adjustbox{valign=c}{%
  \begin{tabular}{@{}c@{}}
    $\langle \pm IX,\pm YI\rangle,
    \langle \pm IX,\pm ZI\rangle,
    \langle \pm IY,\pm XI\rangle,
    \langle \pm IY,\pm ZI\rangle$\\
    $\langle \pm IZ,\pm XI\rangle,
    \langle \pm IZ,\pm YI\rangle,
    \langle IZ,ZI\rangle_\pm,
    \langle XY,YX\rangle_\pm$\\
    $\langle \pm XZ,\pm YX\rangle,
    \langle \pm XY,\pm YZ\rangle $
  \end{tabular}
}
&
\adjustbox{valign=c}{%
  \includegraphics[width=0.28\linewidth]{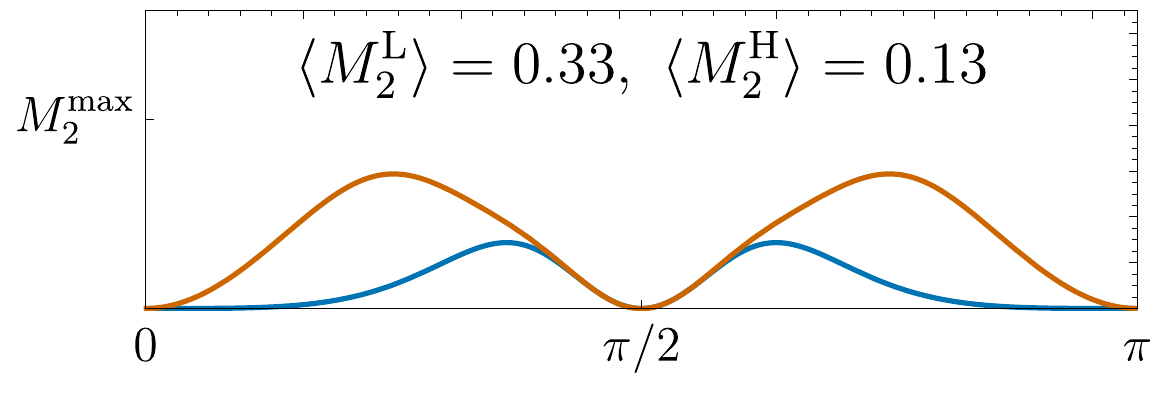}
}
\\
\hline

\end{tabular}
\caption{Magic $M_2$ versus the scattering angle $\theta$ in the helicity basis (blue) and lab basis (orange) for $e^+e^- \to \mu^+\mu^-$ scattering in the ultra-relativistic limit for different stabilizer states. The largest value of the magic can be reached is $\log(9/5)$ for the stabilizer state in G3. }
\label{tab:magic-distribution-etom}
\end{table}

In the ultra-relativistic limit, the scattering processes exhibit 13 distinct patterns of magic distribution for both \textbf{Process (A)} Bhabha scattering and \textbf{Process (B)} (\tab{magic-distribution-eeee}), 12 distinct patterns for \textbf{Process (C)} (\tab{magic-distribution-emem}), and 4 distinct patterns for \textbf{Process (D)}.

Several classes of behaviors can be identified. First, there are cases in which the magic vanishes in the helicity basis, while in the lab basis it follows the pattern induced purely by the basis transformation, as shown in Fig.~\ref{fig:rot}.  These include the states stabilized by $\langle XX,YY\rangle$ for all four processes, the states stabilized by $\langle IZ,-ZI\rangle_\pm$ for \textbf{Process (A)} and \textbf{Process (D)}, the states stabilized by $\langle IZ,ZI\rangle_\pm$ for \textbf{Process (B)}, and the states stabilized by $\langle \pm IZ,\pm ZI\rangle$ for \textbf{Process (C)}. For these cases, the largest value of the magic reaches $\log(16/9)$, which is  the maximal value attainable  by a change of computational basis. (See \fig{rot}.)

A second class consists of cases in which the magic vanishes in both bases. One example is the states stabilized by $\langle -XX,\pm YY\rangle$, which appear for all four processes. These stabilizer states remain unchanged up to a normalization factor under the scattering evolution, preserving zero magic. For the remaining stabilizer states in \textbf{Process (D)} with vanishing magic in both bases, we find that the $s$-channel scattering matrix maps these states to other stabilizer states.

We also find cases where the magic distributions coincide in the helicity and lab bases. In particular, for the states stabilized by $\langle IY,-YI\rangle_\pm$ and $\langle XZ,-YY\rangle_\pm$ in \textbf{Processes (A)}, \textbf{(B)}, and \textbf{(C)}, the final states have identical Pauli spectra, up to a permutation. This indicates that the final states in the two bases are related by a Clifford-type global unitary rotation, which preserves the magic.

Finally, we examine the magic averaged over the scattering angle. Except for two exceptional cases (G12 and G13) in \tab{magic-distribution-eeee}, the angle-averaged magic in the helicity basis is  smaller than that in the lab basis. After averaging further over all stabilizer initial states, the magic $\<{\cal M}_2\>$ in the helicity basis is smaller for all processes considered, as shown in \tab{averaged-magic}. This behavior can be traced back to the sparsity of the scattering amplitudes, induced by helicity selection rules,  in the spin-space. In the ultra-relativistic limit, leptons can be treated as effectively massless particles with definite helicities, and the corresponding helicity amplitudes satisfy selection rules associated with helicity conservation~\cite{Cheung:2015aba}. Consequently, many helicity amplitudes vanish, reducing the amount of magic generated in the helicity basis. For example, consider \textbf{Process (A)} Bhabha scattering. The amplitude in the spin-space in the lab basis 
is given by
\begin{align}
    \mathcal{M}_{\rm L}&=\left(
\begin{array}{cccc}
 2 \cot ^2\frac{\theta }{2} & \sin \theta (\cos \theta
   +1)-2 \cot \frac{\theta }{2} & \sin \theta (\cos
   \theta+1)-2 \cot \frac{\theta }{2} & 2 \\
 2 \cot \frac{\theta }{2} & \frac{15 \cos \theta+\cos 3
   \theta }{4-4 \cos \theta} & -2 \cos ^2\frac{\theta
   }{2} \cos \theta & -2 \cot \frac{\theta }{2} \\
 2 \cot \frac{\theta }{2} & -2 \cos ^2\frac{\theta
   }{2} \cos \theta & \frac{15 \cos \theta+\cos 3 \theta
   }{4-4 \cos \theta} & -2 \cot \frac{\theta }{2} \\
 2 & 2 \cot \frac{\theta }{2}-\sin \theta (\cos \theta
   +1) & 2 \cot \frac{\theta }{2}-\sin \theta (\cos
   \theta+1) & 2 \cot ^2\frac{\theta }{2} \\
\end{array}
\right).\label{eq:amp-lab}
\end{align}
On the other hand, in the helicity basis only six different helicity amplitudes are nonvanishing due to helicity conservation, which leads to a sparser scattering matrix given by 
\begin{align}
\mathcal{M}_{\rm H}=
\left(
\begin{array}{cccc}
 \frac{1}{8} \sin ^4\theta  \csc ^6\frac{\theta }{2} & 0 &
   0 & \cos \theta -1 \\
 0 & 2 \csc ^2\frac{\theta }{2} & 0 & 0 \\
 0 & 0 & 2 \csc ^2\frac{\theta }{2} & 0 \\
 \cos \theta -1 & 0 & 0 & \frac{1}{8} \sin ^4\theta  \csc
   ^6\frac{\theta }{2} \\
\end{array}
\right).
\label{eq:amp-helicity}
\end{align} 
Such sparsity of the scattering amplitude in the helicity basis is a generic feature for all processes considered above in the ultra-relativistic limit, as shown in Appendix \ref{sec:scattering}.

Given that, in the ultra-relativistic limit, the magic is generally smaller in the helicity basis, the non-local magic is expected to more closely follow—or even coincide with—the magic distribution in the helicity basis, which is also observed in \cite{Robin:2025ymq}. Interestingly, for the two exceptional cases in which the angle-averaged magic $\<M_2\>$ is larger in the helicity basis, G12 and G13 in \tab{magic-distribution-eeee}, the corresponding non-local magic is found to vanish.

In \textbf{Process (D)} \(\mu^+\mu^- \to e^+e^-\), we also consider the case where initial states are non-relativistic while the final states are relativistic, due to $m_e \ll m_{\mu}$. In this scenario, we identify 12 distinct magic-distribution patterns, as summarized in \tab{magic-distribution-mmee-ur}. In the lab basis, the maximal possible magic \(M^{\rm max}_2\) attainable for a two-qubit system is realized by the stabilizer states in group G3, as listed in \tab{magic-distribution-mmee-ur}. After further averaging over all stabilizer initial states, we find that the magic \(\langle{\cal M}_2\rangle\) in the helicity basis is also smaller than that in the lab basis, as shown in \tab{averaged-magic}, similar to the ultra-relativistic cases listed above.

For all processes considered above, the non-local magic vanishes for entangled initial states. In \cite{Robin:2025ymq}, the authors also observed that the non-local magic vanishes for entangled initial states, although their analysis was restricted to Møller scattering and Bhabha scattering in the ultra-relativistic limit. 

\begin{table}
    \centering
    \begin{tabular}{|c|c|c|}
    \hline
     Process    &  $\<\mathcal{M}_2\>$ in lab basis
         & $\<\mathcal{M}_2\>$ in helicity basis \\
    \hline
    $e^+e^-\rightarrow e^+e^-$ & $0.310...$ & $0.153...$ \\
    \hline
    $e^-e^-\rightarrow e^-e^-$& $0.310...$ & $0.153...$ \\
    \hline
    $e^-\mu^-\rightarrow e^-\mu^-$& $0.284...$ & $0.105...$ \\
    \hline
    $e^+e^-\rightarrow \mu^+\mu^-$,  $\mu^+\mu^-\rightarrow e^+e^- $& $0.229...$ & $0.078...$ \\
    \hline
    $\mu^+\mu^-\rightarrow e^+e^-$, non-relativistic initial state & 0.263... & 0.219... \\
    \hline
    \end{tabular}
    \caption{The magic $\<\mathcal{M}_2 \>$ averaged over all stabilizer states and scattering angle in the ultra-relativistic cases.}
    \label{tab:averaged-magic}
\end{table}

\begin{table}
\centering
\begin{tabular}{|c|c|c|}
\hline
\multicolumn{2}{|c|}{$\mathcal{S}_{\rm H}$} & $\mu^+\mu^-\to e^+e^-$\\
\hline
G1&\adjustbox{valign=c}{%
  \begin{tabular}{@{}c@{}}
    $\langle IX,XI\rangle_\pm$,
    $\langle XX,YZ\rangle_\pm$
  \end{tabular}
}
&
\adjustbox{valign=c}{%
  \includegraphics[width=0.28\linewidth]{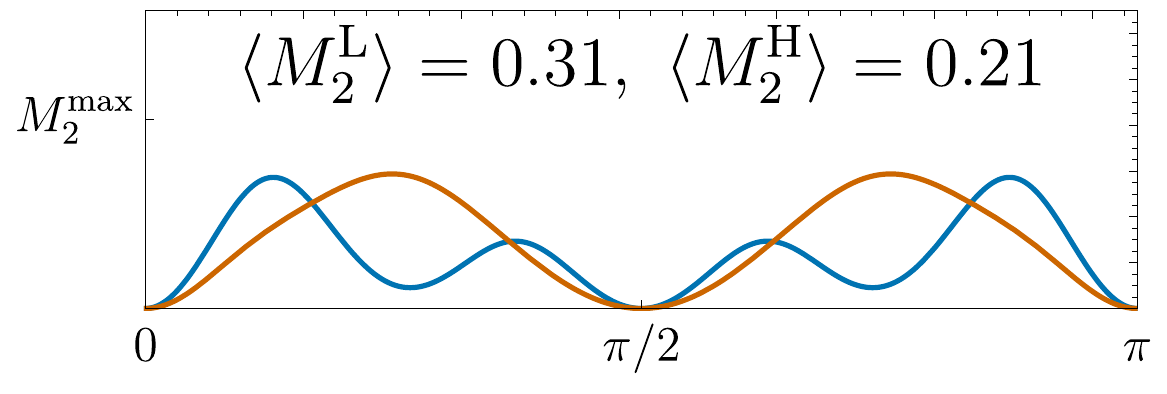}
}
\\
\hline
G2&\adjustbox{valign=c}{%
  \begin{tabular}{@{}c@{}}
    $\langle IX,-XI\rangle_\pm,\langle XX,-YZ\rangle_\pm,\langle -XX,-YY\rangle$
  \end{tabular}
}
&
\adjustbox{valign=c}{%
  \includegraphics[width=0.28\linewidth]{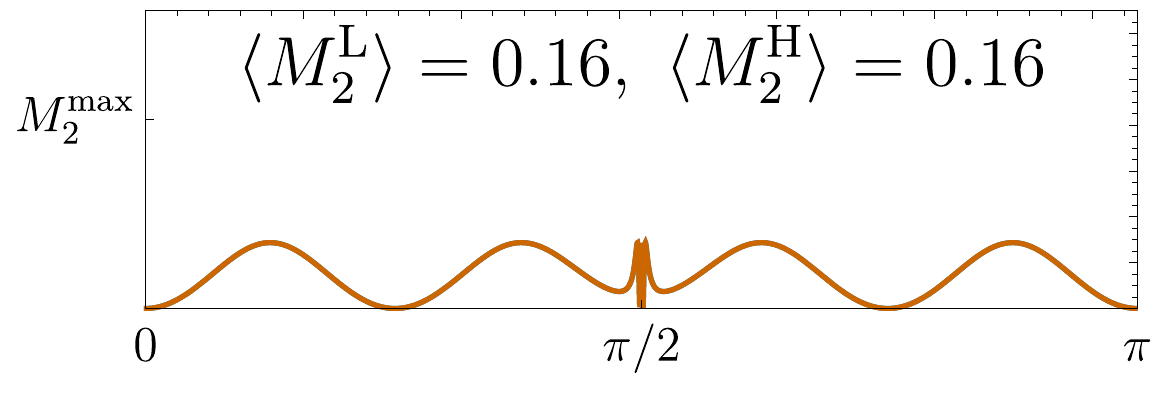}
}
\\
\hline

G3&\adjustbox{valign=c}{%
  \begin{tabular}{@{}c@{}}
    $\langle \pm IX,\pm YI\rangle,\langle \pm IY,\pm XI\rangle$\\
    $\langle \pm IY,\pm ZI\rangle,\langle \pm IZ,\pm YI\rangle$
  \end{tabular}
}
&
\adjustbox{valign=c}{%
  \includegraphics[width=0.28\linewidth]{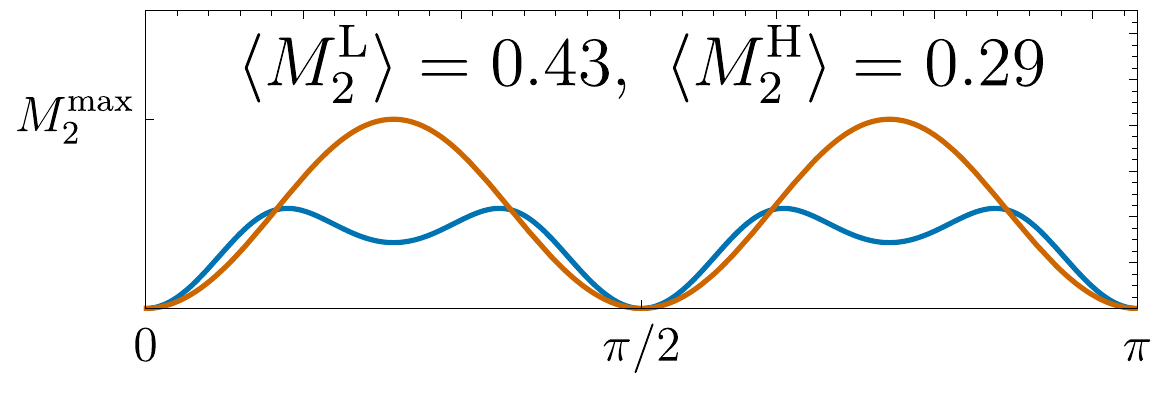}
}
\\
\hline

G4&\adjustbox{valign=c}{%
  \begin{tabular}{@{}c@{}}
    $\langle IX,ZI\rangle_\pm,\langle IZ,-XI\rangle_\pm$\\
    $\langle \pm XZ,-YX\rangle,\langle -XY,\pm YZ\rangle$
  \end{tabular}
}
&
\adjustbox{valign=c}{%
  \includegraphics[width=0.28\linewidth]{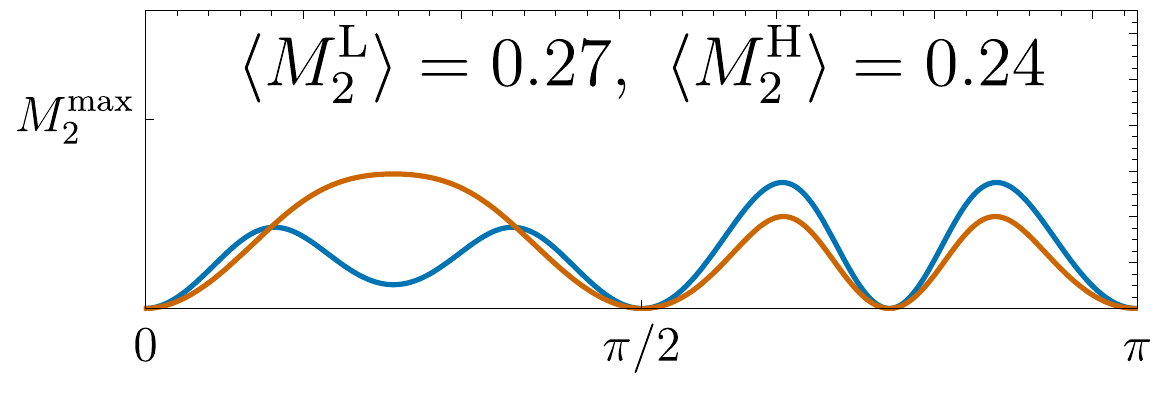}
}
\\
\hline

G5&\adjustbox{valign=c}{%
  \begin{tabular}{@{}c@{}}
    $\langle IX,-ZI\rangle_\pm,\langle IZ,XI\rangle_\pm$\\
    $\langle \pm XZ,YX\rangle,\langle XY,\pm YZ\rangle$
  \end{tabular}
}
&
\adjustbox{valign=c}{%
  \includegraphics[width=0.28\linewidth]{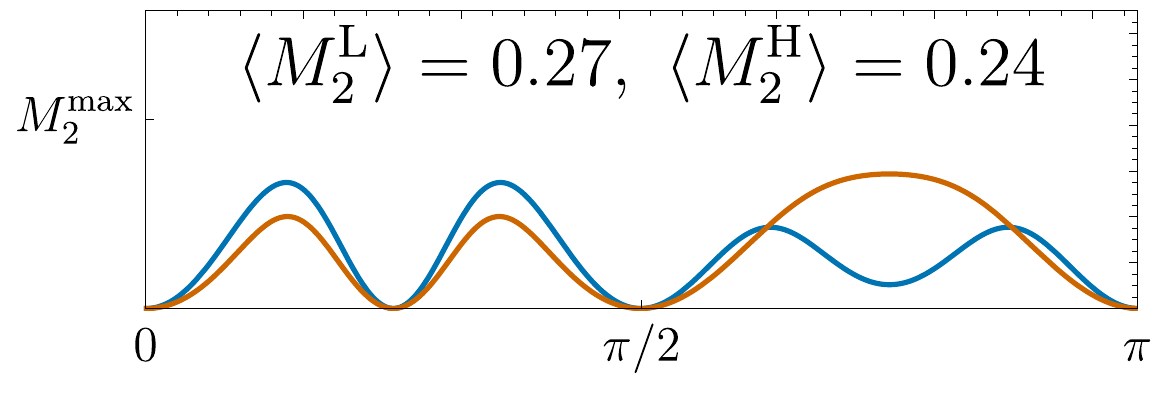}
}
\\
\hline

G6&\adjustbox{valign=c}{%
  \begin{tabular}{@{}c@{}}
    $\langle IY,YI\rangle_\pm,\langle XZ,YY\rangle_\pm,\langle XX,YY\rangle$
  \end{tabular}
}
&
\adjustbox{valign=c}{%
  \includegraphics[width=0.28\linewidth]{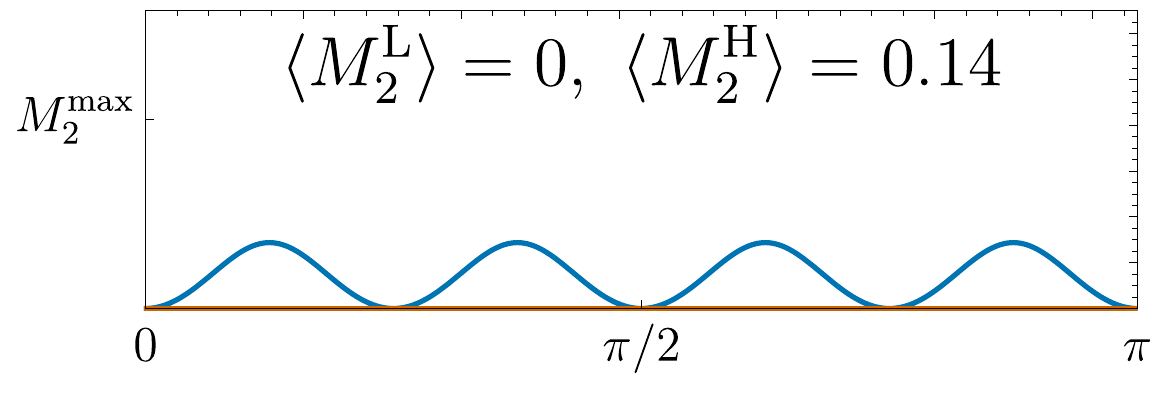}
}
\\
\hline

G7&\adjustbox{valign=c}{%
  \begin{tabular}{@{}c@{}}
    $\langle IY,-YI\rangle_\pm$
  \end{tabular}
}
&
\adjustbox{valign=c}{%
  \includegraphics[width=0.28\linewidth]{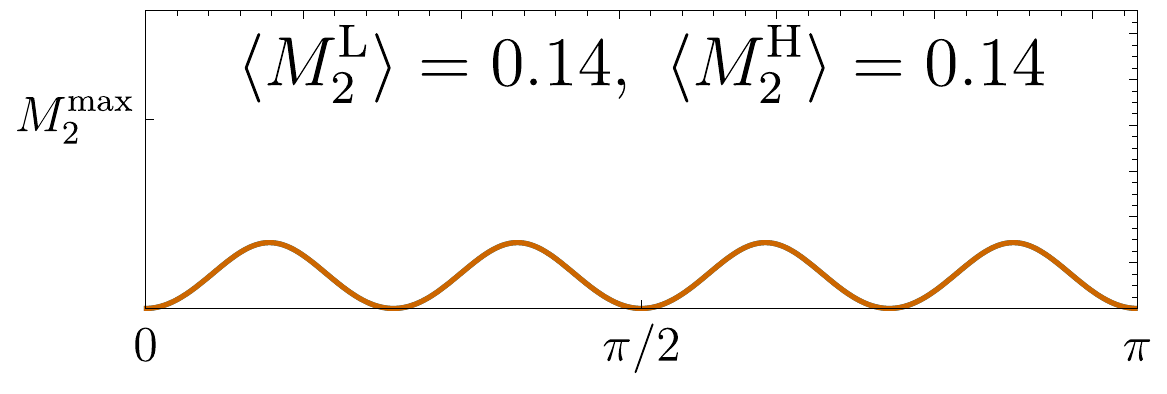}
}
\\
\hline

G8&\adjustbox{valign=c}{%
  \begin{tabular}{@{}c@{}}
    $\langle IZ,ZI\rangle_\pm,\langle XY,YX\rangle_\pm,\langle -XX,YY\rangle$
  \end{tabular}
}
&
\adjustbox{valign=c}{%
  \includegraphics[width=0.28\linewidth]{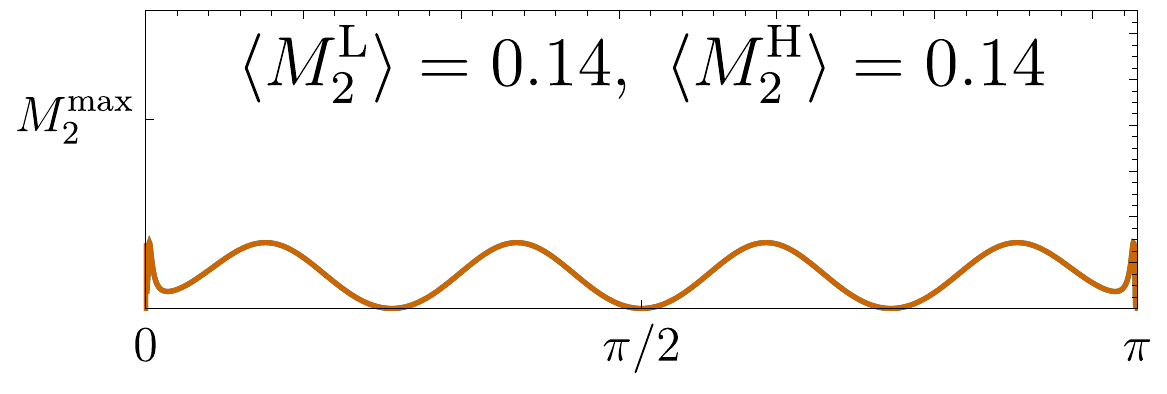}
}
\\
\hline

G9&\adjustbox{valign=c}{%
  \begin{tabular}{@{}c@{}}
    $\langle IZ, ZI\rangle_\pm,\langle XY,- YX\rangle_\pm$
  \end{tabular}
}
&
\adjustbox{valign=c}{%
  \includegraphics[width=0.28\linewidth]{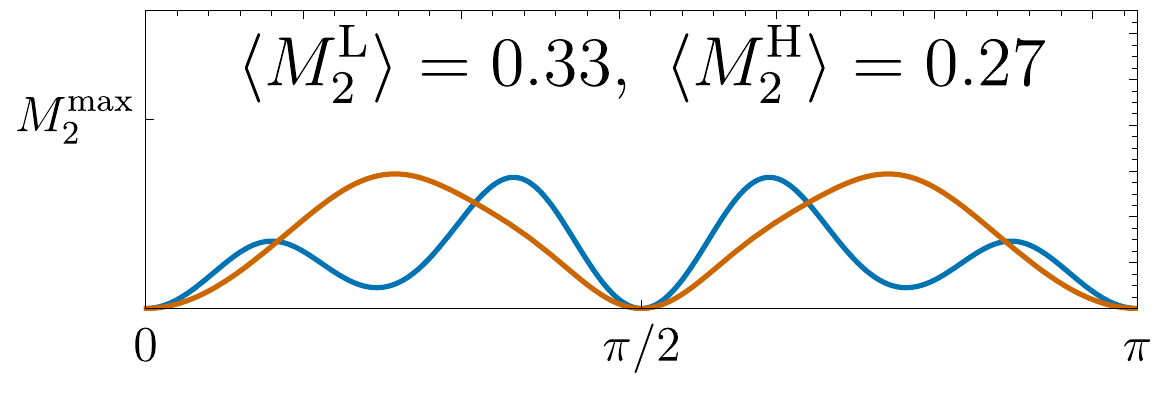}
}
\\
\hline

G10&\adjustbox{valign=c}{%
  \begin{tabular}{@{}c@{}}
    $\langle XX, -YY\rangle$
  \end{tabular}
}
&
\adjustbox{valign=c}{%
  \begin{tabular}{@{}c@{}}
    $M_2^{\rm L}=M_2^{\rm H}=0$
  \end{tabular}
}
\\
\hline

G11&\adjustbox{valign=c}{%
  \begin{tabular}{@{}c@{}}
    $\langle XZ,-YY\rangle$
  \end{tabular}
}
&
\adjustbox{valign=c}{%
  \includegraphics[width=0.28\linewidth]{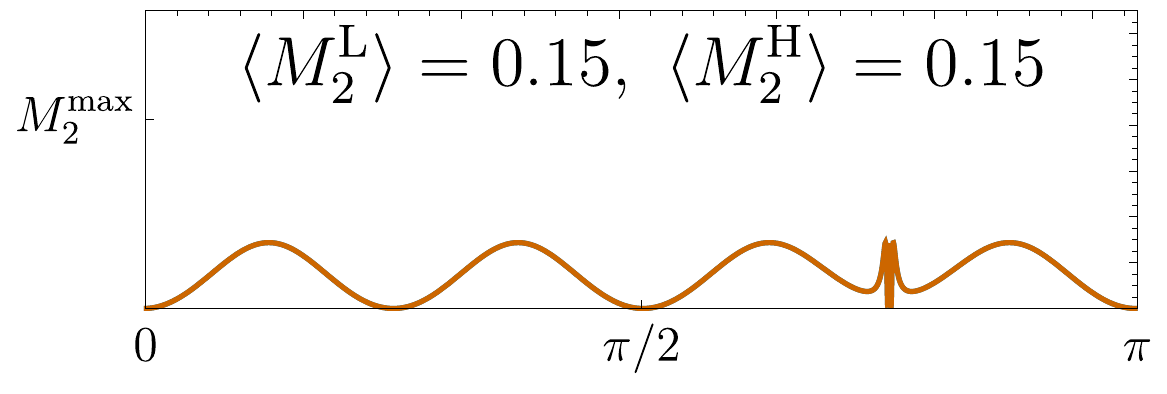}
}
\\
\hline

G12&\adjustbox{valign=c}{%
  \begin{tabular}{@{}c@{}}
    $\langle -XZ,YY\rangle$
  \end{tabular}
}
&
\adjustbox{valign=c}{%
  \includegraphics[width=0.28\linewidth]{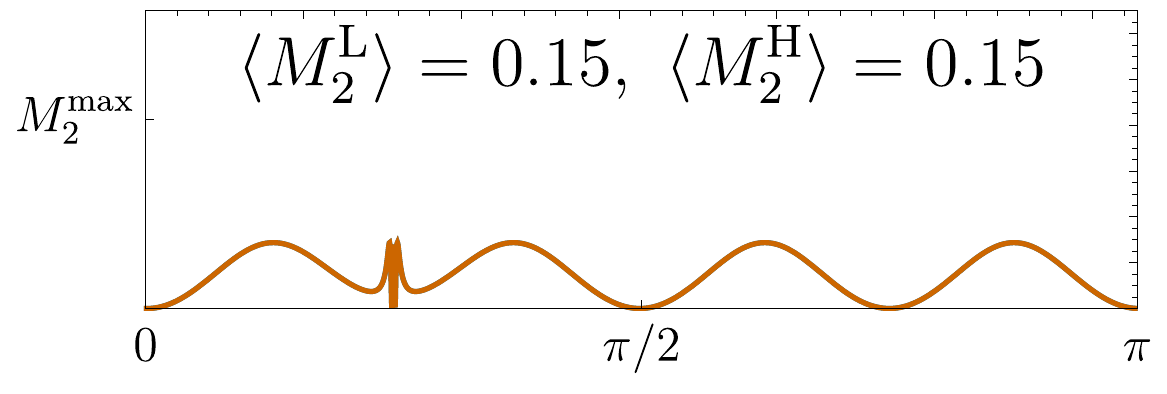}
}
\\
\hline
\end{tabular}
\caption{Magic $M_2$ versus the scattering angle $\theta$ in the helicity basis (blue) and lab basis (orange) for $\mu^+\mu^- \to e^+e^-$ scattering where only the final states are relativistic for different stabilizer states. The maximum value of the magic can be reached is $\log[16/7]$ for the stabilizer state in G3.}
\label{tab:magic-distribution-mmee-ur}
\end{table}
\subsection{Non-relativistic limit}
For \textbf{Process (A)} (Bhabha scattering) in the non-relativistic limit, the magic vanishes identically in the lab basis.  The same behavior is observed for \textbf{Process (C)} \((e^- \mu^- \to e^- \mu^-)\), where both the electron and muon are taken to be non-relativistic. Similarly, for \textbf{Process (D)} \((e^+e^-\to\mu^+\mu^-)\) with non-relativistic final-state muons, the magic in the lab basis also vanishes up to corrections of order \(m_e/m_\mu\), which are numerically negligible.

The origin of this behavior is that the corresponding scattering amplitudes in the spin-space are proportional to the identity in the lab basis, up to corrections of order \(m_e/m_\mu\) for \textbf{Process (D)}. Consequently, the scattering itself does not generate magic in the lab basis. The magic observed in the helicity basis therefore arises entirely from the basis transformation, and the magic distribution for each stabilizer state follows the universal pattern shown in \tab{rotyss}, with small deviations of order \(m_e/m_\mu\) in the case of \textbf{Process (D)}.
Accordingly, the non-local magic vanishes for all three processes.

For \textbf{Process (B)} (M{\o}ller scattering) in the non-relativistic limit, the magic distribution is shown in \tab{magic-af-distribution-eeee-nr}. We find that the magic in the lab basis is generally smaller than that in the helicity basis. As shown in Appendix~\ref{sec:scattering}, the scattering matrix in the lab basis becomes sparse in the non-relativistic limit for this process. Consequently, one expects the non-local magic to more closely track, or even coincide with, the magic distribution in the lab basis.

This expectation is confirmed by the results in \tab{magic-af-distribution-eeee-nr}: for the stabilizer groups G1, G2, G7, and G8, the non-local magic (green) exactly coincides with the magic in the lab basis. Moreover, similar to cases in the ultra-relativistic limit where the non-local magic vanishes for entangled initial states \cite{Robin:2025ymq}, the non-local magic also remains zero for entangled initial states.

\begin{table}
\centering
\begin{tabular}{|c|c|c|}
\hline
\multicolumn{2}{|c|}{$\mathcal{S}_{\rm H}$} & $e^-e^-\to e^-e^-$\\
\hline

G1&\adjustbox{valign=c}{%
  \begin{tabular}{@{}c@{}}
    $\langle IX,XI\rangle_\pm,
    \langle IZ,ZI\rangle_\pm$
  \end{tabular}
}
&
\adjustbox{valign=c}{%
  \includegraphics[width=0.28\linewidth]{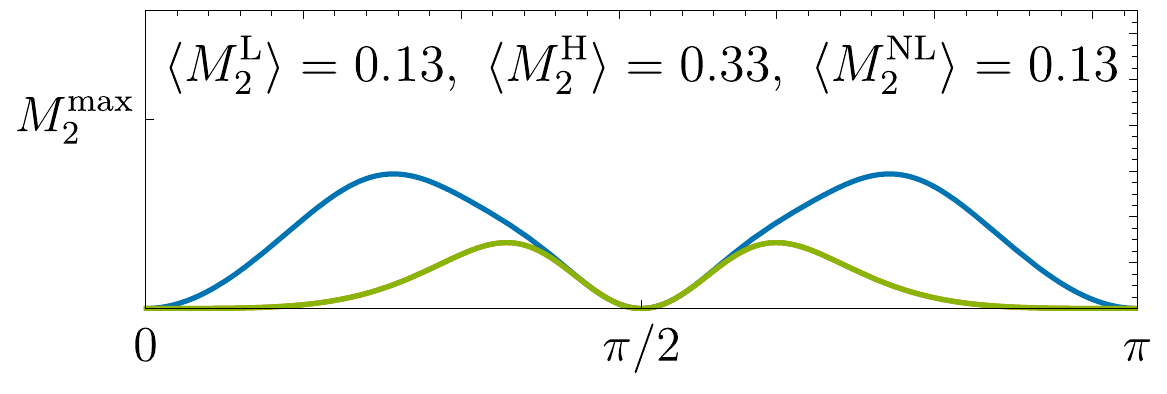}
}
\\
\hline

G2&\adjustbox{valign=c}{%
  \begin{tabular}{@{}c@{}}
    $\langle IX,-XI\rangle_\pm,\langle IZ,-ZI\rangle_\pm$\\
    $\langle XY,-YX\rangle_\pm,\langle XX,-YZ\rangle_\pm$
  \end{tabular}
}
&
\adjustbox{valign=c}{%
  \includegraphics[width=0.28\linewidth]{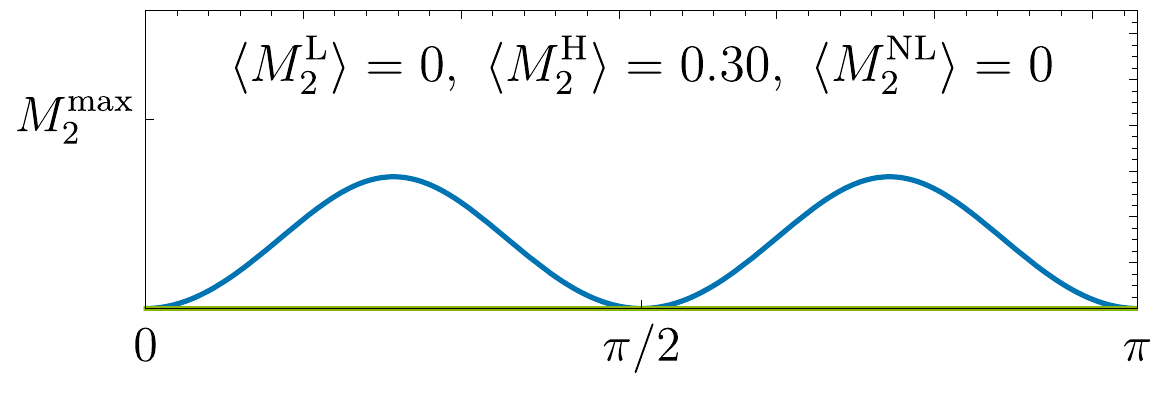}
}
\\
\hline

G3&\adjustbox{valign=c}{%
  \begin{tabular}{@{}c@{}}
    $\langle \pm IX,\pm YI\rangle,
    \langle \pm IY,\pm XI\rangle$\\
    $\langle \pm IY,\pm ZI\rangle,
    \langle \pm IZ,\pm YI\rangle$
  \end{tabular}
}
&
\adjustbox{valign=c}{%
  \includegraphics[width=0.28\linewidth]{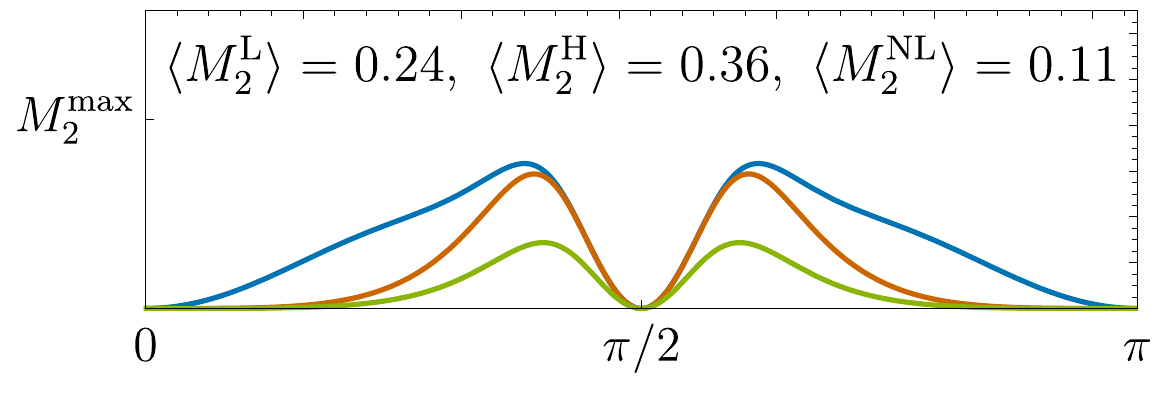}
}
\\
\hline

G4&\adjustbox{valign=c}{%
  \begin{tabular}{@{}c@{}}
    $\langle \pm IX,\pm ZI\rangle,
    \langle \pm IZ,\pm XI\rangle$
  \end{tabular}
}
&
\adjustbox{valign=c}{%
  \includegraphics[width=0.28\linewidth]{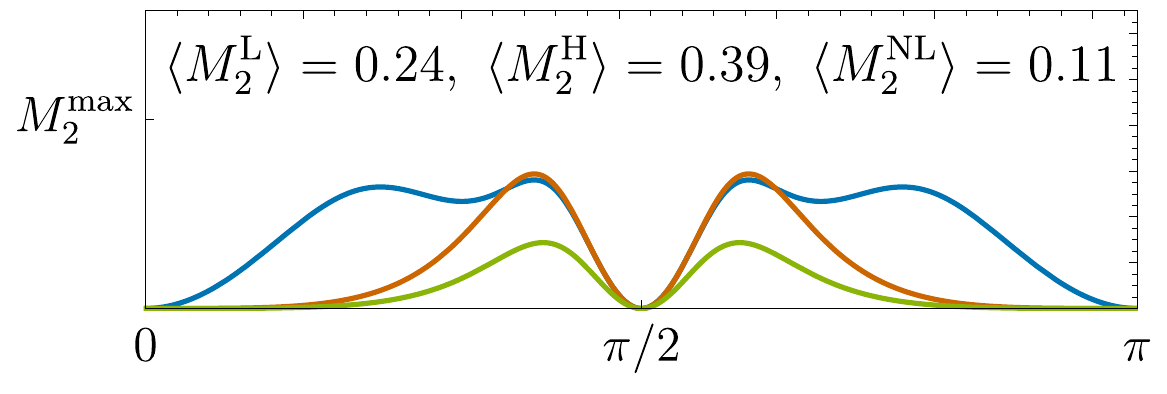}
}
\\
\hline

G5&\adjustbox{valign=c}{%
  \begin{tabular}{@{}c@{}}
    $\langle \pm XZ,\pm YX\rangle,
    \langle \pm XY,\pm YZ\rangle$
  \end{tabular}
}
&
\adjustbox{valign=c}{%
  \includegraphics[width=0.28\linewidth]{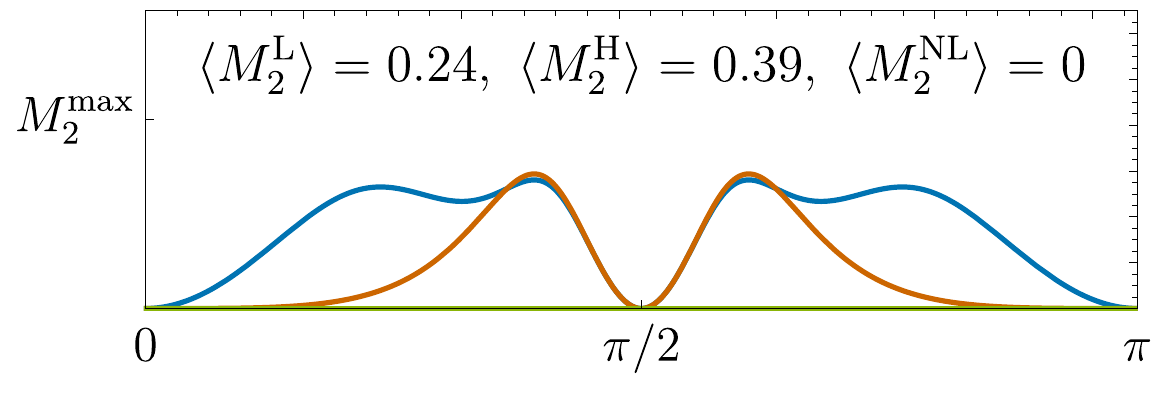}
}
\\
\hline

G6& $\langle IY,YI\rangle_\pm,\langle -XX,\pm YY\rangle$
&
$M_2^{\rm L}=M^{\rm H}_2=M^{\rm NL}_2=0$
\\
\hline

G7&\adjustbox{valign=c}{%
  \begin{tabular}{@{}c@{}}
    $\langle IY,-YI\rangle_\pm$
  \end{tabular}
}
&
\adjustbox{valign=c}{%
  \includegraphics[width=0.28\linewidth]{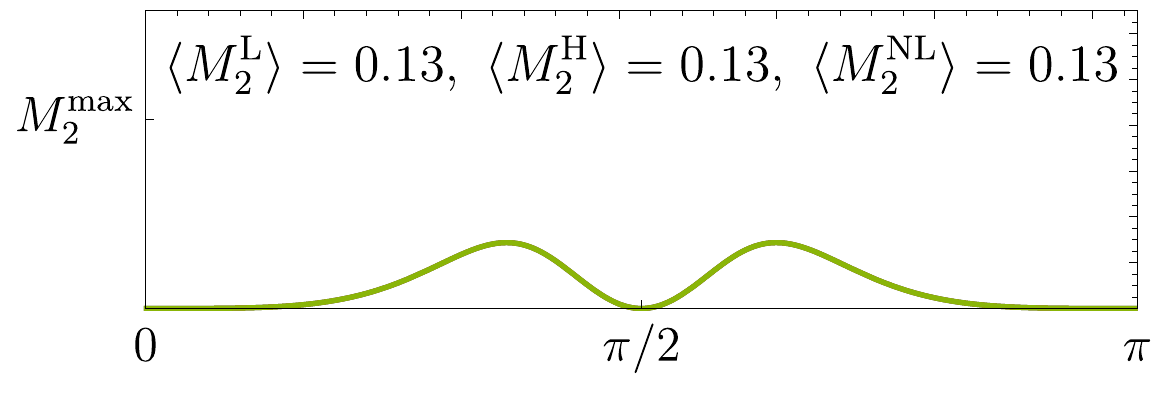}
}
\\
\hline

G8&\adjustbox{valign=c}{%
  \begin{tabular}{@{}c@{}}
    $\langle XX,\pm YY\rangle,\langle XZ,YY\rangle_\pm$
  \end{tabular}
}
&
\adjustbox{valign=c}{%
  \includegraphics[width=0.28\linewidth]{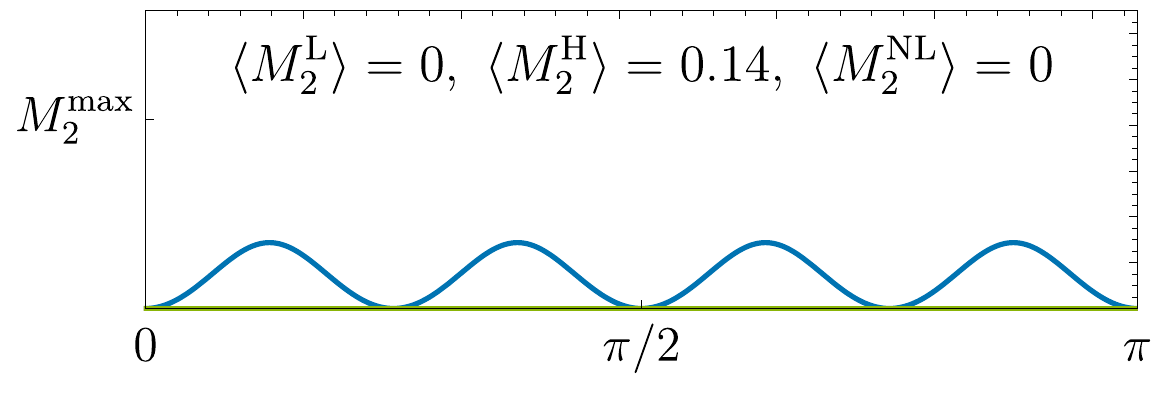}
}
\\
\hline

G9&\adjustbox{valign=c}{%
  \begin{tabular}{@{}c@{}}
    $\langle XY,YX\rangle_\pm,\langle XX,YZ\rangle_\pm$
  \end{tabular}
}
&
\adjustbox{valign=c}{%
  \includegraphics[width=0.28\linewidth]{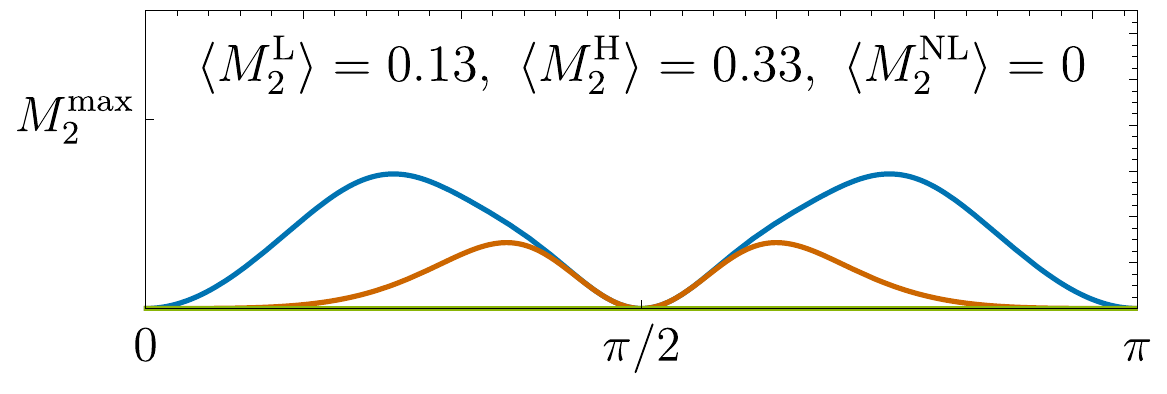}
}
\\
\hline

G10&\adjustbox{valign=c}{%
  \begin{tabular}{@{}c@{}}
    $\langle XZ,-YY\rangle_\pm$
  \end{tabular}
}
&
\adjustbox{valign=c}{%
  \includegraphics[width=0.28\linewidth]{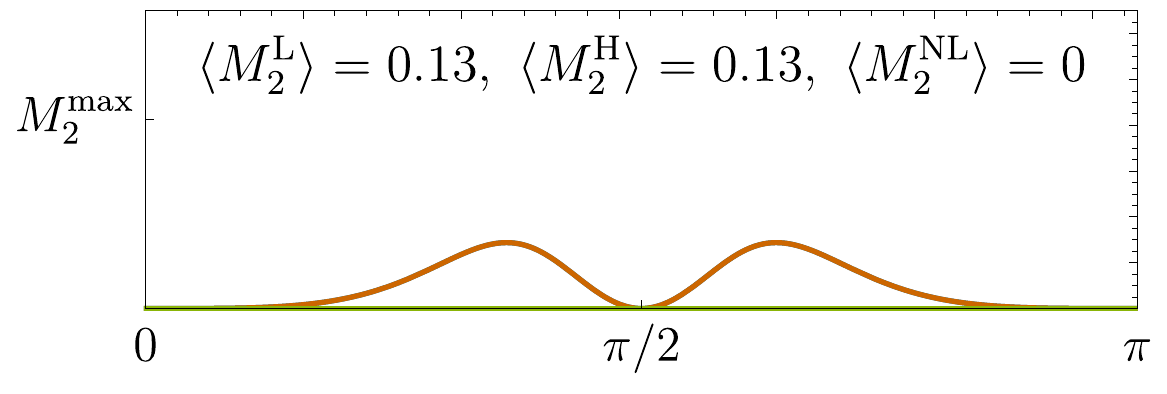}
}
\\
\hline
\end{tabular}
\caption{Magic $M_2$ as a function of the scattering angle $\theta$ in the helicity basis (blue) and the lab basis (orange), together with the non-local magic (green), for M{\o}ller scattering in the non-relativistic limit for different stabilizer states. Whenever the orange curve is not visible, it coincides with the green curve. The maximum magic attainable is approximately $0.633$, achieved by the stabilizer states in G3.}
\label{tab:magic-af-distribution-eeee-nr}
\end{table}

        





\section{Magic production in electroweak processes for different bases}

\label{sec:ew}

Now let us consider the charged lepton scattering in SM, at CM energy $\sqrt{s} \gtrsim m_Z$, which was explored in \cite{Liu:2025bgw}. At tree level, the 2 $\to$ 2 scattering processes are mediated by the photon and the $Z$ boson, as shown in Fig. 
\ref{fig:feynman_diagram}. One can compute the final state magic averaged over the 60 stabilizer states $\{|\psi_{\text{s}}\>_{i} \}$ chosen as the initial state,
\bea
{\cal M}_2(\theta) &\equiv& \frac{1}{60} \sum_{i=1}^{60}   M_2(|\psi_{\text{s}}\>_{i},\theta),
\eea
and consider its dependence on the weak mixing angle $\theta_W$, in terms of $s_W^2= \sin^2\theta_W$. It has been observed in Ref.~\cite{Liu:2025bgw} that up to $\sqrt{s} \sim \ordr (1 \text{ TeV})$, for processes including $e^- e^- \to e^- e^-$ and $e^- \mu^- \to e^- \mu^-$, the SM value of $s_W^2$ in the $\overline{\text{MS}}$ scheme, $\hat{s}_W^2 (\sqrt{s})$, tends to minimize $\<{\cal M}_2\>$ defined in \eq{averageM},
which is ${\cal M}_2(\theta)$ averaged over the scattering angle. Such an observation is established using the lab basis, and now we would like to investigate its basis dependence by comparing with results in the helicity basis.

\begin{figure}
    \centering
    \includegraphics[width=0.8\linewidth]{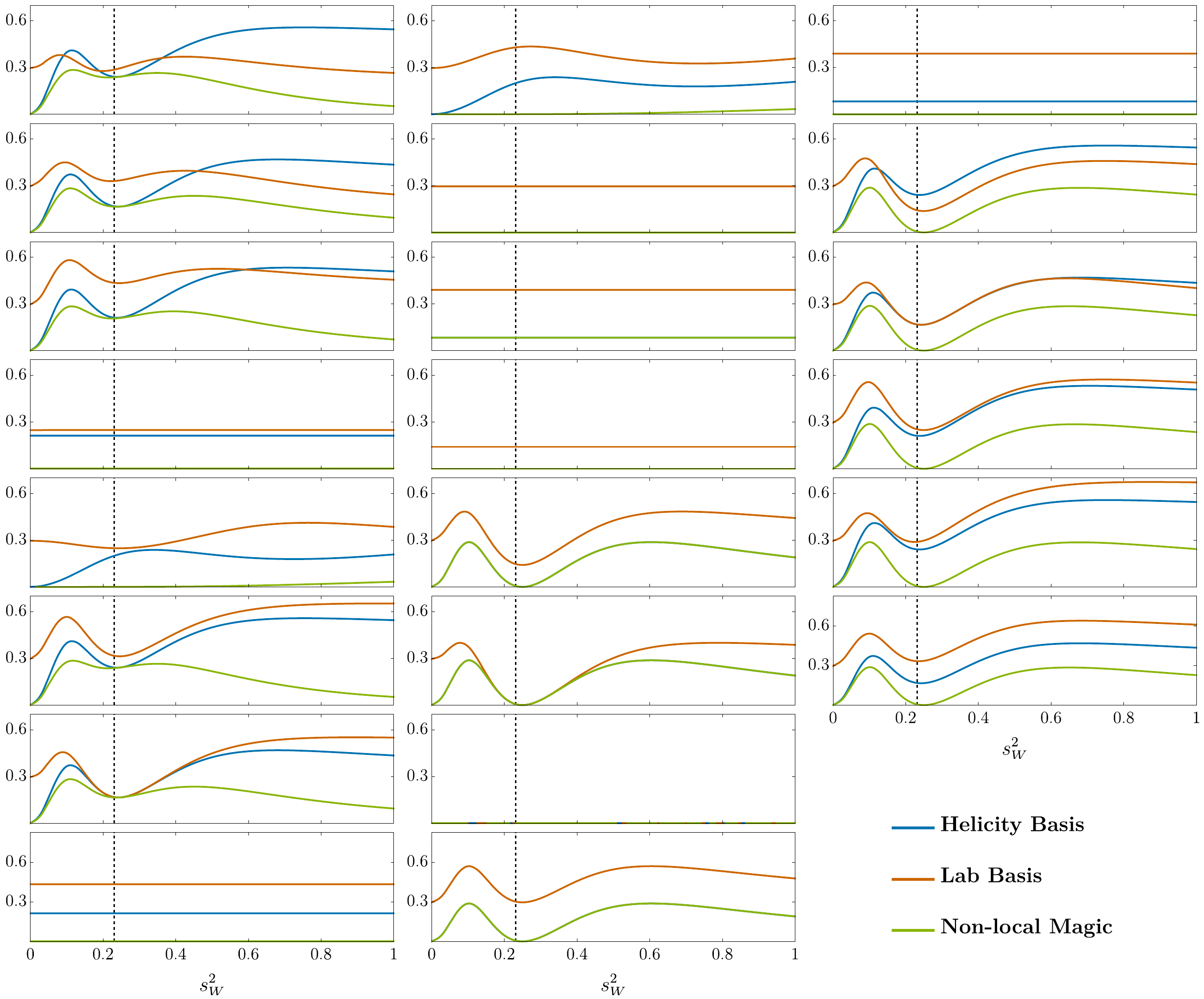}
    \caption{Distinct distributions of the $s_W^2$ dependence of magic $\<M_{2,i}\>$ for the EW process $e^-e^-\to e^-e^-$ at $\sqrt{s}= m_Z$; the dotted, vertical line shows the value of $\hat{s}_W^2 (\sqrt{s})$.}
    \label{fig:EW_91GeV}
\end{figure}

\begin{figure}
    \centering
    \includegraphics[width=0.8\linewidth]{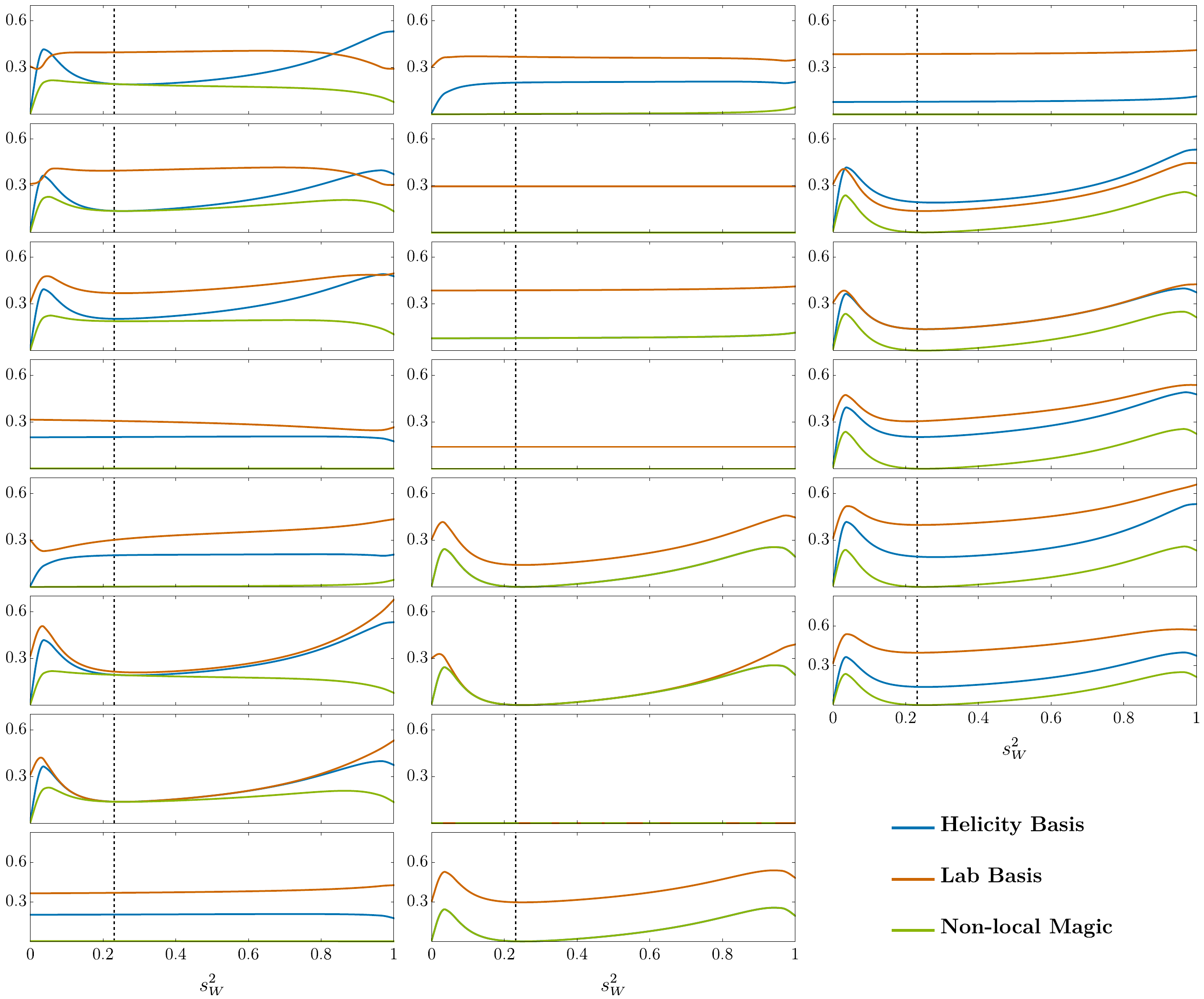}
    \caption{Distinct distributions of $s_W^2$ dependence of magic $\<M_{2,i}\>$ in helicity and lab basis for the EW process $e^-e^-\to e^-e^-$ at $\sqrt{s}=1\, \text{TeV}$, with the dotted, vertical line showing the value of $\hat{s}_W^2 (\sqrt{s})$.}
    \label{fig:EW_1TeV}
\end{figure}

\begin{figure}
    \centering
    \includegraphics[width=0.5\linewidth]{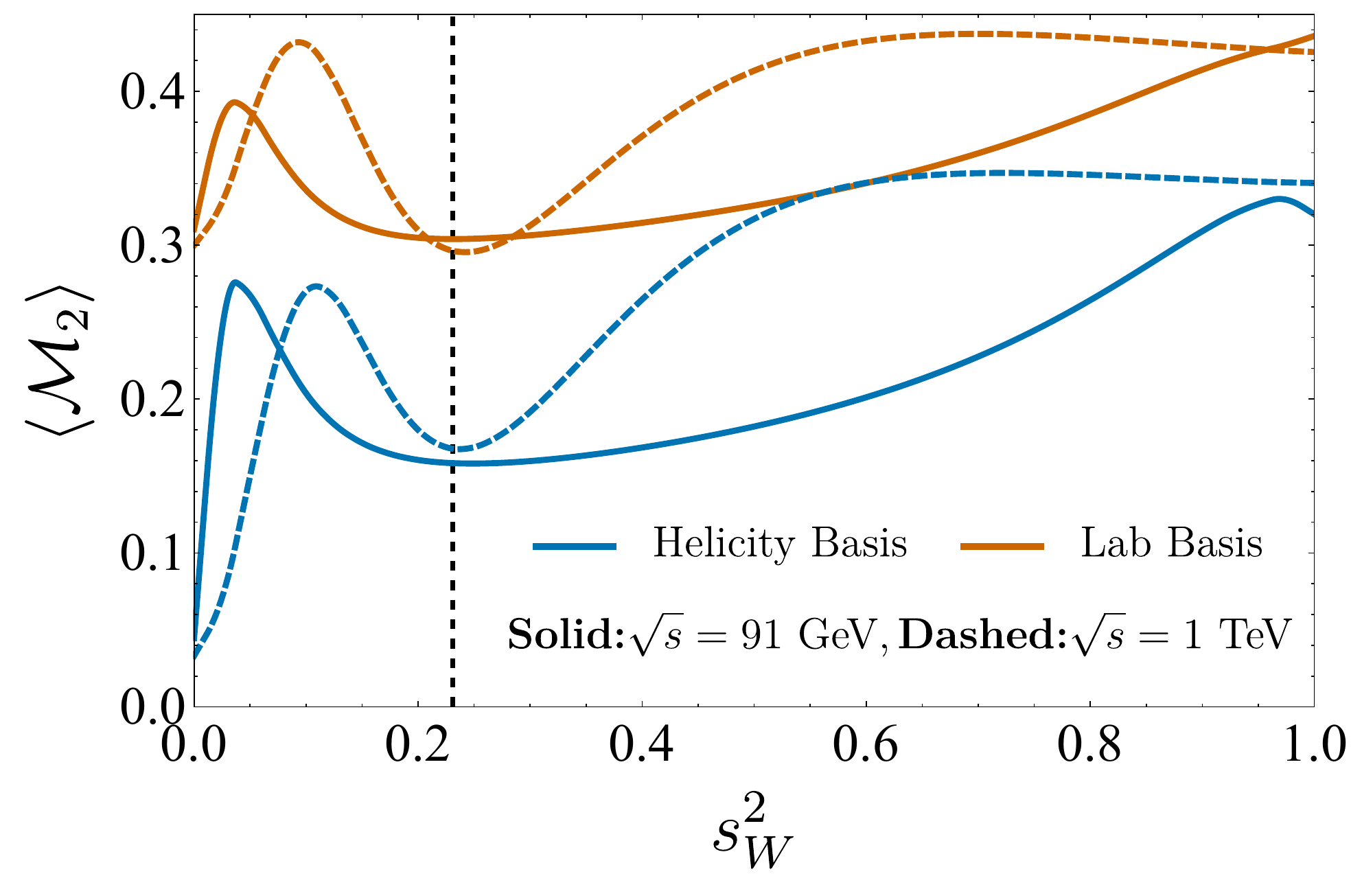}
    \caption{The magic $\<{\cal M}_2\>$, averaged over all stabilizer states and scattering angle $\theta$,  for the EW process $e^-e^-\to e^-e^-$ at $\sqrt{s}= m_Z$ and $1\, \rm{TeV}$. The vertical dashed line shows the value of $\hat{s}_W^2 (\sqrt{s})$.}
    \label{fig:avg_ew}
\end{figure}

For the process $e^- e^- \to e^- e^-$, Figs. \ref{fig:EW_91GeV} and \ref{fig:EW_1TeV} show distinct dependence on $s_W^2$ of the angle-averaged magic $\<M_{2,i}\>$, defined in \eq{averageOtheta},
for different individual stabilizer states $|\psi_{\text{s}}\>_{i}$, computed in the lab and helicity bases, at $\sqrt{s} = m_Z$ or $1 \text{ TeV}$; we also plot the corresponding averaged non-local magic for reference. We observe that in both of these two bases, though some of the distributions are not sensitive to $s_W^2$, most of the others in general tend to realize a minimum near $\hat{s}_W^2 (\sqrt{s})$. In other words, the observation that $\hat{s}_W^2 (\sqrt{s})$ tends to minimize the final state magic holds for many of the individual stabilizer states chosen as the initial state. We have also plotted $\<{\cal M}_2\>$, averaged over both initial states and $\theta$, in Fig. \ref{fig:avg_ew}, where we see a local minimum near $\hat{s}_W^2 (\sqrt{s})$ for both of the bases.

On the other hand, it is clear from Figs. \ref{fig:EW_91GeV} and \ref{fig:EW_1TeV} that more often than not, the final magic $\<M_{2,i}\>$, averaged over the scattering angle $\theta$, is higher in the lab basis compared to the helicity basis (which are usually both higher than the non-local magic). We also see in Fig. \ref{fig:avg_ew} that for all values of $s_W^2$, $\<{\cal M}_2\>$ is significantly higher in the lab basis. As the fermion masses are much smaller than the CM energy that we are considering and thus can be treated as massless, these observations are similar to what we have seen in the previous section for QED, and may be related to the selection rules in the high energy limit.

\begin{figure}
    \centering
    \includegraphics[width=0.5\linewidth]{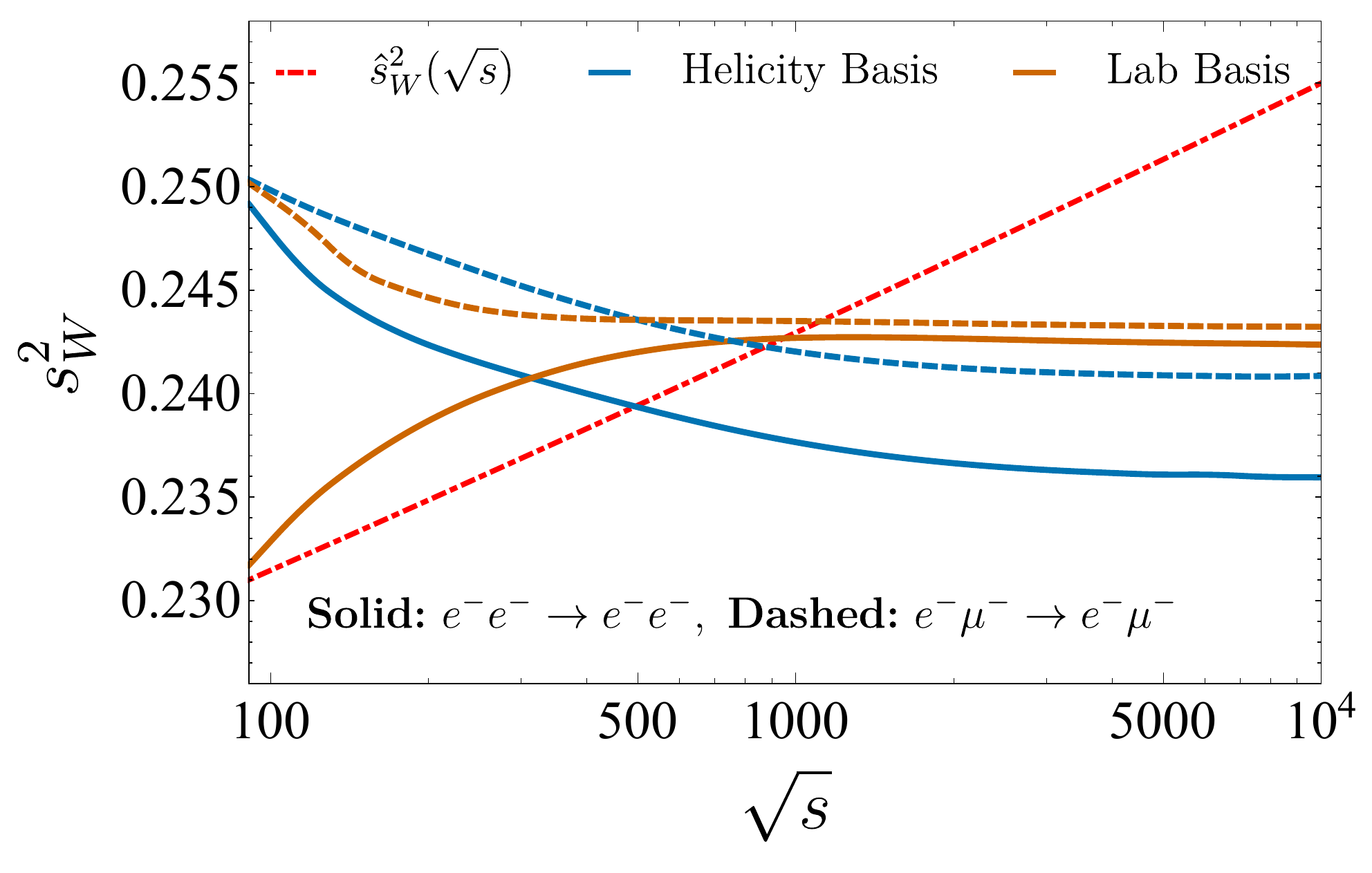}
    \caption{The values of $s_W^2$ minimizing the magic of EW processes $e^-e^-\to e^-e^-$ (solid curves) and $e^-\mu^-\to e^-\mu^-$ (dashed curves) in lab basis (orange) and helicity basis (blue) as a function of $\sqrt{s}$. For comparison, we have plotted $\hat{s}_W^2 (\sqrt{s})$ using the red, dot-dashed line.}
    \label{fig:minimal_sintheta}
\end{figure}

We have then plotted in Fig. \ref{fig:minimal_sintheta} the $s_W^2$ values in both lab and helicity bases that minimize $\<{\cal M}_2\>$ for different CM energy, for both processes $e^- e^- \to e^- e^-$ and $e^- \mu^- \to e^- \mu^-$. We see that in the energy range of $m_Z \lesssim \sqrt{s} \lesssim 1 \, \text{TeV}$, the magic-minimizing $s_W^2$ 
significantly deviates from $s_W^2 = 1/4$, toward smaller values that are closer to $\hat{s}_W^2 (\sqrt{s})$ in SM. This suggests that the fact that $\hat{s}_W^2 (\sqrt{s})$ tends to minimize $\<{\cal M}_2\>$ is not sensitive to the choice of bases, especially for larger values of $\sqrt{s}$. We do see that at $\sqrt{s} = m_Z$, the prediction of magic-minimizing $s_W^2$ 
agrees much better with $\hat{s}_W^2 (\sqrt{s})$ in the lab basis.

\section{Circuit realization for the scattering amplitude}
\label{sec:lcu-circuit}
The SRE characterizes the magic carried by a quantum
state, but it does not directly specify the non-Clifford resources required to
prepare that state.  To give a   perspective of quantum resource to the analysis above,
we consider quantum circuits that implement scattering amplitudes in the two-particle spin space.  Since a scattering amplitude is
not, in general, unitary, we employ the linear-combination-of-unitaries (LCU)
construction~\cite{Childs:2012gwh}, in which the non-unitary operator appears as a block
of a larger unitary gate acting on the system and an ancilla register. For illustration, we consider the Bhabha scattering in the high-energy limit. Let us recall that the scattering matrix for this process in the helicity basis is given by Eq. (\ref{eq:amp-helicity}):
\begin{align}
\mathcal{M}_{\rm H}=
\left(
\begin{array}{cccc}
 \frac{1}{8} \sin ^4\theta  \csc ^6\frac{\theta }{2} & 0 &
   0 & \cos \theta -1 \\
 0 & 2 \csc ^2\frac{\theta }{2} & 0 & 0 \\
 0 & 0 & 2 \csc ^2\frac{\theta }{2} & 0 \\
 \cos \theta -1 & 0 & 0 & \frac{1}{8} \sin ^4\theta  \csc
   ^6\frac{\theta }{2} \\
\end{array}
\right).
\end{align}

We first decompose the scattering matrices into Pauli strings and implement them using the Linear Combination of Unitaries (LCU) framework. For the helicity-basis scattering matrix, we first notice that it admits the diagonalization \cite{Cao:2026nyj} with the unitary $U_{\rm diag} = (H \otimes I) {\rm CNOT}$:
\begin{align}
\mathcal{M}_{\rm H}
=U^\dagger_{\rm diag}
\,M_{\rm diag}\,
U_{\rm diag}, 
\end{align}
where $M_{\rm diag}$ can be decomposed as Pauli strings: 
\begin{align}
M_{\rm diag}
=
\alpha_0 II
+
\alpha_1 ZI +\alpha_2 IZ +\alpha_3 ZZ = \Lambda \sum_{j=0}^{3}
\beta_j\, e^{i\phi_j} P_j,
\quad
\text{with}
\quad
P_j
=
\{II,ZI,IZ, ZZ\}.
\end{align}
We found $\alpha_0 =\frac{1}{8}[11+4\cos\theta+\cos(2\theta)]\csc^2\left(\frac{\theta}{2}\right)$, $\alpha_1 = \alpha_3 = \frac{1}{2}\left(-1+\cos\theta\right)$, $\alpha_2 = \frac{1}{2}\left(-3-\cos\theta\right)$, 
and defined $\Lambda = \sum_{j=0}^{3} |\alpha_j|$, 
$\beta_j = \frac{|\alpha_j|}{\Lambda}$. 
For all scattering angles $\theta$, we have $\phi_j=\{0, \pi, \pi, \pi\}$.

\begin{figure}
    \centering    \includegraphics[width=0.98\linewidth]{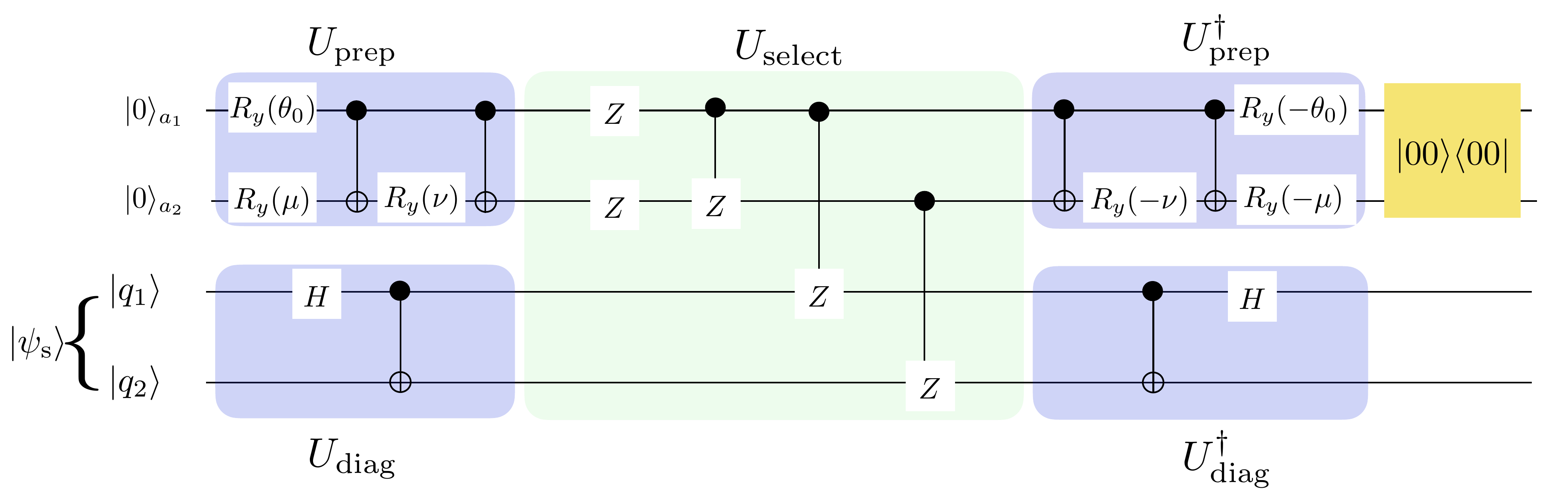}
    \caption{The circuit that realize the operation $\mathcal{M}_{\rm H}/\Lambda$ for Bhabha scattering in the helicity basis.}
    \label{fig:qcircuit}
\end{figure}

We introduce a two-qubit ancilla register and prepare the state
\[
|\chi\rangle
=
\sum_{j=0}^{3}
\sqrt{\beta_j}\,|j\rangle
=\sqrt{\beta_0} |00\rangle+\sqrt{\beta_1}(|10\rangle +|11\rangle)+ \sqrt{\beta_2}|01\rangle=
U_{\rm prep}|00\rangle.
\]
The operation $ U_{\rm prep}$ is shown in \fig{qcircuit} with the following preparation angles:
\begin{align}
\cos\frac{\theta_0}{2}
&=\sqrt{\beta_0+\beta_{2}},
&
\sin\frac{\theta_0}{2}
&=\sqrt{2\beta_{1}},
\\
\cos\frac{\theta_1}{2}
&=\sqrt{\frac{\beta_0}{\beta_0+\beta_2}},
&
\sin\frac{\theta_1}{2}
&=\sqrt{\frac{\beta_2}{\beta_0+\beta_2}},
\label{eq:lcu-prep-angles}
\end{align}
together with
\begin{align}
\mu\equiv\frac{\theta_1+\pi/2}{2},
\qquad
\nu\equiv\frac{\theta_1-\pi/2}{2}.
\label{eq:lcu-multiplex-angles}
\end{align}
Throughout this section, $R_y(\varphi)\equiv e^{-i\varphi Y/2}$ denotes the
standard circuit gate.

The Linear Combination of Unitaries (LCU) construction encodes the operator terms into a controlled unitary. We define the SELECT operator as
\begin{equation}
U_{\rm select}
=
\sum_{j=0}^{3}
e^{i\phi_j} |j\rangle\langle j|
\otimes
P_j,
\end{equation}
which applies the unitary $e^{i\phi_j} P_j$ to the system register, conditioned on the ancilla state $|j\rangle$. The SELECT operator factorizes exactly as
\begin{align}
U_{\rm select}
&=
\Phi_a\,
CZ(a_1, q_1)\,
CZ(a_2, q_2),
&
\Phi_a
&\equiv
\operatorname{diag}(1,-1,-1,-1)
=Z(a_1)Z(a_2)\,CZ(a_1, a_2),
\end{align}
where the index of the qubit the operation is acting on is explicitly shown in the parenthesis. The first index is for the control qubit, while the second for target qubit for the $CZ$ gates. We thus found that the $U_{\rm select}$ can be implemented purely by Clifford operation. The full LCU unitary is
\begin{equation}
U_{\rm LCU}
=\left(U^\dagger_{\rm prep} \otimes U^\dagger_{\rm diag}\right)
U_{\rm select}
\left(U_{\rm prep} \otimes U_{\rm diag}\right).
\end{equation}

When projecting to the $|00\rangle$ state of the ancilla qubits, the action of $U_{\rm LCU}$ yields
\[|\psi\rangle_s
\;\longrightarrow\; \frac{\mathcal{M}_{\rm H}}{\Lambda} |\psi_s\rangle.
\]

As $U_{\rm diag}$ can be implemented with Clifford operation and $U_{\rm select}$ can be implemented with purely Clifford gates, the $T$-count of the complete LCU circuit for $\mathcal{M}_{\rm H}$ arises entirely from approximating the required single-qubit rotations in $U_{\rm prep}$, which is $3\log_2(1/\delta)+\mathcal O(\log\log(1/\delta))$ for each rotation gate~\cite{Ross:2014okw}.
Thus allocating
an overall error budget $\epsilon$ uniformly among them gives the
construction-specific estimate
\begin{align}
T_{\rm H}^{\rm(circ)}
=18\log_2\!\left(\frac{6}{\epsilon}\right)
+\mathcal O\!\left(\log\log\frac1\epsilon\right).
\label{eq:lcu-t-count-helicity}
\end{align}

Similarly, the lab-basis scattering matrix admits the decomposition
\begin{align}
\mathcal{M}_{\rm L}
=
\sum_{j=0}^{7}
\alpha^{l}_j(\theta)P_j,
\qquad
P_j
=
\{II,ZZ,XX,YY,IY,YI,XZ,ZX\},
\end{align}
which contains eight Pauli strings and would naively require three ancilla qubits in an LCU implementation. The fact that the two bases are related by a unitary transformation implies a two-ancilla implementation of $\mathcal M_{\rm L}$,
\begin{align}
\mathcal M_{\rm L}
&=R_f(\theta)\,\mathcal M_{\rm H}R_i^\dagger,
&
R_i
&=I\otimes R_y(\pi),
&
R_f(\theta)
&=R_y(\theta)\otimes R_y(\theta+\pi).
\label{eq:lcu-basis-map}
\end{align} 
Since multiplication by a unitary operator can be implemented coherently without increasing the LCU branch complexity, the implementation of $\mathcal{M}_{\rm L}$ should require the same number of ancilla qubits as $\mathcal{M}_{\rm H}$. 
For a generic scattering angle, $R_f(\theta)$ adds two more generic $R_y$
rotations. We therefore have
\begin{align}
T_{\rm L}^{\rm(circ)}-T_{\rm H}^{\rm(circ)}
&=6\log_2\!\left(\frac1\epsilon\right)
+\mathcal O\!\left(\log\log\frac1\epsilon\right).
\label{eq:lcu-t-count-difference}
\end{align}
Indeed the circuit realization in the lab basis at a generic scattering angle has a higher estimated $T$-gate count than in the helicity basis. It is worth pointing out that,  at $\theta=\pi/2$ and $\theta=\pi$,
$R_f(\theta)$ is Clifford, and the additional $T$-gate overhead from the basis transformation vanishes.  At
$\theta=\pi$, the PREP angles are also Clifford, so the complete displayed
helicity circuit has zero $T$ gates. These are consistent with what we observed in \tab{magic-distribution-eeee}.

\section{Conclusions}

In this work we have examined how the magic produced in high energy collisions of $2\to2$ charged-lepton scattering depends on the choice of spin-projection axis, by comparing the lab basis used in Refs.~\cite{Liu:2025qfl,Liu:2025bgw} with the helicity basis. The motivation is rooted in a basic property of the stabilizer R\'enyi entropy: because $M_2$ is constructed from the expectation values of Pauli strings, it is invariant only under Clifford operations and not under a general rotation of the spin-projection axis. The magic assigned to a scattering final state is therefore dependent of the reference frame. Indeed, most physical observables such as spin, energy, and momentum are frame-dependent. Nevertheless it is important to understand how different choices of coordinate systems affect the observables.


We first established, in Sec.~\ref{sec:basis}, how much magic is generated from a rotation of coordinate system. Acting on the 60 stabilizer states, a single rotation $R_y(\theta)$ produces only four distinct angular patterns, and the magic it generates is bounded by $\log(16/9)$---strictly below the two-qubit maximum $\log(16/7)$ of \eq{M2def}. Since the transformation between the lab and helicity bases is precisely such a rotation accompanied by a Clifford operation, a portion of the difference in magic between the two bases is purely kinematic, set by the scattering angle rather than by the dynamics of the interaction.

For the QED processes studied in Sec.~\ref{sec:qed}, the magic is generally smaller in the helicity basis in the ultra-relativistic limit.  We attribute this to the helicity selection rules~\cite{Cheung:2015aba}, which force many helicity amplitudes to vanish in the massless limit, leaving the final state supported on a restricted set of helicity configurations. In the non-relativistic limit the ordering reverses: the lab basis tends to produce less magic, since the final states approach well-defined spin configurations and hence a reduced degree of superposition. In several cases the two bases yield identical magic distributions, which we traced to final states that differ only by a global Clifford rotation and therefore share the same Pauli spectrum. These observations bear directly on the basis-independent, or non-local, magic~\cite{Robin:2025ymq}, which is expected to follow---or coincide with---whichever basis carries the smaller magic: the helicity basis at high energy and the lab basis in the non-relativistic regime.

For the electroweak processes of Sec.~\ref{sec:ew}, we find that the observation of Ref.~\cite{Liu:2025bgw}---that the measured weak mixing angle $\hat{s}_W^2$ sits at a value which closely minimizes the magic production---persists in both bases. While individual stabilizer states show varying sensitivity to $s_W^2$, the magic-minimizing value deviates substantially from $s_W^2 = 1/4$ toward the Standard Model value in either basis, and the agreement improves with increasing center-of-mass energy. At $\sqrt{s} = m_Z$ the lab basis reproduces $\hat{s}_W^2$  more accurately, but the qualitative conclusion is insensitive to the frame. This robustness strengthens the interpretation that the electroweak sector tends to generate minimal quantum resources, and shows that the determination is not an artifact of the lab basis.

Finally, the basis dependence has a concrete computational counterpart, developed in Sec.\ref{sec:lcu-circuit}. The helicity-basis scattering matrix for high energy Bhabha scattering is considerably sparser than its lab-basis counterpart, and within the LCU framework this sparseness translates into a lower $T$-gate cost; because the two matrices are related by a unitary, both can be implemented with the same number of ancilla qubits. The basis that carries less magic is thus also the one that is cheaper to simulate on a fault-tolerant quantum computer, making explicit the link between the magic content of a process and the resources required to realize it.

Several directions remain open. Our reference frames are fixed on an event-by-event basis, so that the amplitudes carry no azimuthal dependence; a more general frame with the transverse axes held fixed would introduce dependence on both $\theta$ and $\phi$ and deserves separate study. It would also be worthwhile to extend the analysis beyond tree level, to non-abelian interactions, and to the full Standard Model, and to sharpen the connection between non-local magic and frame-independent statements about the quantum resources generated by fundamental interactions.

\section*{Acknowledgements}
Z. Y. thanks Zhi-Heng Shen and Hao-Lin Zhang for useful discussions. Y.-Y. L. is supported by the National Key
R\&D Program of China (Grant Nos. 2025YFA1614200), the National Science Foundation
of China under Grant Nos. 12522509, 12305107, and IHEP under Grant No. E55153U1. I. L. is supported
in part by the U.S. Department of Energy under contracts DE-AC02-06CH11357 (Argonne), DE-SC0023522 (Northwestern), DE-SC0010143 (Northwestern), and No. 89243024CSC000002 (QuantISED Program). Z. Y. is supported by IHEP under Grant No. E5515AU1.

\appendix
\section{Two-qubit Stabilizer States}
\label{sec:state-list}
In this appendix, we present 60 2-qubit stabilizer states in the computational basis $\{|00\rangle\,|01\rangle,|10\rangle,|11\rangle\}$, see Table \ref{tab:state_list}

\begin{table}
\centering
\begin{tabular}{|c|c|c|c|c|c|}
\hline
$\mathcal{S}$ & $\lvert\psi_s\rangle$ &
$\mathcal{S}$ & $\lvert\psi_s\rangle$ &
$\mathcal{S}$ & $\lvert\psi_s\rangle$
\\
\hline

$\langle IX,XI\rangle$ & $(1,1,1,1)$ &
$\langle IY,ZI\rangle$ & $(1,i,0,0)$ &
$\langle XY,YX\rangle$ & $(1,0,0,i)$
\\

$\langle IX,-XI\rangle$ & $(1,-1,1,-1)$ &
$\langle IY,-ZI\rangle$ & $(1,-i,0,0)$ &
$\langle XY,-YX\rangle$ & $(0,1,i,0)$
\\

$\langle -IX,XI\rangle$ & $(1,1,-1,-1)$ &
$\langle -IY,ZI\rangle$ & $(0,0,1,i)$ &
$\langle -XY,YX\rangle$ & $(0,1,-i,0)$
\\

$\langle -IX,-XI\rangle$ & $(1,-1,-1,1)$ &
$\langle -IY,-ZI\rangle$ & $(0,0,1,-i)$ &
$\langle -XY,-YX\rangle$ & $(1,0,0,-i)$
\\

$\langle IX,YI\rangle$ & $(1,1,i,i)$ &
$\langle IZ,XI\rangle$ & $(1,0,1,0)$ &
$\langle XZ,YY\rangle$ & $(1,1,1,-1)$
\\

$\langle IX,-YI\rangle$ & $(1,-1,i,-i)$ &
$\langle IZ,-XI\rangle$ & $(0,1,0,1)$ &
$\langle XZ,-YY\rangle$ & $(1,1,-1,1)$
\\

$\langle -IX,YI\rangle$ & $(1,1,-i,-i)$ &
$\langle -IZ,XI\rangle$ & $(1,0,-1,0)$ &
$\langle -XZ,YY\rangle$ & $(1,-1,1,1)$
\\

$\langle -IX,-YI\rangle$ & $(1,-1,-i,i)$ &
$\langle -IZ,-XI\rangle$ & $(0,1,0,-1)$ &
$\langle -XZ,-YY\rangle$ & $(1,-1,-1,-1)$
\\

$\langle IX,ZI\rangle$ & $(1,1,0,0)$ &
$\langle IZ,YI\rangle$ & $(1,0,i,0)$ &
$\langle XZ,YX\rangle$ & $(1,i,1,-i)$
\\

$\langle IX,-ZI\rangle$ & $(1,-1,0,0)$ &
$\langle IZ,-YI\rangle$ & $(0,1,0,i)$ &
$\langle XZ,-YX\rangle$ & $(1,i,-1,i)$
\\

$\langle -IX,ZI\rangle$ & $(0,0,1,1)$ &
$\langle -IZ,YI\rangle$ & $(1,0,-i,0)$ &
$\langle -XZ,YX\rangle$ & $(1,-i,1,i)$
\\

$\langle -IX,-ZI\rangle$ & $(0,0,1,-1)$ &
$\langle -IZ,-YI\rangle$ & $(0,1,0,-i)$ &
$\langle -XZ,-YX\rangle$ & $(1,-i,-1,-i)$
\\

$\langle IY,XI\rangle$ & $(1,i,1,i)$ &
$\langle IZ,ZI\rangle$ & $(1,0,0,0)$ &
$\langle XY,YZ\rangle$ & $(1,1,i,-i)$
\\

$\langle IY,-XI\rangle$ & $(1,-i,1,-i)$ &
$\langle IZ,-ZI\rangle$ & $(0,1,0,0)$ &
$\langle XY,-YZ\rangle$ & $(1,1,-i,i)$
\\

$\langle -IY,XI\rangle$ & $(1,i,-1,-i)$ &
$\langle -IZ,ZI\rangle$ & $(0,0,1,0)$ &
$\langle -XY,YZ\rangle$ & $(1,-1,i,i)$
\\

$\langle -IY,-XI\rangle$ & $(1,-i,-1,i)$ &
$\langle -IZ,-ZI\rangle$ & $(0,0,0,1)$ &
$\langle -XY,-YZ\rangle$ & $(1,-1,-i,-i)$
\\

$\langle IY,YI\rangle$ & $(1,i,i,-1)$ &
$\langle XX,YY\rangle$ & $(0,1,1,0)$ &
$\langle XX,YZ\rangle$ & $(1,i,i,1)$
\\

$\langle IY,-YI\rangle$ & $(1,-i,i,1)$ &
$\langle XX,-YY\rangle$ & $(1,0,0,-1)$ &
$\langle XX,-YZ\rangle$ & $(1,i,-i,-1)$
\\

$\langle -IY,YI\rangle$ & $(1,i,-i,1)$ &
$\langle -XX,YY\rangle$ & $(1,0,0,1)$ &
$\langle -XX,YZ\rangle$ & $(1,-i,i,-1)$
\\

$\langle -IY,-YI\rangle$ & $(1,-i,-i,-1)$ &
$\langle -XX,-YY\rangle$ & $(0,1,-1,0)$ &
$\langle -XX,-YZ\rangle$ & $(1,-i,-i,1)$
\\

\hline
\end{tabular}
\caption{The 60 stabilizer states represented using maximal abelian subgroups of $\mathcal{P}_2$. The explicit expressions of $\lvert\psi_s\rangle$ are up to an overall normalization factor.}
\label{tab:state_list}
\end{table}

\section{Scattering Amplitudes in the Spin-space}
\label{sec:scattering}
\subsection{QED in Ultra-relativistic Limit}
We present the scattering matrix $\mathcal{M}_{\rm L}$ and $\mathcal{M}_{\rm H}$ in the lab basis and in the helicity basis, respectively. The lab basis ordered by the spin orientation of the two particles as \{$\ket{\uparrow\uparrow}, \ket{\uparrow\downarrow},\ket{\downarrow\uparrow}, \ket{\downarrow\downarrow}$\}, and the helicity basis is ordered by the helicities of the two particles as \{$\ket{++}, \ket{+-},\ket{-+}, \ket{--}$\}. For all the cases discussed below, the zero entries of $\mathcal{M}_{\rm H}$ correspond to the helicity selection rules.

\textbf{Process (A)} Bhabha scattering: 
\begin{align}
\label{eq:bbnrM}
    \mathcal{M}_{\rm L}&=\left(
\begin{array}{cccc}
 2 \cot ^2\frac{\theta }{2} & \sin \theta (\cos \theta
   +1)-2 \cot \frac{\theta }{2} & \sin \theta (\cos
   \theta+1)-2 \cot \frac{\theta }{2} & 2 \\
 2 \cot \frac{\theta }{2} & \frac{15 \cos \theta+\cos 3
   \theta }{4-4 \cos \theta} & -2 \cos ^2\frac{\theta
   }{2} \cos \theta & -2 \cot \frac{\theta }{2} \\
 2 \cot \frac{\theta }{2} & -2 \cos ^2\frac{\theta
   }{2} \cos \theta & \frac{15 \cos \theta+\cos 3 \theta
   }{4-4 \cos \theta} & -2 \cot \frac{\theta }{2} \\
 2 & 2 \cot \frac{\theta }{2}-\sin \theta (\cos \theta
   +1) & 2 \cot \frac{\theta }{2}-\sin \theta (\cos
   \theta+1) & 2 \cot ^2\frac{\theta }{2} 
\end{array}
\right), 
\end{align}
\begin{align}
\mathcal{M}_{\rm H}=
\left(
\begin{array}{cccc}
 \frac{1}{8} \sin ^4\theta  \csc ^6\frac{\theta }{2} & 0 &
   0 & \cos \theta -1 \\
 0 & 2 \csc ^2\frac{\theta }{2} & 0 & 0 \\
 0 & 0 & 2 \csc ^2\frac{\theta }{2} & 0 \\
 \cos \theta -1 & 0 & 0 & \frac{1}{8} \sin ^4\theta  \csc
   ^6\frac{\theta }{2} \\
\end{array}
\right).
\end{align} 

\textbf{Process (B)} $e^-e^-\to e^-e^-$: 
\begin{align}
   \mathcal{M}_{\rm L}= \left(
\begin{array}{cccc}
 (\cos 2 \theta +7) \cot \theta  \csc \theta  & -4 \csc \theta  & -4 \csc \theta
    & 2 \cos \theta  \\
 (\cos 2 \theta +3) \csc \theta  & 2 \csc ^2\frac{\theta }{2} &
   -\frac{4}{\cos \theta +1} & 2 \sin \theta -4 \csc \theta  \\
 (\cos 2 \theta +3) \csc \theta  & -\frac{4}{\cos \theta +1} & 2 \csc
   ^2\frac{\theta }{2} & 2 \sin \theta -4 \csc \theta  \\
 2 \cos \theta  & 4 \csc \theta  & 4 \csc \theta  & (\cos 2 \theta +7) \cot
   \theta  \csc \theta  \\
\end{array}
\right),
\end{align}
\begin{align}
   \mathcal{M}_{\rm H}=\left(
\begin{array}{cccc}
 8 \csc ^2\theta  & 0 & 0 & 0 \\
 0 & 2 \cot ^2\frac{\theta }{2} & 2 \tan ^2\frac{\theta }{2} &
   0 \\
 0 & 2 \tan ^2\frac{\theta }{2} & 2 \cot ^2\frac{\theta }{2} &
   0 \\
 0 & 0 & 0 & 8 \csc ^2\theta  \\
\end{array}
\right).
\end{align}

\textbf{Process (C)} $e^-\mu^-\to e^-\mu^-$: 
\begin{align}
   \mathcal{M}_{\rm L}= \left(
\begin{array}{cccc}
 \cos \theta +\frac{4}{\cos \theta -1}+3 & 2 \cot \frac{\theta }{2} & 2
   \cot \frac{\theta }{2} & -\cos \theta -1 \\
 \sin \theta -2 \cot \frac{\theta }{2} & -2 \cot ^2\frac{\theta
   }{2} & 2 & 2 \cot \frac{\theta }{2}-\sin \theta  \\
 \sin \theta -2 \cot \frac{\theta }{2} & 2 & -2 \cot ^2\frac{\theta
   }{2} & 2 \cot \frac{\theta }{2}-\sin \theta  \\
 -\cos \theta -1 & -2 \cot \frac{\theta }{2} & -2 \cot \frac{\theta
   }{2} & \cos \theta +\frac{4}{\cos \theta -1}+3 \\
\end{array}
\right),
\end{align}
\begin{align}
   \mathcal{M}_{\rm H}=\left(
\begin{array}{cccc}
 \frac{4}{\cos \theta -1} & 0 & 0 & 0 \\
 0 & -2 \cot ^2\frac{\theta }{2} & 0 & 0 \\
 0 & 0 & -2 \cot ^2\frac{\theta }{2} & 0 \\
 0 & 0 & 0 & \frac{4}{\cos \theta -1} \\
\end{array}
\right).
\end{align}

\textbf{Process (D)} $e^+e^-\to \mu^+\mu^-$: 
\begin{align}
   \mathcal{M}_{\rm L}=
\left(
\begin{array}{cccc}
 -\frac{\sin ^2\theta }{l} & \frac{\sin \theta  \cos \theta  (\lambda 
   l-1)}{\lambda  l} & \frac{\sin \theta  \cos \theta  (\lambda  l-1)}{\lambda  l} &
   \frac{\sin ^2\theta }{l} \\
 \frac{\sin \theta  \cos \theta }{l} & \frac{\cos (2 \theta ) (1-\lambda  l)-3
   \lambda  l-1}{2 \lambda  l} & \frac{\sin ^2\theta  (\lambda  l-1)}{\lambda  l} &
   -\frac{\sin \theta  \cos \theta }{l} \\
 \frac{\sin \theta  \cos \theta }{l} & \frac{\sin ^2\theta  (\lambda  l-1)}{\lambda
    l} & \frac{\cos (2 \theta ) (1-\lambda  l)-3 \lambda  l-1}{2 \lambda  l} &
   -\frac{\sin \theta  \cos \theta }{l} \\
 \frac{\sin ^2\theta }{l} & \frac{\sin \theta  \cos \theta  (1-\lambda  l)}{\lambda
    l} & \frac{\sin \theta  \cos \theta  (1-\lambda  l)}{\lambda  l} & -\frac{\sin
   ^2\theta }{l} \\
\end{array}
\right),
\end{align}
\begin{align}
   \mathcal{M}_{\rm H}=\left(
\begin{array}{cccc}
 -\cos \theta -1 & -\frac{\sin \theta }{l} & -\frac{\sin \theta }{l} & \cos (\theta
   )-1 \\
 \frac{\sin \theta }{\lambda  l} & 0 & 0 & -\frac{\sin \theta }{\lambda  l} \\
 \frac{\sin \theta }{\lambda  l} & 0 & 0 & -\frac{\sin \theta }{\lambda  l} \\
 \cos \theta -1 & \frac{\sin \theta }{l} & \frac{\sin \theta }{l} & -\cos \theta
   -1 \\
\end{array}
\right).
\end{align}
To ensure a non-vanishing normalization in the calculation of the magic, we expand the scattering matrix to $\mathcal{O}(1/l)$, with $l \equiv |p_e|/m_e$ and $\lambda \equiv m_e/m_\mu$. 

\subsection{QED in Non-relativistic limit}
\textbf{Process (B)} $e^-e^-\to e^-e^-$:
\begin{align}
   \mathcal{M}_{\rm L}= \left(
\begin{array}{cccc}
 4 \cot \theta  \csc \theta  & 0 & 0 & 0 \\
 0 & \csc ^2\frac{\theta }{2} & -\frac{2}{\cos \theta +1} & 0 \\
 0 & -\frac{2}{\cos \theta +1} & \csc ^2\frac{\theta }{2} & 0 \\
 0 & 0 & 0 & 4 \cot \theta  \csc \theta  \\
\end{array}
\right),
\end{align}
\begin{align}
   \mathcal{M}_{\rm H}=\left(
\begin{array}{cccc}
 4 \csc ^2\theta -2 & 2 \cot \theta  & 2 \cot \theta  & 2 \\
 \tan \frac{\theta }{2}-\cot \frac{\theta }{2} & \cos \theta 
   \csc ^2\frac{\theta }{2} & \tan ^2\frac{\theta }{2}-1 & 2
   \cot \theta  \\
 \tan \frac{\theta }{2}-\cot \frac{\theta }{2} & \tan
   ^2\frac{\theta }{2}-1 & \cos \theta  \csc ^2\frac{\theta
   }{2} & 2 \cot \theta  \\
 2 & \tan \frac{\theta }{2}-\cot \frac{\theta }{2} & \tan
   \frac{\theta }{2}-\cot \frac{\theta }{2} & 4 \csc ^2\theta
   -2 \\
\end{array}
\right).
\end{align}

\textbf{Process (D)} $\mu^+\mu^-\to e^+e^-$ in lab basis:
\begin{align}
\mathcal A_{\uparrow\uparrow\to\uparrow\uparrow}
&=
\mathcal A_{\downarrow\downarrow\to\downarrow\downarrow}
=
\frac12\Big[(\lambda-1)\cos2\theta-\lambda-3\Big],
\\[1ex]
\mathcal A_{\uparrow\uparrow\to\uparrow\downarrow}
&=
\mathcal A_{\uparrow\downarrow\to\uparrow\uparrow}
=
\mathcal A_{\uparrow\downarrow\to\downarrow\downarrow}
=
\mathcal A_{\downarrow\downarrow\to\uparrow\downarrow}
=
-\mathcal A_{\uparrow\uparrow\to\downarrow\uparrow}\nonumber\\
&
=
-\mathcal A_{\downarrow\uparrow\to\uparrow\uparrow}=
-\mathcal A_{\downarrow\uparrow\to\downarrow\downarrow}
=
-\mathcal A_{\downarrow\downarrow\to\downarrow\uparrow}
=
-(\lambda-1)\sin\theta\cos\theta,
\\[1ex]
\mathcal A_{\uparrow\uparrow\to\downarrow\downarrow}
&=
\mathcal A_{\downarrow\downarrow\to\uparrow\uparrow}
=
-(\lambda-1)\sin^2\theta,
\\[1ex]
\mathcal A_{\uparrow\downarrow\to\uparrow\downarrow}
&=
\mathcal A_{\downarrow\uparrow\to\downarrow\uparrow}
=
-\frac12\Big[(\lambda-1)\cos2\theta+\lambda+1\Big],
\\[1ex]
\mathcal A_{\uparrow\downarrow\to\downarrow\uparrow}
&=
\mathcal A_{\downarrow\uparrow\to\uparrow\downarrow}
=
\frac12\Big[(\lambda-1)\cos2\theta+\lambda+1\Big].
\end{align}

\textbf{Process (D)} $\mu^+\mu^-\to e^+e^-$ in helicity basis:
\begin{align}
\mathcal A_{++\to++}
&=
\mathcal A_{++\to--}
=
\mathcal A_{--\to++}
=
\mathcal A_{--\to--}
=
-\frac12\Big[(\lambda-1)\cos2\theta+\lambda+1\Big],
\\[1ex]
\mathcal A_{+-\to++}
&=
\mathcal A_{--\to+-}
=
-\mathcal A_{-+\to++}
=
-\mathcal A_{--\to-+}
=(\lambda-1)\sin\theta\cos\theta,
\\
\mathcal A_{++\to+-}
&=
\mathcal A_{+-\to--}
=
\sin\theta\Big[(\lambda-1)\cos\theta-1\Big],
\\
\mathcal A_{++\to-+}
&=
\mathcal A_{-+\to--}
=
-\sin\theta\Big[(\lambda-1)\cos\theta+1\Big],
\\
\mathcal A_{+-\to-+}
&=
\mathcal A_{-+\to+-}
=
2\sin^2\frac{\theta}{2}
\Big[(\lambda-1)\cos\theta+\lambda\Big],
\\
\mathcal A_{+-\to+-}
&=
\mathcal A_{-+\to-+}
=
2\cos^2\frac{\theta}{2}
\Big[(\lambda-1)\cos\theta-\lambda\Big].
\end{align}

\bibliography{ref}

\end{document}